%% file: ms.tex
\documentclass[modern]{aastex62}

\input{cortex}

\input{style}

\usepackage{etoolbox}
\makeatletter 
\patchcmd{\env@cases}{\quad}{\qquad\qquad}{}{}
\makeatother

\defcitealias{PaperI}{Paper I}

\usepackage{enumitem}

\begin{document}

\title{%
    \vspace{-3em}
    \textbf{
        Mapping stellar surfaces\\
        II: An interpretable Gaussian process model for light curves
    }
}

\author[0000-0002-0296-3826]{Rodrigo Luger}\altaffiliation{Flatiron Fellow}
\email{rluger@flatironinstitute.org}
\affil{Center~for~Computational~Astrophysics,~Flatiron~Institute,~New~York,~NY}
\affil{Virtual~Planetary~Laboratory, University~of~Washington, Seattle, WA}
\author[0000-0002-9328-5652]{Daniel Foreman-Mackey}
\affil{Center~for~Computational~Astrophysics,~Flatiron~Institute,~New~York,~NY}
\author[0000-0002-3385-8391]{Christina Hedges}
\affil{Bay~Area~Environmental~Research~Institute,~Moffett~Field,~CA}
\affil{NASA~Ames~Research~Center,~Moffett~Field,~CA}


\features{open-source figures \codeicon; equation unit tests: \input{tests/tally}}

\begin{abstract}
    The use of Gaussian processes (GPs) as models for astronomical time series datasets has recently
    become almost ubiquitous, given their ease of use and flexibility.
    GPs excel in particular at marginalization
    over the stellar signal in cases where the
    variability due to starspots rotating in and out of view is treated
    as a nuisance, such as in exoplanet transit modeling.
    However, these effective models are less useful in cases where the starspot signal is of primary interest since it is not obvious how the parameters of the GP model are related to the physical properties of interest, such
    as the size, contrast, and latitudinal distribution of the spots.
    Instead, it is common practice to explicitly model the effect
    of individual starspots on the light curve and attempt to infer their
    properties via optimization or posterior inference. Unfortunately,
    this process is
    degenerate, ill-posed, and often computationally intractable when
    applied to stars with more than a few spots and/or to ensembles of many
    light curves.
    In this paper, we derive a closed-form expression for the
    mean and covariance of a Gaussian process model that describes
    the light curve of a rotating, evolving stellar surface
    conditioned on a given distribution of
    starspot sizes, contrasts, and latitudes.
    We demonstrate that this model is correctly calibrated,
    allowing one to robustly infer physical parameters of interest from one
    or more stellar light curves, including
    the typical radii and the mean and variance of the latitude
    distribution of starspots.
    Our GP has far-ranging implications for understanding the variability and
    magnetic activity of stars from both light curves and radial velocity (RV)
    measurements, as well as for robustly modeling correlated noise in both transiting
    and RV exoplanet searches.
    Our implementation is efficient, user-friendly, and open source, available
    as the \Python package \starryprocess.
    \href{https://github.com/rodluger/starry_process}{\color{linkcolor}\faGithub}
\end{abstract}

\section{Introduction}
\label{sec:intro-gps}

Over the past two decades, Gaussian processes \citep[GPs;][]{RasmussenWilliams2005} have gained traction as
a leading tool for modeling correlated signals in astronomical datasets.
In particular, GPs are commonly used to model
stellar variability in photometric time series
\citep[e.g.,][]{BrewerStello2009,Aigrain2016,Luger2016,ForemanMackey2017,Angus2018}
and radial velocity measurements \citep[e.g.,][]{Rajpaul2015,Jones2017,Perger2020}.
GPs are popular models for these applications because they allow marginalization
over a stochastic noise process specified only by a kernel describing its autocorrelation structure.
There are several popular open source implementations that allow
efficient evaluation of GPs, and these have been widely demonstrated
to be useful effective models for the time series when the stochastic
variability due to the star is primarily a nuisance
\citep[e.g.,][]{Ambikasaran2015,ForemanMackey2017,Gilbertson2020}.


A major source of stellar variability in both light curves and radial velocity
datasets is the modulation induced by magnetically-driven surface features like starspots
rotating in and out of view. While GPs excel at marginalizing over stellar rotational
variability, they have been less useful when the goal is to make
inferences about the actual source of this variability, such as the
properties of starspots and the magnetic processes that generate them.
\textbf{While it is straightforward to derive
    posterior constraints on the hyperparameters of an effective GP model for
    observations of a star, it is not clear what those constraints actually
    tell us about the stellar surface.}
Specifically, in all but a few restricted cases, there is no \emph{first principles}
relationship between the descriptive parameters of a typical GP model
(see \S\ref{sec:gp-intro})
and the physical properties of the stellar surface that is being observed.
For instance, it may be tempting to interpret the GP amplitude hyperparameter
as some measure of the spot contrast or the total number of spots,
or the GP timescale hyperparameter as the
spot lifetime, but there are no guarantees these interpretations will
hold in general. After all, the choice of kernel is quite often \emph{ad hoc},
providing an \emph{effective}---as opposed to \emph{interpretable}---description
of the physics.
There are two important exceptions to this: asteroseismic studies, in which the
the GP hyperparameters can offer direct insight into
the behavior of complex pulsation modes and thus physical properties of the
stellar interior \citep[e.g.,][]{BrewerStello2009,ForemanMackey2017}; and
stellar rotation period studies, in which the period hyperparameter $P$ can usually
be associated with the rotation period of the star \citep[e.g.,][]{Angus2018}.%
\footnote{An exception to this is in the presence of strong differential
    rotation, in which case many periods may be present in the data, or
    when spots evolve coherently, which can also introduce weak periodicities
    in the light curve.}
For spot-induced variability, on the other hand,
GPs are usually used
when the variability itself is a nuisance parameter. For example, if the
goal is to constrain the properties of a transiting exoplanet or to
search for a planetary signal in a radial velocity dataset, a GP might be
used to remove (or, better yet, to marginalize over) the stellar variability
\citep[e.g.,][]{Haywood2014,Rajpaul2015,Luger2017b}.
In this case, the physics behind the variability is irrelevant, so an
effective model of this sort may be sufficient.

However, understanding the properties of stellar surfaces and starspots
in particular is a crucial step
toward understanding stellar magnetism, which plays a fundamental part
in stellar interior structure and evolution. Stellar magnetic fields
control the spin-down of stars over time, on which the field of
gyrochronology is founded \citep[][]{Barnes2001,Angus2019}. They affect
wave propagation in stellar interiors and must be properly
understood to interpret asteroseismic measurements \citep[e.g.,][]{Fuller2015}.
Strong magnetic fields are also likely the driving force behind chemical
peculiarity in Ap/Bp stars~\citep{Turcotte2003,Sikora2018}, as well as
radius inflation in M dwarfs \citep{Gough1966,Ireland2018}.
Stellar magnetohydrodynamics (MHD) is therefore an active area of
research, with many open questions \citep[e.g.,][]{Miesch2009}.
Because of the nonlinearity of the MHD equations and the vast range
of scales on which magnetic processes operate,
there is still significant theoretical uncertainty concerning how dynamos
operate in stars of different masses, how magnetic fields affect stellar rotation,
and how star spots form \citep{Yadav2015,Weber2016}.
Observational constraints on starspots and other magnetically-controlled
surface features are therefore extremely valuable to understanding various problems in
stellar astrophysics.

Moreover, even when the stellar signal is considered a nuisance, a
physically-driven variability model may be a better choice than an
effective model in some cases, particularly when the signal of interest is small
compared to the systematics. A specific example of this is in transmission
spectroscopy of transiting exoplanets, where the contribution from
unocculted spots and faculae to the spectrum
can be an order of magnitude larger than that of the planet atmosphere
\citep{Rackham2018}. In this case, failure to explicitly model the effect of starspots
can lead to spurious features in the planet spectrum.
A similar situation arises in extreme precision radial velocity (EPRV)
searches for planets, where the stellar signal can be orders of magnitude
larger than the planetary signal. While effective models of
variability have often been successful at disentangling the planetary
and stellar contributions \citep[e.g.,][]{Rajpaul2015}, these models can struggle
when the (a priori unknown) orbital period of the planet
is close to an alias of the rotational period of the star \citep{Vanderburg2016,Damasso2019,Robertson2020}.
In this case, a physically-driven model of variability would likely
perform better.


When the goal is to learn about the stellar surface, the common approach
in the literature has not been to use GPs, but to explicitly
\emph{forward model} the surface. Such a model allows one to
compute a stellar light curve or spectral timeseries conditioned
on certain surface properties, a procedure that must then be inverted
in order to constrain the surface given a dataset.
We discussed this approach for rotational light curves of stars
in \citet{PaperI} \citepalias[hereafter][]{PaperI},
where we argued a unique solution to the surface map of
the star is not possible without the use of
aggressive (and often \emph{ad hoc}) priors. \textbf{The degeneracies at play
    make it effectively impossible for one to know the exact configuration
    of starspots and other features on the surface of a star from its rotational
    light curve alone.}

However, it is hardly ever the case that this is actually our end goal.
After all, physics can be used to predict properties of stellar surfaces at a fairly
high level: i.e., typical spot sizes, active spot latitudes, or approximate
timescales on which spots evolve
\citep[e.g.,][]{Schuessler1996,Solanki2006,Cantiello2019}.
We are hardly ever interested in the
\emph{particular} properties of a \emph{particular} spot, as we wouldn't really
know what to do with that information! Instead, we often treat
(whether explicitly or not)
the properties of a starspot as a draw from some parent distribution
controlling (say) the average and spread in the radii of the spots.
The parameters controlling this distribution are the ones that we can
predict with physics; they are therefore also the ones we are usually
interested in.

Thus, if it were possible to derive robust posterior constraints
on the properties of each of the spots on a star, we could then
\emph{marginalize} (integrate) over them to infer the properties
describing the distribution of all the spots as a whole.
We could do this using the forward model approach described above, by
modeling the properties of each of the spots and computing the
corresponding light curves. Then, we could solve the ``inverse'' problem
via a posterior sampling scheme, such as Markov Chain Monte Carlo (MCMC),
while including a few
\emph{hyperparameters} controlling the distribution of those properties
across all spots: i.e., a one-level hierarchical model.
The marginal posteriors for the hyperparameters, then,
would encode what we actually wish to know.
In practice, however, the degeneracies and often extreme multi-modality
of the distributions of individual spot properties would make this
quite hard (and expensive) to perform.
If only we could use the elegant machinery of Gaussian processes to
perform this marginalization for us!

\textbf{In this paper, we derive an exact, closed-form expression for
    the Gaussian approximation to the marginal likelihood of a light curve
    conditioned on the statistical properties of starspots,} which allows us
define an interpretable Gaussian process for stellar light curves.
Our GP analytically marginalizes over the degenerate and often unknowable
distributions of properties of individual starspots, revealing the
constraints imposed on the bulk spot properties without the need to
explicitly model or sample over properties of individual spots. It inherits
the speed, ease-of-use, and all other properties of traditionally-used
GPs, with the added benefit of direct physical interpretability of its
hyperparameters.

While our GP can be used to model light curves of individual
stars, it is particularly useful for \textbf{ensemble analyses of light curves of
    many similar stars.}
As we showed in \citetalias{PaperI}, the joint information content of
the light curves of many similar stars can be harnessed to constrain
statistical properties of the surfaces of those stars, even in the
presence of degeneracies that preclude knowledge about the surfaces
of individual stars.
By ``similar'', we do not mean stars that \emph{look} similar, but
whose spot properties are drawn from
the same parent distribution. The parameters of this parent distribution
are the ones we can constrain; the are also usually the physically
interesting ones, such as the typical spot sizes or typical active
latitudes and the variance in those quantites across the
population.
Ensembles may thus comprise light curves of stars in
a narrow spectral type, metallicity, and rotation period bin,
which we might reasonably expect to have \emph{statistically}
similar surfaces.
We encourage readers to read \citetalias{PaperI} to
better understand this and other points regarding the information theory
behind stellar rotational light curves.

\vspace{1em}

The present paper is organized as follows:
we present an overview of the derivation of the GP in \S\ref{sec:gp}
and a suite of tests on synthetic data to show the model is calibrated
in \S\ref{sec:calibration}. We discuss our results and the limitations
of our model in \S\ref{sec:discussion} and present straightforward
extensions of the GP, including its application to time-variable
surfaces, in \S\ref{sec:extensions}.
In \S\ref{sec:conclusions} we summarize our results and discuss
topics we will address in future papers in this series.

Most of the math behind the algorithm
is presented in the Appendix, followed by a series of supplementary figures
(discussed in \S\ref{sec:calibration}). Appendix~\ref{sec:notation} discusses the
notation we adopt throughout the paper and Table~\ref{tab:variables}
lists the main symbols and variables, with links to their definitions.
The algorithm developed in this paper is fully implemented in the \starryprocess code, which is
available on \href{https://github.com/rodluger/starry_process}{GitHub}
and is described in more detail in \citet{JOSSPaper}.

Finally, we note that all of the figures in this paper were auto-generated
using the Azure Pipelines continuous integration (CI) service, which
ensures they are up to date with the latest version of the
\starryprocess code. In particular, icons next to each of the figures \codeicon \,
link to the exact script used to generate them to ensure the reproducibility
of our results. As in \citetalias{PaperI}, the principal equations are
accompanied by
``unit tests'': \textsf{pytest}-compatible test scripts associated
with the principal equations that pass (fail) if the equation is correct (wrong),
in which case a clickable \testpassicon \, (\testfailicon) is shown next to the equation
label.
In most cases, the validity of an equation is gauged by comparison to
a numerical solution. Like the figure scripts, the equation unit tests are
run on Azure Pipelines upon every commit of the code.

\section{A Gaussian Process for starspots}
\label{sec:gp}

In this section, we provide a brief overview of Gaussian processes
(\S\ref{sec:gp-intro}) and spherical harmonics (\S\ref{sec:gp-ylms}),
followed by an outline of the derivation of our interpretable GP
(\S\ref{sec:gp-gp}).
This derivation boils down to computing the mean and covariance of the stellar
flux conditioned on certain physical properties of the star and its starspot
distribution.
In \S\ref{sec:gp-inc} and \S\ref{sec:gp-norm} we derive useful extensions
of the model. \textbf{For convenience, we summarize the results of this entire section in
    \S\ref{sec:gp-summary}.}
Most of the math is folded into the Appendix for readability;
readers may want to refer to Appendix~\ref{sec:notation} in particular for
a discussion of the notation and conventions we adopt.

\subsection{Brief overview of Gaussian processes}
\label{sec:gp-intro}

Despite whatever mystique the words ``Gaussian process'' may evoke, a GP
is nothing but a Gaussian distribution in many (formally infinite)
dimensions. Specifically, it is a Gaussian distribution over
\emph{functions} spanning a continuous domain (in our case, the time domain).
Similar to a multivariate Gaussian, which is described by
a $(K \times 1)$ vector $\pmb{\mu}$ characterizing
the mean of the process and a $(K \times K)$
matrix $\pmb{\Sigma}$ characterizing its covariance,
a GP is fully specified by a mean
function $m(t)$ and a kernel function $k(t, t')$.
To say that a random vector-valued variable $\mathbbb{f}$
defined on a $(K \times 1)$ time array $\mathbf{t}$
is ``distributed as a GP'' means that we may write
\begin{align}
    \label{eq:fnormal}
    \mathbbb{f} \sim \mathcal{N}\left( \pmb{\mu}, \pmb{\Sigma} \right)
    \quad,
\end{align}
where the elements of the mean and covariance are given by
$\mu_i = m(t_i)$ and $\Sigma_{i,j} = k(t_i, t_j)$, respectively.%
\footnote{%
    In this paper, we will use blackboard font (i.e., $\mathbbb{f}$) to
    denote random variables and serif font (i.e., $\mathbf{f}$) to denote
    particular realizations of those variables.
    See Appendix~\ref{sec:notation} for a detailed explanation of our notation.
}
Because of this relationship to multivariate Gaussians,
GPs are easy to sample from.%
\footnote{Given a 1-d array \texttt{mean} and a 2-d array \texttt{cov} in \Python,
    sampling from the corresponding GP (if it exists)
    can be done in a single line of code by calling
    \texttt{numpy.random.multivariate\_normal(mean, cov)}.}
But, as we alluded to earlier, the real showstopper is the application of GPs to inference problems.
Multivariate Gaussian distributions have a closed-form (marginal) likelihood
function, so it is easy to compute the probability
of one's data conditioned on a given value of $\pmb{\mu}$ and $\pmb{\Sigma}$
(i.e., the ``likelihood''; see Equation~\ref{eq:log-like} below).
This can in turn be maximized
to infer the optimal values of the model parameters
or used in a
numerical sampling scheme to compute the probability of those parameters
given the data (i.e., the ``posterior'').
Thanks to modern computer architectures, linear algebra packages, and
GP algorithms,
evaluating the GP likelihood may typically be done in a fraction of a second
for a reasonably-sized dataset (i.e., $K \lesssim 10^4$ datapoints).

Another big advantage of GPs is their flexibility. GPs are often dubbed
a class of ``non-parametric'' models, given that nowhere in the specification
of the GP is there an explicit functional form for $\mathbbb{f}$. Rather, a GP
is a stochastic
process whose draws can in principle take on \emph{any} functional form,
subject, however, to certain smoothness and correlation criteria
of tunable strictness
that are fully
encoded in the covariance $\pmb{\Sigma}$.
In many applications, particularly when modeling stellar light curves,
it is customary to restrict the problem by
assuming that the process is \emph{stationary}, such that we may write
\begin{align}
    \label{eq:kernel}
    \Sigma_{i,j} & = k(t_i, t_j)
    \nonumber                                    \\
                 & = k(\left| t_i - t_j \right|)
    \nonumber                                    \\
                 & \equiv k(\Delta t)
    \quad.
\end{align}
A stationary process is one that is independent of phase (or, in this case,
the actual value of the time $t$); rather, it depends only on the \emph{difference}
between the phases of two data points. The kernel of a stationary process is
therefore a one-dimensional function, typically chosen from a set of
standard functions with desirable smoothness and spectral properties.

The GP we derive in this paper is stationary and admits a representation
as a one-dimensional kernel function. However, as we show in
\S\ref{sec:gp-norm}, the common practice of normalizing stellar light curves to
their mean or median value breaks this stationarity. For this reason,
it is more convenient to derive and present our GP covariance as a
$(K \times K)$ matrix $\pmb{\Sigma}$ and our GP mean as a
$(K \times 1)$ vector $\pmb{\mu}$ for arbitrary $K$ instead of
as a kernel and a mean function. Note, importantly, that these representations
are equivalent given the definitions above.

\subsection{Spherical harmonics}
\label{sec:gp-ylms}

\begin{figure}[t!]
    \begin{centering}
        \includegraphics[width=\linewidth]{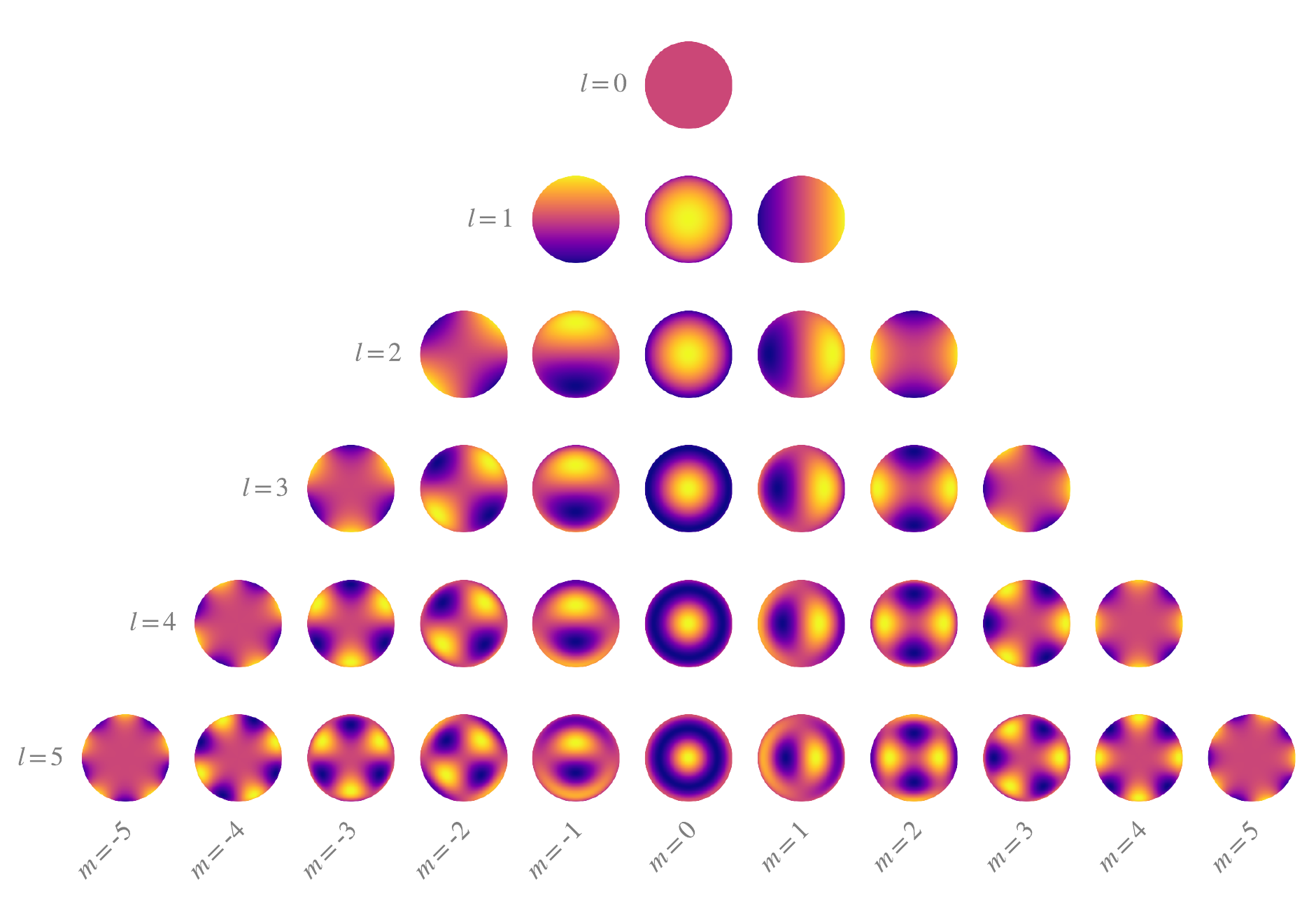}
        \oscaption{ylms}{%
            The real spherical harmonics in the polar frame up to
            $l = 5$, where dark colors correspond to negative intensity
            and bright colors to positive intensity.
            Rows correspond to the degree $l$ and columns to
            the order $m$. The set of all spherical harmonics forms a
            complete, orthogonal basis on the sphere.
            \label{fig:ylms}
        }
    \end{centering}
\end{figure}

Before we dive into the computation of our GP, it is useful to introduce the spherical harmonics,
a set of orthogonal functions on the surface of the sphere which we will use
to describe the intensity field on the surface of a star
(Figure~\ref{fig:ylms}). As we will see below,
the spherical harmonics are a particularly convenient basis in which to
describe starspot distributions%
\footnote{There are, of course, drawbacks to using this basis:
    in particular, the spherical harmonics are smooth, continuous functions that
    struggle (at finite degree $l$) to capture high resolution features such as
    small starspots. We discuss this point at length in \S\ref{sec:tinyspots},
    where we show that our model is useful even when applied to stars with spots
    smaller than the effective resolution of the GP.}, as they will allow us to compute
moments of the intensity distribution analytically. Of more immediate concern,
\citet{Luger2019} showed that there is a linear relationship between the
spherical harmonic expansion of a stellar surface and the total disk-integrated
flux $\mathbf{f}$ (i.e., the light curve)
one would observe as the star rotates about a fixed axis.
If the stellar surface intensity is described by a spherical harmonic
coefficient vector $\mathbf{y}$ (up to a certain degree $l_\mathrm{max}$),
the flux is given by
\begin{align}
    \label{eq:fAy}
    \mathbf{f} = \mathbf{1} + \pmb{\mathcal{A}}(I, P, \mathbf{u}) \, \mathbf{y}
    \quad,
\end{align}
where $\mathbf{1}$ is the ones vector and
$\pmb{\mathcal{A}}$ is the \starry
design matrix, a purely linear operator that transforms from the spherical
harmonic basis to the flux basis; it is
a function of the stellar inclination $I$, the stellar
rotation period $P$, and the stellar limb darkening coefficients $\mathbf{u}$,
as well as the observation times (see Appendix~\ref{sec:starry} for details).

\subsection{Computing the GP}
\label{sec:gp-gp}

Let
$\mathbbb{f} = \left( \mathbb{f}_0 \,\, \mathbb{f}_1 \,\, \cdots \,\,  \mathbb{f}_{K-1} \right)^\top$
denote a random vector of $K$ flux measurements at times
$\mathbf{t} = \left( t_0 \,\,  t_1 \,\,  \cdots \,\, t_{K-1} \right)^\top$,
defined in units such that a star with no spots on it will have
unit flux.%
\footnote{
    Note, importantly, that these are not the units we observe in! See \citetalias{PaperI}
    and \S\ref{sec:gp-norm} below.
}
Conditioned on the stellar inclination $I$, the rotational period $P$,
a set of limb darkening coefficients $\mathbf{u}$, and
on certain properties of the starspots $\pmb{\theta}_\bullet$
(including their number, sizes, positions, and contrasts),
we wish to compute the mean $\pmb{\mu}(I, P, \mathbf{u}, \pmb{\theta}_\bullet)$ and
covariance $\pmb{\Sigma}(I, P, \mathbf{u}, \pmb{\theta}_\bullet)$
of $\mathbbb{f}$. Together, these specify a multidimensional Gaussian
distribution, which we assume fully describes%
\footnote{The \emph{true} distribution of stellar light curves conditioned
    on $I$, $P$, $\mathbf{u}$, and $\pmb{\theta}_\bullet$ is not Gaussian, so our
    assumption is formally wrong. But, as the saying goes, \emph{all models are
        wrong; some are useful}. As we will show later, this turns out to be
    an extremely useful assumption.}
how our flux measurements
are distributed:
\begin{align}
    \mathbbb{f}\left(I, P, \mathbf{u}, \pmb{\theta}_\bullet\right) \sim
    \mathcal{N}\Big(
    \pmb{\mu}\left(I, P, \mathbf{u}, \pmb{\theta}_\bullet\right),
    \,
    \pmb{\Sigma}\left(I, P, \mathbf{u}, \pmb{\theta}_\bullet\right)
    \Big)
\end{align}
As with any random variable, the mean and covariance may be computed from
the expectation values of $\mathbbb{f}$ and
$\mathbbb{f}\,\mathbbb{f}^\top$, respectively:
\begin{align}
    \label{eq:mean}
    \pmb{\mu}(I, P, \mathbf{u}, \pmb{\theta}_\bullet)
     & = \mathrm{E} \Big[ \mathbbb{f} \, \Big| \, I, P, \mathbf{u}, \pmb{\theta}_\bullet \Big]
    \\
    \label{eq:cov}
    \pmb{\Sigma}(I, P, \mathbf{u}, \pmb{\theta}_\bullet)
     & = \mathrm{E} \Big[ \mathbbb{f} \, \mathbbb{f}^\top \, \Big| \, I, P, \mathbf{u}, \pmb{\theta}_\bullet \Big] - \pmb{\mu}(I, P, \mathbf{u}, \pmb{\theta}_\bullet) \pmb{\mu}^\top(I, P, \mathbf{u}, \pmb{\theta}_\bullet)
    \quad.
\end{align}
Given the linear relationship between flux and spherical harmonic
coefficients (Equation~\ref{eq:fAy}),
we may write the mean and covariance of our GP as
\begin{align}
    \label{eq:mean_f}
    \pmb{\mu}(P, \mathbf{u}, \pmb{\theta}_\bullet)
     & = \mathbf{1} + \pmb{\mathcal{A}}(I, P, \mathbf{u}) \, \pmb{\mu}_{\mathbf{y}}(\pmb{\theta}_\bullet)
    \\
    \label{eq:cov_f}
    \pmb{\Sigma}(P, \mathbf{u}, \pmb{\theta}_\bullet)
     & = \pmb{\mathcal{A}}(I, P, \mathbf{u}) \, \pmb{\Sigma}_{\mathbf{y}}(\pmb{\theta}_\bullet) \, \pmb{\mathcal{A}}^\top(I, P, \mathbf{u})
    \quad,
\end{align}
where
\begin{align}
    \label{eq:mean_y}
    \pmb{\mu}_{\mathbf{y}}(\pmb{\theta}_\bullet)
     & = \mathrm{E} \Big[ \mathbbb{y} \, \Big| \, \pmb{\theta}_\bullet \Big]
    \\
    \label{eq:cov_y}
    \pmb{\Sigma}_{\mathbf{y}}(\pmb{\theta}_\bullet)
     & = \mathrm{E} \Big[ \mathbbb{y} \, \mathbbb{y}^\top \, \Big| \, \pmb{\theta}_\bullet \Big] - \pmb{\mu}_{\mathbf{y}}(\pmb{\theta}_\bullet) \pmb{\mu}_{\mathbf{y}}^\top(\pmb{\theta}_\bullet)
\end{align}
are the mean and covariance of the distribution over spherical harmonic coefficient
vectors $\mathbbb{y}$.
The bulk of the math in this paper (Appendix~\ref{sec:integrals})
is devoted to computing
the expectations in the expressions above, which
are given by the integrals
\begin{align}
    \label{eq:exp_y}
    \mathrm{E} \Big[ \mathbbb{y} \, \Big| \, \pmb{\theta}_\bullet \Big]
     & =
    \int \mathbbb{y}(\mathbbb{x} ) \, p(\mathbbb{x} \, \big| \, \pmb{\theta}_\bullet)\mathrm{d}\mathbbb{x}
    \\
    \label{eq:exp_yy}
    \mathrm{E} \Big[ \mathbbb{y} \, \mathbbb{y}^\top \, \Big| \, \pmb{\theta}_\bullet \Big]
     & =
    \int \mathbbb{y}(\mathbbb{x} ) \mathbbb{y}^\top(\mathbbb{x} ) \, p(\mathbbb{x} \, \big| \, \pmb{\theta}_\bullet)\mathrm{d}\mathbbb{x}
    \quad,
\end{align}
where $\mathbbb{x}$ is a random vector-valued variable corresponding to a particular
distribution of features on the surface
and $p(\mathbbb{x} \, \big| \, \pmb{\theta}_\bullet)$ is its probability density
function (PDF).
In the Appendix we show that for suitable choices of $\pmb{\theta}_\bullet$,
$\mathbbb{y}(\mathbbb{x})$,
and $p(\mathbbb{x} \, \big| \, \pmb{\theta}_\bullet)$, the integrals in the expressions
above have closed form solutions that may be evaluated quickly.
While we present a few different ways of specifying $\pmb{\theta}_\bullet$,
our default representation of the GP hyperparameters is
\begin{align}
    \label{eq:thetaspot}
    \pmb{\theta}_\bullet
     & =
    \left(
    n
    \,\,\,
    c
    \,\,\,
    \mu_\phi
    \,\,\,
    \sigma_\phi
    \,\,\,
    r
    \right)^\top
    \quad,
\end{align}
where $n$ is the number of starspots, $c$ is their contrast
(defined as the intensity difference between the spot and the
background intensity, as a fraction of the background intensity),
$\mu_\phi$ and $\sigma_\phi$ are the mode and standard deviation
of the spot latitude distribution, respectively, and $r$ is the radius
of the spots.
For simplicity, the PDFs for the spot radius, the spot contrast, and the number
of spots are chosen to be delta functions centered at $r$, $c$, and $n$, respectively
(Appendices \ref{sec:size} and \ref{sec:contrast}),
while
the spot longitude is assumed to be uniformly distributed
(Appendix \ref{sec:lon}).
Finally,
the PDF for the latitude $\phi$ of the spots is chosen to be a Beta distribution in
$\cos\phi$ with (normalized) parameters $a$ and $b$,
which have a one-to-one correspondence to the mode $\mu_\phi$ and
standard deviation $\sigma_\phi$ of the distribution in $\phi$
(Appendix \ref{sec:lat}). This allows us to model starspot distributions
with ``active latitudes'' of tunable width that are symmetric about
the equator. The distribution is flexible enough to also model equatorial
spots and isotropically-distributed spots. Stars with multiple active
latitudes can easily be modeled as a sum of Gaussian processes
(\S\ref{sec:mixture}). These choices for the spatial distribution of
spots are based on the Sun, whose spots emerge in azimuthally-symmetric
belts at roughly the same latitude in both hemispheres, then
migrate toward the equator over the course of the 11-year cycle
\citep{Solanki2006}.

In this paper, we assume that the parameters $\pmb{\theta}_\bullet$
described above are the \emph{physically interesting} ones. That is,
given a light curve $\mathbf{f}$ or an ensemble of $M$ light curves of
statistically similar stars
$\left( \mathbf{f}_0 \,\, \mathbf{f}_1 \,\, \cdots \,\,  \mathbf{f}_{M-1} \right)^\top$,
we wish to infer the statistical properties of the starspots, encoded in
the entries of the vector $\pmb{\theta}_\bullet$.
This is typically a tall order, since it requires marginalizing over all
the nuisance parameters, which include the nitty-gritty details of the
size, contrast, and location of \emph{every spot} (and, if $M > 1$, on \emph{every star}
in the ensemble). Fortunately, however, the Gaussian process we constructed
does just that. Specifically, given the mean and covariance of the process,
we are able to directly evaluate the log marginal likelihood of the $m^\mathrm{th}$
dataset
conditioned on a specific value of $\pmb{\theta}_\bullet$ (as well as $I$,
$P$, and $\mathbf{u}$):
\begin{align}
    \label{eq:log-like}
    \ln \mathcal{L}_m\left(I, P, \mathbf{u}, \pmb{\theta}_\bullet\right)
    =
     & -\frac{1}{2}
    \mathbf{r}_m^\top\left(I, P, \mathbf{u}, \pmb{\theta}_\bullet\right)
    \big[
        \pmb{\Sigma}\left(I, P, \mathbf{u}, \pmb{\theta}_\bullet\right)
        +
        \mathbf{C}_m
        \big]^{-1}
    \mathbf{r}_m\left(I, P, \mathbf{u}, \pmb{\theta}_\bullet\right)
    \nonumber       \\[0.75em]
     & -
    \frac{1}{2}
    \ln \Big|
    \pmb{\Sigma}\left(I, P, \mathbf{u}, \pmb{\theta}_\bullet\right)
    +
    \mathbf{C}_m
    \Big|
    -
    \frac{K}{2}
    \ln \left( 2 \pi \right)
    \quad,
\end{align}
where
\begin{align}
    \mathbf{r}_m\left(I, P, \mathbf{u}, \pmb{\theta}_\bullet\right)
     & \equiv
    \mathbf{f}_m - \pmb{\mu}\left(I, P, \mathbf{u}, \pmb{\theta}_\bullet\right)
\end{align}
is the residual vector,
$\mathbf{C}_m$ is the data covariance
(which in most cases is a diagonal matrix whose entries
are the squared uncertainty corresponding to each data point in the light curve),
$| \cdots |$ denotes the determinant, and $K$ is the number of data points in
each light curve.%
\footnote{
    In Equation~(\ref{eq:log-like}) we implicitly assume all stars in the
    ensemble are observed at the same set of times $\mathbf{t}$.
    If this is not the case, the mean and covariance of the GP for each
    star must be computed from Equations~(\ref{eq:mean}) and (\ref{eq:cov})
    with the flux design matrix $\pmb{\mathcal{A}}(\mathbf{t}_m)$ evaluated
    at the particular observation times $\mathbf{t}_m$.
}
In an ensemble analysis, the joint marginal likelihood of all datasets is
simply the product of the individual likelihoods, so in log space we have
\begin{align}
    \ln \mathcal{L}\left(I, P, \mathbf{u}, \pmb{\theta}_\bullet\right)
     & =
    \sum_{m} \ln \mathcal{L}_m\left(I, P, \mathbf{u}, \pmb{\theta}_\bullet\right)
    \quad.
\end{align}
The marginal likelihood may be interpreted as the probability of the data
given the model. Typically, we are interested in the reverse: the probability
of the \emph{model} given the \emph{data}, i.e., the posterior probability
distribution. In later sections we present a comprehensive suite of
posterior inference exercises demonstrating that our GP model is correctly
calibrated, allowing one to efficiently infer statistical properties of starspots
from light curves with minimal bias.

\subsection{Marginalizing over inclination}
\label{sec:gp-inc}
As we mentioned in the previous section,
the equations for the mean and covariance of our GP
(Equations~\ref{eq:mean_f} and \ref{eq:cov_f}, respectively) are conditioned
on specific values of the stellar and spot properties. To obtain the posterior distribution
for these parameters, we must typically resort to numerical sampling techniques,
which often scale steeply with the number of parameters. It is therefore generally
desirable to keep the total number of parameters small, especially when
employing the GP in an ensemble setting.
In such a setting, we might have light curves from $M$ stars, all of which
we believe to have similar spot properties (perhaps because they have
similar spectral types and rotation periods, for example).
The total number of parameters in our problem is therefore
\begin{align}
    N = 4 M + 5
    \quad,
\end{align}
since each of the stars will have their own set of 4 stellar properties
(an inclination, a period, and usually two limb darkening coefficients)
but will all share the same 5 spot properties
$\pmb{\theta}_\bullet$ (by assumption).
For a reasonably sized ensemble of $M=100$
stars, we would have to sample over $N = 405$ parameters.
While large, this number is certainly not absurd, especially by modern standards.
However, it does pose
a problem when considering how complex the posterior distribution for the
spot mapping problem can be. In addition to strong nonlinear degeneracies
between some of the parameters (such as the contrast $c$ and the
number of spots $n$), the posterior is often multimodal, especially in the
stellar inclinations. While modern sampling schemes such as Hamiltonian
Monte Carlo and Nested Sampling may in principle be able to deal with these
issues, in practice it can be very difficult to obtain convergence in a
reasonable amount of time.

One workaround is to fix the values of the stellar parameters. This could be done,
for instance, to the rotational period $P$, which can often be estimated with
fairly good accuracy from a periodogram. The limb darkening coefficients could
be fixed at theoretical values, or perhaps their values could be shared among
all stars (and sampled over), given the similarity assumption above.

The inclination, however, is a different matter. Absent prior information
for a particular star (such as a measurement of its projected rotational
velocity $v\sin I$ or the knowledge that it hosts a transiting planet), it
is simply not possible to reliably estimate the inclination in a
pre-processing step. Any light curve statistic one might argue should scale
with inclination---such as the amplitude of the variability---is invariably
degenerate with the spot parameters $\pmb{\theta}_\bullet$. If one knew
$\pmb{\theta}_\bullet$, then perhaps a decent point estimate of $I$ could
be obtained, but in that case the analysis wouldn't be needed in the first place!

Fortunately, there is a better way to reduce the number of parameters in
the problem: we can explicitly marginalize over the stellar inclination.
That is, we may write the mean and covariance of our GP as
\begin{align}
    \label{eq:mu_marg}
    \pmb{\mu}(P, \mathbf{u}, \pmb{\theta}_\bullet)
     & = \mathrm{E} \Big[ \mathbbb{f} \, \Big| \, P, \mathbf{u}, \pmb{\theta}_\bullet \Big]
    \nonumber                                                                                                                                                                                                        \\
     & = \mathbf{1} + \mathbf{e}_I
    \\[0.5em]
    \label{eq:cov_marg}
    \pmb{\Sigma}(P, \mathbf{u}, \pmb{\theta}_\bullet)
     & = \mathrm{E} \Big[ \mathbbb{f} \, \mathbbb{f}^\top \, \Big| \, P, \mathbf{u}, \pmb{\theta}_\bullet \Big] - \pmb{\mu}(P, \mathbf{u}, \pmb{\theta}_\bullet) \pmb{\mu}^\top(P, \mathbf{u}, \pmb{\theta}_\bullet)
    \nonumber                                                                                                                                                                                                        \\
     & =
    \mathbf{E}_I
    -
    \mathbf{e}_I \,
    \mathbf{e}_I^\top
\end{align}
where we define the inclination first moment integral
\begin{align}
    \label{eq:eI}
    \mathbf{e}_I
     & \equiv
    \int
    \pmb{\mathcal{A}}(\mathbb{I}, P, \mathbf{u}) \,
    \mathrm{E} \Big[ \mathbbb{y} \, \Big| \, \pmb{\theta}_\bullet \Big]
    p(\mathbb{I}) \, \mathrm{d}\mathbb{I}
\end{align}
and the inclination second moment integral
\begin{align}
    \label{eq:EI}
    \mathbf{E}_I
     & \equiv
    \int
    \pmb{\mathcal{A}}(\mathbb{I}, P, \mathbf{u}) \,
    \mathrm{E} \Big[ \mathbbb{y} \, \mathbbb{y}^\top \, \Big| \, \pmb{\theta}_\bullet \Big] \,
    \pmb{\mathcal{A}}^\top(\mathbb{I}, P, \mathbf{u})
    p(\mathbb{I}) \, \mathrm{d}\mathbb{I}
    \quad,
\end{align}
and $\mathbb{I}$ is the random variable corresponding to the inclination.
The expectations inside the integrals in the expressions for
$\mathbf{e}_I$ and $\mathbf{E}_I$
are given by
Equations~(\ref{eq:exp_y}) and (\ref{eq:exp_yy}), respectively, and
are computed in Appendix~\ref{sec:integrals}.
If we are able to perform the integrals in those expressions,
we can \emph{dramatically} reduce the number of
parameters in our ensemble problem.
As we show in Appendix~\ref{sec:gp-inc}, if we assume that stellar
inclinations are distributed isotropically, these integrals
do in fact have closed-form solutions.

Finally, for future reference, it is useful to note that the mean of the GP
is constant:
\begin{align}
    \label{eq:scalar-mean}
    \pmb{\mu}(P, \mathbf{u}, \pmb{\theta}_\bullet)
     & =  (1 + e_I)\mathbf{1}
    \nonumber                 \\
     & \equiv
    \mu \, \mathbf{1}
    \quad,
\end{align}
since by construction our GP is longitudinally isotropic
(see Appendix~\ref{sec:inc-mom1}).

\subsection{Normalization correction}
\label{sec:gp-norm}

In \citetalias{PaperI} we discussed a subtle but important point
about stellar light curves: the common procedure of normalizing
light curves to their mean or median level changes the covariance
structure of the data, since it correlates all the observations
in a nontrivial way.
When normalizing a light
curve by the mean,%
\footnote{In practice, the expressions derived here also work well
    for median-normalized light curves, since the distribution of the GP sample median
    is usually close to the distribution of the sample mean.}
the operation we perform is
\begin{align}
    \label{eq:ftilde}
    \tilde{\mathbbb{f}\hspace{0.2em}} & = \frac{\mathbbb{f}}{\left<\mathbb{f}\right>}
    \quad,
\end{align}
where $\tilde{\mathbbb{f}\hspace{0.2em}}$ is the normalized, unit-mean light curve,
$\mathbbb{f}$ is the measured light curve (in detector counts), and
$\left<\mathbb{f}\right>$ is the \emph{sample} mean: i.e., the average value of
a given star's light curve (which we model as a sample from our GP).
This may be close to but is in general different from the \emph{process} mean,
$\pmb{\mu}(P, \mathbf{u}, \pmb{\theta}_\bullet)$, since the mean of
a draw from the GP is itself normally distributed with a variance that scales
with the GP variance.%
\footnote{
    Importantly, the sample mean and process mean will be different even in the
    absence of measurement error! In other words, the mean flux of a given
    star (i.e., the sample mean)
    will in general be different from the mean flux \emph{across all stars} with similar
    surface properties (the process mean).
}

When modeling normalized light curves, we must correct our expression for
the covariance matrix $\pmb{\Sigma}$ of the GP.
Computing the new covariance matrix is tricky, especially
because the normalized process is \emph{not} strictly Gaussian: the distribution
of normalized light curves
has heavy tails due to the fact that $\tilde{\mathbbb{f}\hspace{0.2em}}$ diverges as
the sample mean approaches zero. In fact, because of these tails, the covariance
of the normalized process is formally \emph{infinite}, since the probability of
drawing a sample whose mean is arbitrarily close to zero is finite.

If this is all starting to sound like a bad idea, that's because it is!
A much safer approach is to resist the temptation to normalize the light curve
and instead model the (unknown) amplitude of the data as a multiplicative
latent variable. However,
this would require an extra parameter \emph{for every light curve}, so the
computational savings we achieved by marginalizing out the inclination
would be gone. Fortunately, in practice, the variance of a stellar light curve
is usually small compared to its mean: stellar variability amplitudes are
typically at the level of a few percent or lower. When this is the case,
the probability of drawing a GP sample whose mean is close to zero is
extremely small, and we can make use of the approximate expression derived
in \citet{Luger2021} for the covariance of a normalized Gaussian process:
\begin{proof}{test_normgp}
    \label{eq:SigmaTilde}
    \tilde{\pmb{\Sigma}}
    & \approx
    \frac{A}{\mu^2} \pmb{\Sigma} +
    z \Big(
    (A + B) \, (\mathbf{1} - \mathbf{q}) \, (\mathbf{1} - \mathbf{q})^\top
    - A \, \mathbf{q} \, \mathbf{q}^\top
    \Big)
    \quad,
\end{proof}
where
\begin{align}
    \label{eq:z}
    z & \equiv \frac{\left< \Sigma \right>}{\mu^2}
\end{align}
is the ratio of the average element in $\pmb{\Sigma}$
to the square of the mean of the Gaussian process,
$\mathbf{q}$ is the ratio of the average of each row in $\pmb{\Sigma}$
to the average element in $\pmb{\Sigma}$, and $A$, $B$ are
order unity and zero scalars, respectively,
given by the optimally-truncated diverging series
\begin{align}
    \label{eq:baseline_alpha}
    A
     & \equiv
    \sum\limits_{i=0}^{i_\mathrm{max}}
    \frac{(2i + 1)!}{2^i \, i!}
    z^i
    \\[1em]
    \label{eq:baseline_beta}
    B
     & \equiv
    \sum\limits_{i=0}^{i_\mathrm{max}}
    \frac{2i(2i + 1)!}{2^i \, i!}
    z^i
    \quad,
\end{align}
where $i_\mathrm{max}$ is the largest value for which the series coefficient at $i_\mathrm{max}$ is
smaller than the coefficient at $i_\mathrm{max} - 1$. In the expressions above, it is
assumed that the mean $\pmb{\mu}$ is constant, i.e., $\pmb{\mu} = \mu\, \mathbf{1}$.
Since our Gaussian process is azimuthally isotropic (i.e., no preferred
longitude), that is the case throughout this paper.


What Equation~(\ref{eq:SigmaTilde}) allows us to do is effectively marginalize over
the unknown normalization by modeling the \emph{normalized} flux as a draw
from a Gaussian process:
\begin{align}
    \tilde{\mathbbb{f}\hspace{0.2em}}
    \left(P, \mathbf{u}, \pmb{\theta}_\bullet\right)
    \sim
    \mathcal{N}\left(
    \mathbf{1},
    \tilde{\pmb{\Sigma}} \left(P, \mathbf{u}, \pmb{\theta}_\bullet\right)
    \right)
    \quad.
\end{align}
This is appropriate as long as $z \ll 1$, for which the true
distribution of $\tilde{\mathbbb{f}\hspace{0.2em}}$ is approximately Gaussian. In practice,
we recommend employing this trick only for $z \lesssim 0.02$, for which the
error in the approximation to the covariance is less than $10^{-6}$.
In cases where the light curve variability exceeds about ten percent, we recommend
modeling the multiplicative amplitude in each light curve as a latent variable, as discussed
above.

\begin{figure}[t!]
    \begin{centering}
        \includegraphics[width=\linewidth]{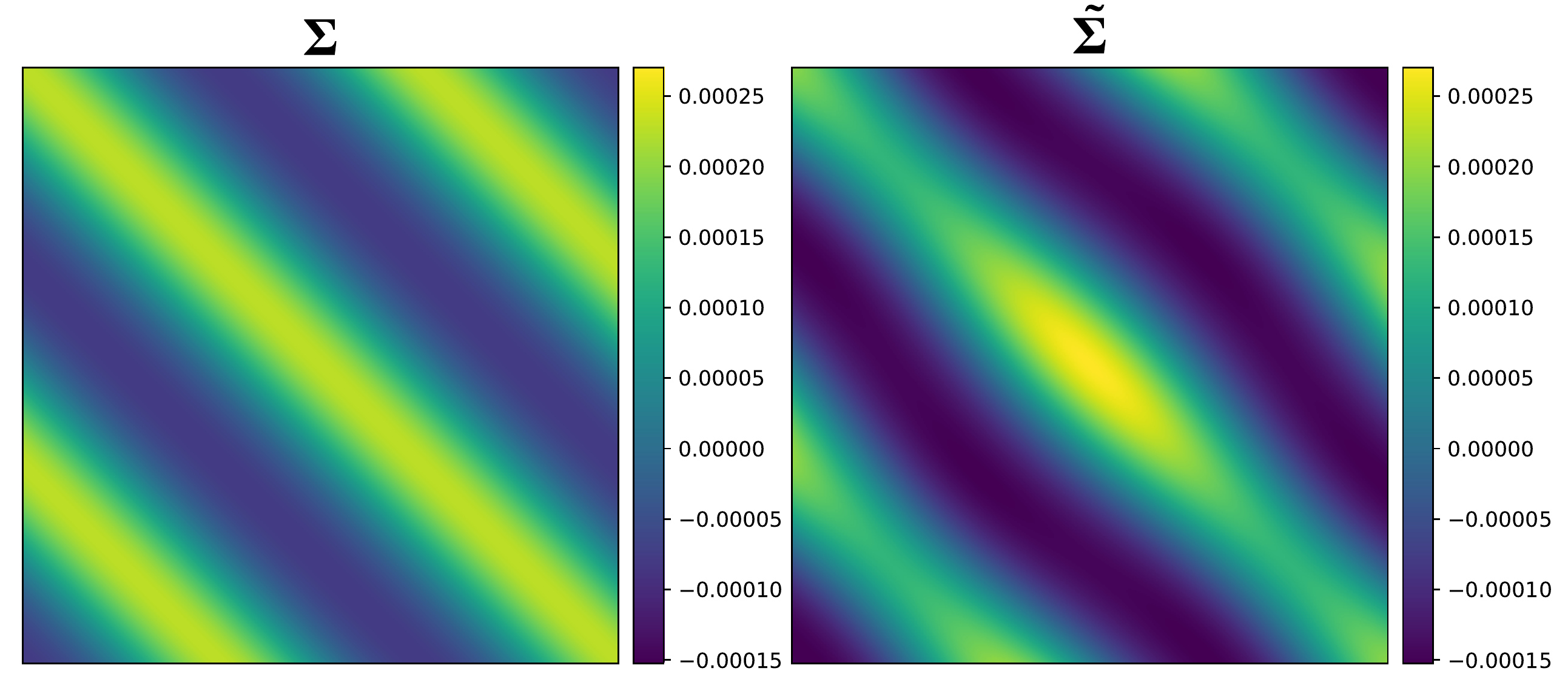}
        \oscaption{normgp}{%
            An example of a flux covariance matrix $\pmb{\Sigma}$ for a
            \starry process (left) and the corresponding covariance of the
            normalized process (right), computed from Equation~(\ref{eq:SigmaTilde}).
            In addition to an offset and an overall scaling relative to the
            original covariance matrix, the covariance of the normalized process
            is discernibly non-stationary.
            \label{fig:normgp}
        }
    \end{centering}
\end{figure}

Figure~\ref{fig:normgp} shows an example of a covariance matrix normalized
according to the procedure outlined above. The principal difference between
the normalized covariance and the original covariance is an overall
scaling and a small offset. However, the normalization also results in
the process becoming non-stationary: the covariance between two points in
a light curve is now slightly dependent on their phases.

\subsection{Summary}
\label{sec:gp-summary}
As the computation of the GP relies on many interdependent equations
scattered throughout the previous sections and the Appendix, it
is useful to summarize the procedure for the case where we marginalize
over the inclination (\S\ref{sec:gp-inc}) and the light curves are
normalized to their means (\S\ref{sec:gp-norm}), which is likely to be the primary
use case for our algorithm.

We model the mean-normalized flux $\tilde{\mathbbb{f}\hspace{0.2em}}$
(Equation~\ref{eq:ftilde}) as a Gaussian process:
\begin{align}
    \tilde{\mathbbb{f}\hspace{0.2em}}
    \left(P, \mathbf{u}, \pmb{\theta}_\bullet\right)
    \sim
    \mathcal{N}\left(
    \mathbf{1},
    \tilde{\pmb{\Sigma}} \left(P, \mathbf{u}, \pmb{\theta}_\bullet\right)
    \right)
    \quad.
\end{align}
The hyperparameters of the GP are
the stellar rotation period $P$,
the vector of limb darkening
coefficients $\mathbf{u}$, and the vector of parameters describing the
spot distribution
\begin{align}
    \pmb{\theta}_\bullet
     & =
    \left(
    n
    \,\,\,
    c
    \,\,\,
    \mu_\phi
    \,\,\,
    \sigma_\phi
    \,\,\,
    r
    \right)^\top
    \quad,
\end{align}
consisting of the number of spots $n$, their contrast $c$ (the fractional
intensity difference between the background and the spot),
the mode $\mu_\phi$ and standard deviation $\sigma_\phi$ of the latitude
distribution, and the radius of the spots $r$.
The quantity $\tilde{\pmb{\Sigma}}$ is the covariance of the normalized process
(Equation~\ref{eq:SigmaTilde}), which is a straightforward correction to the
true covariance of the process, accounting for
changes in scale and phase introduced by the common process of
normalizing light curves to a mean of unity. It depends on the
true (constant) mean $\mu$
and true covariance $\pmb{\Sigma}$,
given by Equations~(\ref{eq:mu_marg}) and (\ref{eq:cov_marg}),
respectively.
Those expressions in turn depend on the inclination expectation integrals
$e_I$ (Appendix~\ref{sec:inc-mom1}) and $\mathbf{E}_I$
(Appendix~\ref{sec:inc-mom2}). Those, in turn, depend on the first and
second moments of the distribution of spherical harmonic coefficient vectors,
$\mathrm{E}\left[ \mathbbb{y} \, \big| \, \pmb{\theta}_\bullet \right]$
and
$\mathrm{E}\left[ \mathbbb{y} \, \mathbbb{y}^\top \, \big| \, \pmb{\theta}_\bullet \right]$,
given by Equations~(\ref{eq:exp_y_sep}) and (\ref{eq:exp_yy_sep}), respectively.
To compute those, we must evaluate four nested integrals
(Equations~\ref{eq:e1}--\ref{eq:e4} for the first moment
and \ref{eq:E1}--\ref{eq:E4} for the second moment), corresponding to integrals
over the radius, latitude, longitude, and contrast distributions, respectively.
The computation of these integrals is discussed at length in Appendix~\ref{sec:integrals}.

While lengthy (and quite tedious), all of the computations described above rely
on equations whose solutions have a closed form.%
\footnote{The exception to this is the
    normalization correction (\S\ref{sec:gp-norm}), which depends on a
    fast-to-evaluate series and thus adds negligible overhead to the computation.}
Moreover, most of the terms in the expectation vectors and matrices may
be computed recursively, and many may be pre-computed, as they do not
depend on user inputs.
It is therefore possible to evaluate $\tilde{\pmb{\Sigma}}$ in an
efficient manner. In the companion paper \citep{JOSSPaper}, we discuss our
implementation of the algorithm in a user-friendly \Python package.

\subsection{An example}
\label{sec:examples}

\begin{figure}[t!]
    \begin{centering}
        \includegraphics[width=\linewidth]{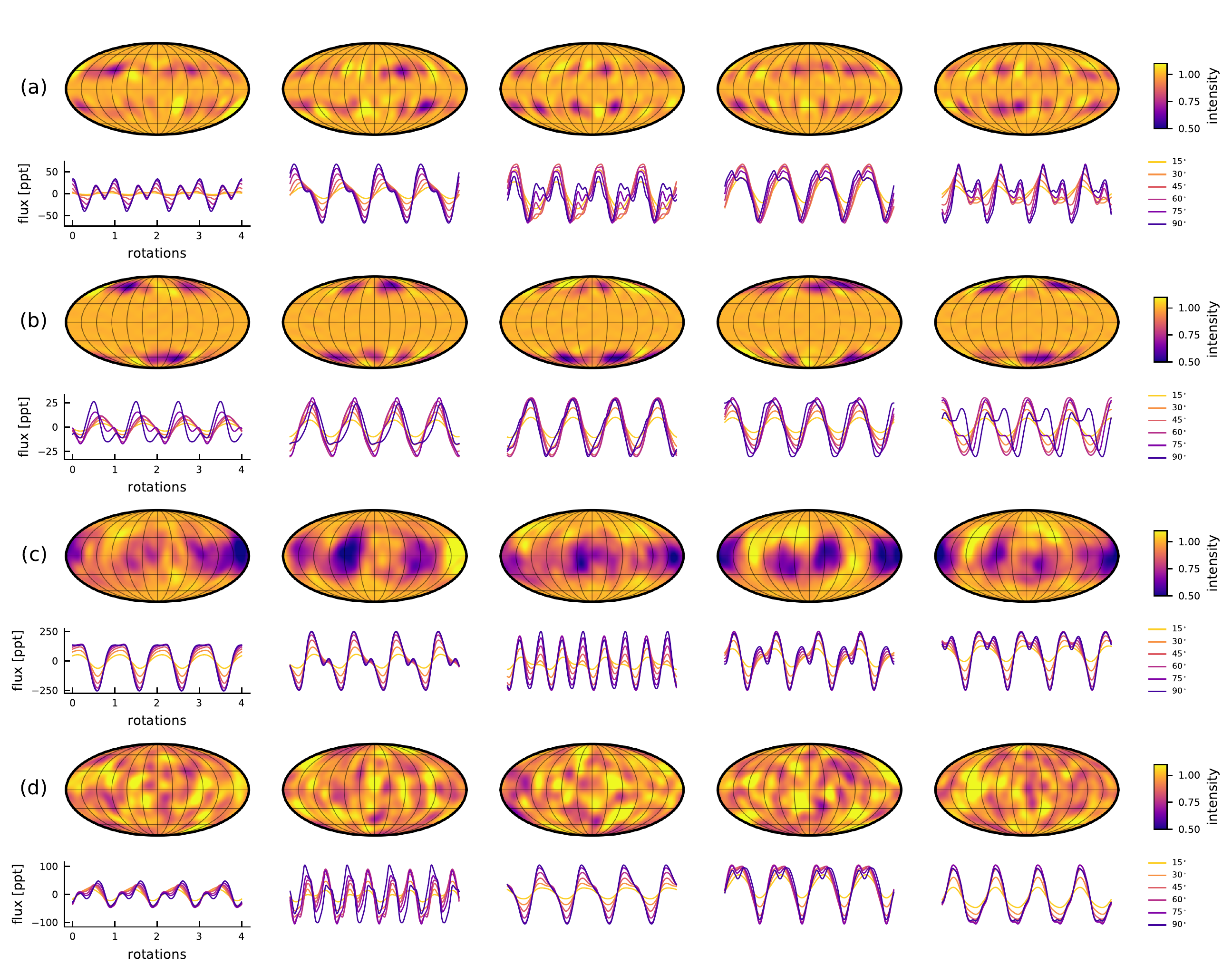}
        \oscaption{samples}{%
            Five random samples from the GP prior (columns) conditioned
            on four different values of $\pmb{\theta}_\bullet$ (rows).
            The samples are shown on the surface of
            the star in a Mollweide projection (upper panels) alongside
            their corresponding light curves viewed over four rotation
            cycles at several different inclinations (lower panels).
            From top to bottom, the GP corresponds to a star with
            (a) small mid-latitude spots; (b) small circumpolar spots;
            (c) large equatorial spots; and (d) small isotropic spots.
            See text for details.
            \label{fig:samples}
        }
    \end{centering}
\end{figure}

A concrete example of the GP derived above is presented in Figure~\ref{fig:samples},
where we show random samples from the process
evaluated up to spherical harmonic degree $l_\mathrm{max} = 15$ and
conditioned on different values
of the hyperparameter vector $\pmb{\theta}_\bullet$.
Each column corresponds to a different random draw from the GP, while each
row corresponds to a different value of $\pmb{\theta}_\bullet$.
The images are intensity maps of the stellar surface seen in an equal-area
Mollweide projection, in units such that a spotless star would have intensity
equal to 1 everywhere. Below them are the corresponding light curves (in units of
parts per thousand deviation from the mean) over four
rotation cycles, seen at inclinations varying from $15^\circ$ (yellow) to
$90^\circ$ (dark blue), and assuming no limb darkening (i.e., $\mathbf{u} = \mathbf{0}$).
From top to bottom, the hyperparameter vectors $\pmb{\theta}_\bullet$
for each row are
\begin{subequations}
    \begin{align}
        \left( n \,\,\, c \,\,\, \mu_\phi \,\,\, \sigma_\phi \,\,\, r\right)^\top
         & =\left( 10.0 \,\,\,\,\,\,\,\,\, 0.10 \,\,\,\,\,\,\,\,\, 30.0 \,\,\,\,\,\,\,\,\, 5.00 \,\,\,\,\,\,\,\,\, 10.0 \right)^\top \\
         & =\left( 10.0 \,\,\,\,\,\,\,\,\, 0.10 \,\,\,\,\,\,\,\,\, 60.0 \,\,\,\,\,\,\,\,\, 5.00 \,\,\,\,\,\,\,\,\, 10.0 \right)^\top \\
         & =\left( 10.0 \,\,\,\,\,\,\,\,\, 0.05 \,\,\,\,\,\,\,\,\, 0.00 \,\,\,\,\,\,\,\,\, 5.00 \,\,\,\,\,\,\,\,\, 30.0 \right)^\top \\
         & =\left( 20.0 \,\,\,\,\,\,\,\,\, 0.10 \,\,\,\,\,\,\,\,\, 0.00 \,\,\,\,\,\,\,\,\, 40.0 \,\,\,\,\,\,\,\,\, 10.0 \right)^\top
        \quad.
    \end{align}
\end{subequations}
These correspond to
(a) 10 spots of radius $10^\circ$ centered at
$30^\circ \pm 5^\circ$ latitude with a contrast of $10\%$;
(b) 10 spots of radius $10^\circ$ centered at
$60^\circ \pm 5^\circ$ latitude with a contrast of $10\%$;
(c) 10 spots of radius $30^\circ$ centered at
$0^\circ \pm 5^\circ$ latitude with a contrast of $5\%$;
and
(d) 20 spots of radius $10^\circ$ centered at
$0^\circ \pm 40^\circ$ latitude (a good approximation to a perfectly
isotropic distribution; see Appendix~\ref{sec:lat}) with a contrast of $10\%$.

The surface maps in the figure show dark, compact features of roughly the expected
size and contrast and at the expected latitudes. However, there are important
differences between these maps and what we would obtain by procedurally adding discrete
circular spots to a gridded stellar surface:
\begin{enumerate}
    \item \emph{The spots are not circular.} This is most evident in row (c),
          where some spots are distinctly asymmetric.
    \item \emph{There is significant variance in contrast from one spot to another},
          even though our model
          implicitly treats $c$ as constant. Within spots, the contrast is also
          not constant, even though (again) our model implictly treats it as such.
    \item \emph{There is ringing in the background.} This is apparent to some extent in
          row (b), where there are small fluctuations in the brightness at low
          latitudes where no spots are present.
    \item \emph{There aren't exactly 10 (or 20) spots in those maps.} This is
          most obvious in row (c), where only a few large distinct spots,
          plus maybe a few smaller ones, are visible.
    \item \emph{There are bright spots in addition to dark spots.} This may be
          the most glaring issue. We explicitly model spots as being dark, and yet
          there are (almost) just as many bright spots, particularly in row (d).
          While bright spots (such as plages) certainly exist in reality, we did not
          explicitly ask for them here!
\end{enumerate}
While these may appear to be critical shortcomings of our model, it is
important to keep in mind that a model consisting of discrete, circular, constant
contrast spots is likely just as far (or perhaps even farther) from the truth. In fact,
points (1) and (2) above suggest our model is more flexible than the discrete
spot model and thus (potentially) better suited to modeling real stellar surfaces.
Points (3), (4), and (5), on the other hand, are more concerning, since they are due,
respectively, to truncation error in the spherical harmonic expansion,
to an intrinsic limitation of our Gaussian approximation,
and to the fact that Gaussian distributions are symmetric about the mean: a positive
deviation is just as likely as a negative deviation of the same magnitude.%
\footnote{
    Even still, the model favors dark spots over bright spots because the GP mean
    itself is lower than unity, in practice making
    positive deviations \emph{from unity} less likely than negative
    deviations. This is why there are usually more dark spots than bright spots in
    the samples shown in the figure.
}
However, as we have argued before, the true power of the GP is in its
applicability to inference problems. In other words, while our GP has some
undesirable features when used as a prior sampling distribution, the real test
of the GP is when it is faced with data in an inference setting.
As long as the data is sufficiently informative, it does not matter that the
prior has finite support for unphysical configurations, as those
will be confidently rejected.

In \S\ref{sec:calibration} below, we exhaustively test the performance of our
GP as an inference tool when used to model synthetic light curves. We will
show that, despite the issues raised above, the GP is in general unbiased and
correctly estimates the posterior variance when used to infer the spot
properties $\pmb{\theta}_\bullet$.

But before we dive into calibration tests, it is worth pausing for a moment to take another
look at Figure~\ref{fig:samples}. While we focused on the shortcomings of
the GP as a prior in the discussion above, it is important to appreciate
that it even works in the first place!
A Gaussian process, after all, is a non-parameteric
\emph{process} describing a smooth and continuous function only via its
covariance structure. The GP knows nothing about the existence of discrete
spots---only how any two points on the surface are correlated.
Because spherical harmonics are smooth functions with support over the
entire sphere, the GP also does not know about features restricted
to certain latitudes; in fact, in most applications of GPs to mapping
problems in astronomy \citep[such as in models of the cosmic microwave
    background; e.g.,][]{Wandelt2012}, the process is assumed to be isotropic,
with no preferred angular direction.
However, by prescribing the correct structure to the covariance matrix,
we are able to approximately model compact spot-like
features with given sizes and restricted to particular latitudes.

\section{Calibration}
\label{sec:calibration}

\subsection{Why we need calibration tests}
\label{sec:why-calibrate}

In the previous section we derived a closed form solution to the
Gaussian approximation to the distribution of stellar surfaces (and their corresponding
light curves) conditioned on a vector $\pmb{\theta}_\bullet$
of spot hyperparameters. As we mentioned, the real test of this GP is
in how well it performs as a likelihood function for stellar light curves.

\begin{figure}[p!]
    \begin{centering}
        \includegraphics[width=\linewidth]{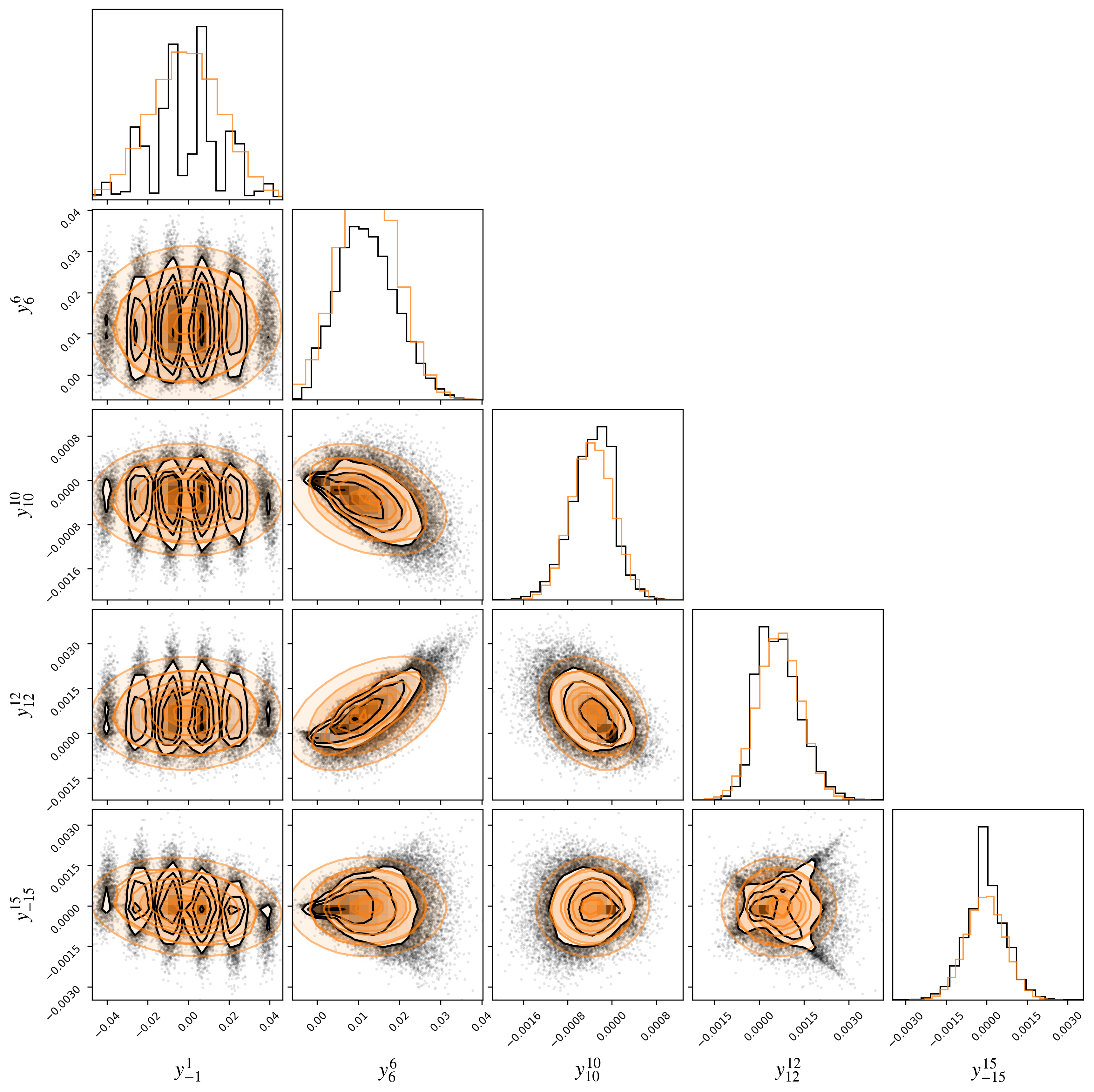}
        \oscaption{nongaussianity}{%
            Corner plot showing the joint distributions of select spherical
            harmonic coefficients corresponding to the $l=15$ expansion of
            $5\times 10^{4}$ stellar surface maps drawn from a certain
            distribution of discrete circular spots (black points and contours).
            At $l=15$ there are 256 coefficients in total;
            we chose five of the coefficients with the most odd-looking
            distributions to illustrate the non-Gaussianity of the process.
            In addition to non-linear correlations, skewness, and the
            existence of points of very high curvature, some of the distributions
            are also multi-modal.
            The orange contours show slices of the Gaussian approximation to
            the joint distribution; this is the approximation adopted in
            this paper.
            \label{fig:nongaussianity}
        }
    \end{centering}
\end{figure}

It is not immediately obvious that the GP approach should work, because the
true marginal likelihood function
$p ( \tilde{\mathbbb{f}\hspace{0.2em}} \, \big| \, P, \mathbf{u}, \pmb{\theta}_\bullet )$
is certainly \emph{not} Gaussian. To see why, let us generate $10^{4}$ stellar
surfaces sampled from the true distribution we are trying to model: that is,
a surface with $n=5$ discrete circular spots of fixed, uniform contrast $c=0.1$
and radius $r=20^\circ$ at latitudes $\mu_\phi \pm \sigma_\phi = 30^\circ \pm 5^\circ$.
Let us then expand each surface in spherical harmonics and visualize the
distribution of coefficients $\mathbbb{y}$. Figure~\ref{fig:nongaussianity}
shows the joint distribution for five of the coefficients with the most
non-Gaussian marginal distributions (selected by eye). Different slices
through this distribution
(in black) are skewed, strongly peaked, non-linearly correlated, and even bimodal.
Our approach in this paper is to model this distribution as a Gaussian
(orange contours). While this may be a good approximation in certain regions
of parameter space, it is certainly a poor approximation in others.
In this section, we will show that, fortunately, the non-Gaussianity of the
distribution is not in general an issue when doing inference with our GP,
as the resulting posteriors are correctly calibrated.

\subsection{Setup}
\label{sec:calibration-setup}

\begin{table}[t!]
    \begin{center}
        \begin{longtable}{W{c}{1cm} W{l}{6cm} W{l}{2cm}}
            \label{tab:synthetic}
            \\
            \toprule
            \multicolumn{1}{c}{\textbf{Symbol}}
             &
            \multicolumn{1}{c}{\textbf{Description}}
             &
            \multicolumn{1}{c}{\textbf{Value}}
            \\
            \midrule
            \endhead
            \bottomrule                                 \\
            \caption{%
                Default parameters used to generate synthetic light curves in
                the calibration tests.
            }
            \endfoot
            $\mathbb{n}$
             & number of spots
             & $\sim\mathcal{N}(20, 0^2)$
            \\
            $\mathbb{c}$
             & spot contrast
             & $\sim\mathcal{N}(0.05, 0^2)$
            \\
            $\bbphi$
             & spot latitude
             & $\sim\mathcal{N}(30^\circ, {5^\circ}^2)$
            \\
            $\bblambda$
             & spot longitude
             & $\sim\mathcal{U}(0^\circ, 360^\circ)$
            \\
            $\mathbb{r}$
             & spot radius
             & $\sim\mathcal{N}(15^\circ, {0^\circ}^2)$
            \\
            $\mathbb{I}$
             & stellar inclination
             & $\sim\sin$
            \\
            $P$
             & rotational period
             & $1$ day
            \\
            $\mathbf{u}$
             & limb darkening coefficients
             & $\left( 0 \,\,\,\, 0 \right)^\top$
            \\
            $\sigma_f$
             & photometric uncertainty
             & $10^{-3}$
            \\
            $K$
             & number of cadences per light curve
             & $10^3$
            \\
            $\Delta t$
             & time baseline
             & $4$ days
            \\
            $M$
             & number of light curves in ensemble
             & $50$
            \\
        \end{longtable}
    \end{center}
\end{table}

We seek to demonstrate that our model is correctly calibrated by testing it
on synthetic data, which we generate as follows. For each of $M$ synthetic
light curves in a given ensemble,
we create a rectangular
$(150 \times 300)$ latitude-longitude grid of surface intensity values, all
intialized
to zero. We then add $\mathbb{n}$ spots to this grid, each of fractional contrast
$\mathbb{c}$ and radius $\mathbb{r}$ centered at latitude $\bbphi$ and longitude $\bblambda$,
by decreasing the intensity at all points within an angular distance $\mathbb{r}$
(measured along the surface of the sphere) by an amount $\mathbb{c}$.
In order to compute the corresponding light curve, we expand the surface
in spherical harmonics, although at much higher degree
($l_\mathrm{max}^{(0)} = 30$)
than the degree we will use in the inference step ($l_\mathrm{max} = 15$)
to minimize potential ringing effects or other artefacts in the synthetic
data. For reference, the chosen degree $l_\mathrm{max} = 30$ is large
enough to resolve features on the order of
$\nicefrac{180^\circ}{30} = 6^\circ$ across, but small enough that
the algorithm for computing the light curve is numerically stable.%
\footnote{%
    A more principled approach would perhaps be to geneate light curves using
    a completely different model, such as by discretizing the surface at very
    high resolution and computing the flux via a weighted sum of the visible
    pixels. However, this would take orders of magnitude longer than the adopted
    approach and would still be subject to artefacts due to the discretization
    scheme. We have gone to great lengths in \citet{Luger2019} to show that
    our flux computation from spherical harmonics is both accurate and precise,
    so we are confident that our synthetic light curves correctly represent the
    assumed spot distributions.
}
We compute the light curve at $K$ points equally spaced over a baseline $\Delta t$
using the \starry algorithm
(Appendix \ref{sec:starry}), assuming an inclination $\mathbb{I}$, a rotational period $P$,
and limb darkening coefficients $\mathbf{u}$. Finally, we divide the
light curve by the mean and add Gaussian noise with standard deviation $\sigma_f$
to emulate photon noise.%
\footnote{
    In theory, we should do this in the reverse order: we should add photon
    noise and \emph{then} normalize the light curve to the mean, as that is the order in which
    those steps occur in reality. However, if we did that, we would have to normalize $\sigma_f$
    in our inference step, such that the variance of each of the normalized light curves
    in the ensemble would be different, requiring us to invert a different matrix for
    each light curve when computing the log likelihood (see Equation~\ref{eq:log-like}).
    This would significantly increase the computational cost of our tests. Fortunately, in
    practice, the difference between these two approaches has negligible effect on
    our results, so we opt for the faster of the two methods. Note, of course,
    that when applying our GP to real data, we won't have this choice!
}
The default values/distributions of each of the parameters mentioned above are given in
Table~\ref{tab:synthetic}. Some, like the number of spots, their contrast,
etc., are drawn from fiducial distributions, while others, like the
photometric uncertainty, the rotational period, and the limb darkening
coefficients,
are fixed across all $M$ light curves in an ensemble. These fixed values
are not realistic, but they greatly speed up the inference step, since they
allow us to invert a single covariance matrix for all light curves when computing
the log likelihood.

\begin{figure}[p!]
    \begin{centering}
        \includegraphics[width=0.9\linewidth]{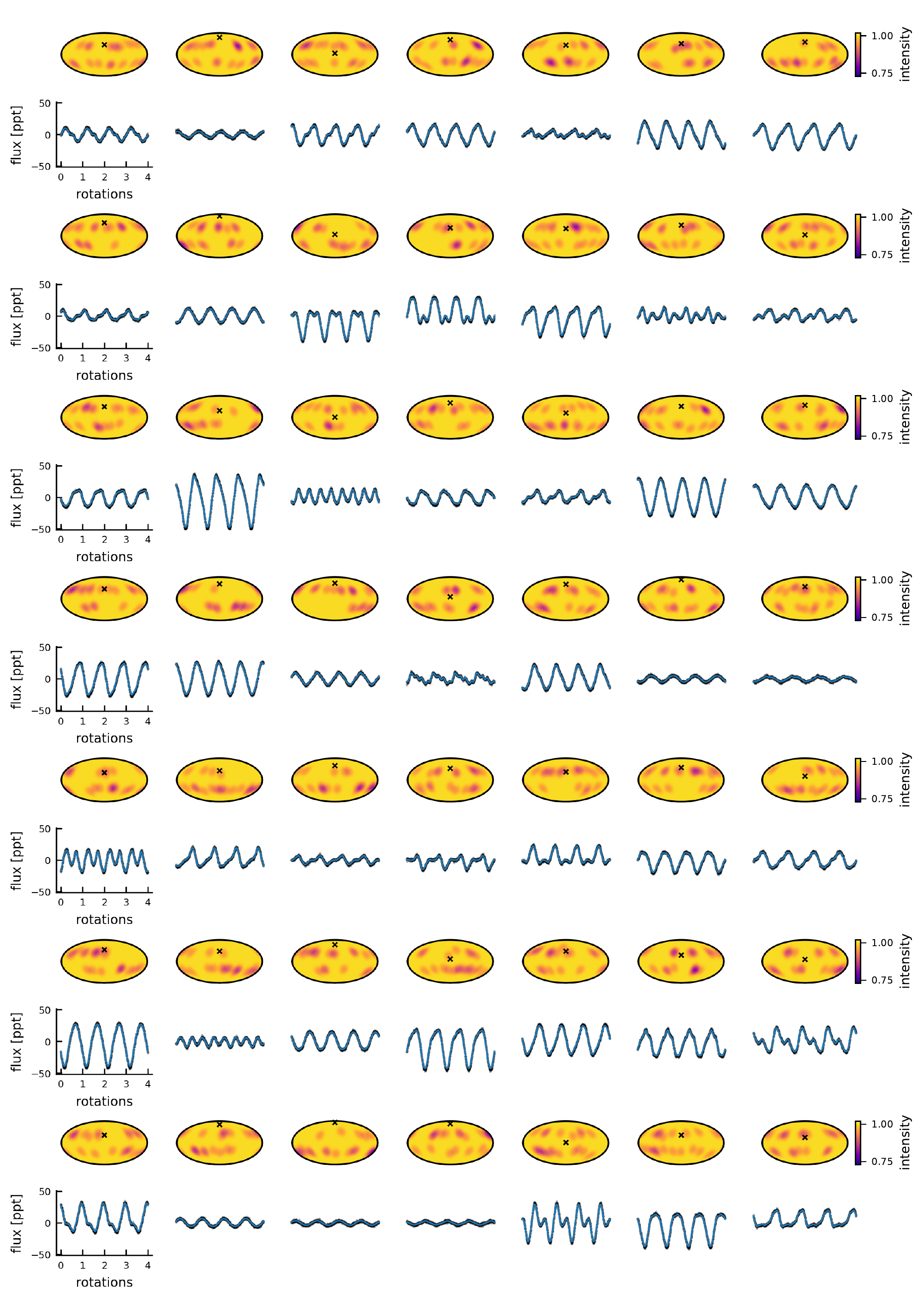}
        \oscaption{calibration}{%
            A synthetic dataset generated by adding discrete spots to a stellar
            surface with parameters given by the default values listed in
            Table~\ref{tab:synthetic}. The surface maps for 49 synthetic stars are
            shown in a Mollweide equal-area projection above their corresponding
            light curves. All maps and light curves are plotted on the same scale.
            The $\times$ on each map indicates the sub-observer latitude assumed
            when generating the light curve. The blue curves correspond to the
            exact light curve, while the black dots are the observed light curve.
            \label{fig:calibration_default_data}
        }
    \end{centering}
\end{figure}

\vfill
\pagebreak

Figure~\ref{fig:calibration_default_data} shows a synthetic dataset generated
from the default values listed in Table~\ref{tab:synthetic}. While the
light curves correspond to surfaces with the same statistical spot properties,
they all look qualitatively different: the mapping from starspot properties
to flux is nontrivial. In the inference step below, we assume we observe
only these 50 light curves (the figure only shows 49 of them),
with no knowledge of the inclination of any individual
star, and attempt to infer the spot properties.

\subsection{Inference}
\label{sec:calibration-inference}
We use our \Python-based implementation of the GP \citep{JOSSPaper}
to perform inference on the synthetic datasets. For simplicity, we assume we
know the value of the period $P$, which is fixed at unity for all stars, as well
as the value of the limb darkening coefficients (fixed at zero for the default
run). In practice these will not be known exactly; we discuss this further in
\S\ref{sec:other-marg}. Since we explicitly marginalize over the inclinations
of individual stars, the only quantities we must constrain are the five
parameters in the spot parameter vector $\pmb{\theta}_\bullet$
(Equation~\ref{eq:thetaspot}). We experimented with three different methods
for doing posterior inference on our synthetic datasets: No-U Turn Sampling,
a variant of Hamiltonian Monte Carlo
\citep[NUTS;][]{Duane1987,Hoffman2011}, automatic differentiation
variational inference \citep[ADVI;][]{Kucukelbir2016,Blei2016}, and nested sampling \citep{Skilling2004,Skilling2006}.
We obtained the best performance using the nested sampling algorithm
implemented in the \textsf{dynesty} package \citep{Speagle2020}, so that is what
we will use below.

Our sampling parameters are the number of spots $n$, their contrast $c$,
their radius $r$, and the Beta distribution parameters $a$ and $b$ describing their
distribution in latitude. As we discuss in Appendix~\ref{sec:lat},
the parameters $a$ and $b$ are easier to sample in than
the mode $\mu_\phi$ and standard deviation $\sigma_\phi$, provided we
account for the Jacobian of the transformation (Equation~\ref{eq:J})
in our log probability function, which maps a uniform prior on $a$ and $b$
to a uniform prior on $\mu_\phi$ and $\sigma_\phi$.

We place uniform priors on all five quantities, with support in
$1 \leq n \leq 50$, $0 < c \leq 1$,
$10^\circ \leq r \leq 30^\circ$,
$0 \leq a \leq 1$, and
$0 \leq b \leq 1$. Note that while $n$ formally represents an integer,
its effect on the GP is a scaling of the covariance (see Equation~\ref{eq:exp_yy_sep});
as such, it has support over all real numbers within the bounds listed
above. We \emph{could} restrict it to integer values, but this would make
sampling quite tricky. Moreover, in practice it is useful to allow for
noninteger values to add some flexibility to the model; we discuss this
in more detail in \S\ref{sec:caveats}.

We use Equation~(\ref{eq:log-like}) as our log likelihood
term, adding the log of the absolute value of Equation~(\ref{eq:J}) to
enforce a uniform prior on $\mu_\phi$ and $\sigma_\phi$. As we mentioned
above, the fact that $P$, $\mathbf{u}$, and $\sigma_f$ are shared among
all $M$ light curves means that $\pmb{\Sigma} + \mathbf{C}_m$ is the same for all
of them, greatly speeding up the likelihood evaluation since we need
only invert (or factorize) it a single time per sample. Our
covariance is the covariance of the \emph{normalized} process,
given by Equation~(\ref{eq:SigmaTilde}). Since we only consider light curves
with variability limited to a few percent or less, the approximation for
$\tilde{\pmb{\Sigma}}$ is always valid. Finally, we restrict our spherical
harmonic expansion to $l_\mathrm{max} = 15$ as a compromise between
resolution, computational speed, and numerical stability \citep[see][]{JOSSPaper}.

We use the standard implementation of the nested sampler,
\textsf{dynesty.NestedSampler}, with all arguments set to their
default values
(multi-ellipsoidal decomposition for bounds determination \citep{Feroz2009},
uniform sampling within the bounds,
500 live points,
and no gradients), to perform our inference step.
Convergence---defined as when the estimate of the remaining evidence $\Delta\ln \mathcal{Z}$ drops
below $0.5$---is usually attained after $5{,}000$ to $10{,}000$ samples and
within a couple hours on a typical machine for most of the trials we perform.

Below we describe several calibration \emph{runs}: experiments where we generate
an ensemble of light curves from synthetic stars with given properties
(\S\ref{sec:calibration-setup}) and attempt to infer their statistical spot
properties.

\subsection{Default run}
\label{sec:inference-default}

\begin{figure}[t!]
    \begin{centering}
        \includegraphics[width=\linewidth]{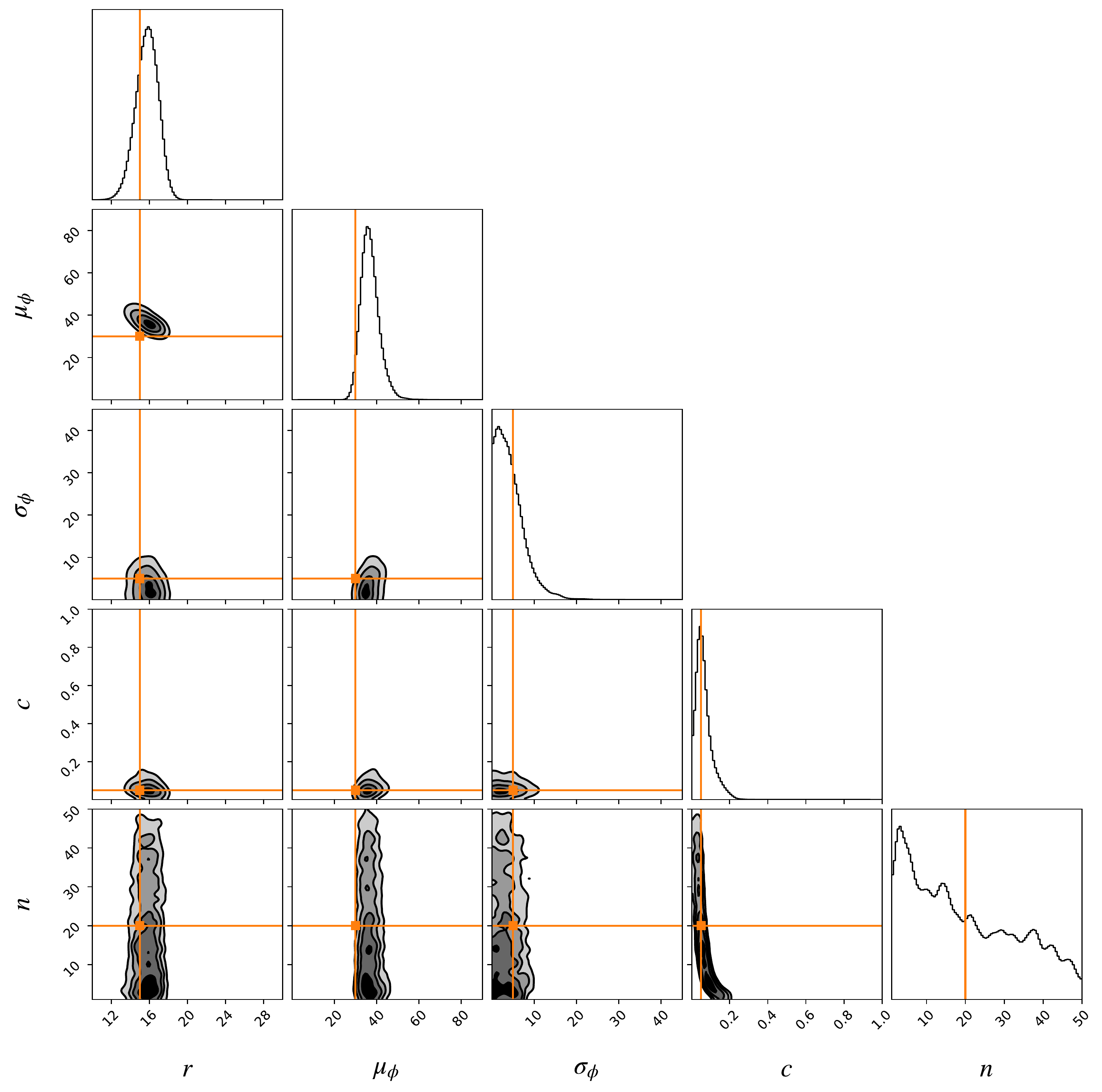}
        \oscaption{calibration_default}{%
            Posterior distributions for the spot parameters $\pmb{\theta}_\bullet$
            (radius $r$ in degrees, latitude mode $\mu_\phi$ and standard deviation $\sigma_\phi$ in degrees,
            fractional contrast $c$, and number of spots $n$)
            for the default run (Table~\ref{tab:synthetic} and Figure~\ref{fig:calibration_default_data}).
            The axes span the entire prior volume, and the orange lines and markers
            indicate the true (input) values.
            \label{fig:calibration_default_corner}
        }
    \end{centering}
\end{figure}

\begin{figure}[t!]
    \begin{centering}
        \includegraphics[width=\linewidth]{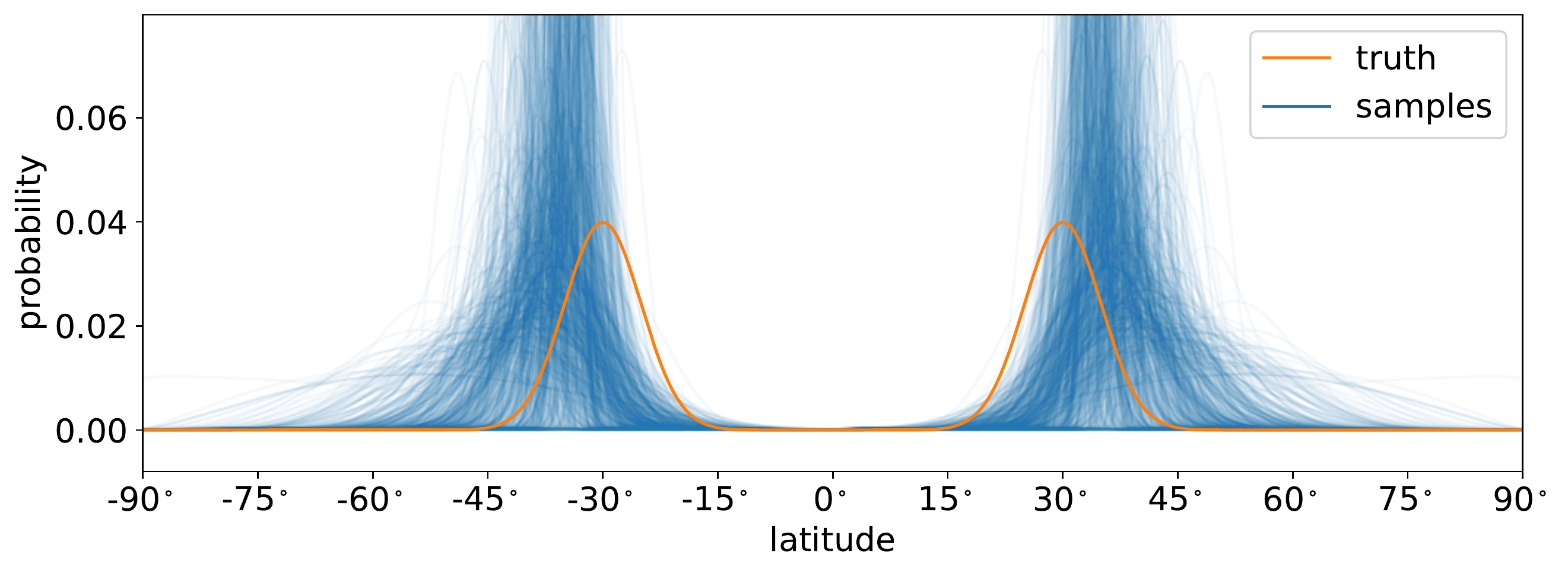}
        \oscaption{calibration_default}{%
            Posterior distributions for the spot latitudes for the default run. Each blue curve corresponds
            to the Beta distribution PDF (Equation~\ref{eq:phi-pdf}) for the spot latitude
            with parameters drawn from the posterior in $\mu_\phi$ and $\sigma_\phi$
            (Figure~\ref{fig:calibration_default_corner}); the (hyper)distribution of blue curves
            quantifies our beliefs about how spots are distributed on any given star. The
            orange curve is the true distribution used to generate the spots (see Table~\ref{tab:synthetic}).
            \label{fig:calibration_default_latitude}
        }
    \end{centering}
\end{figure}

\begin{figure}[t!]
    \begin{centering}
        \includegraphics[width=\linewidth]{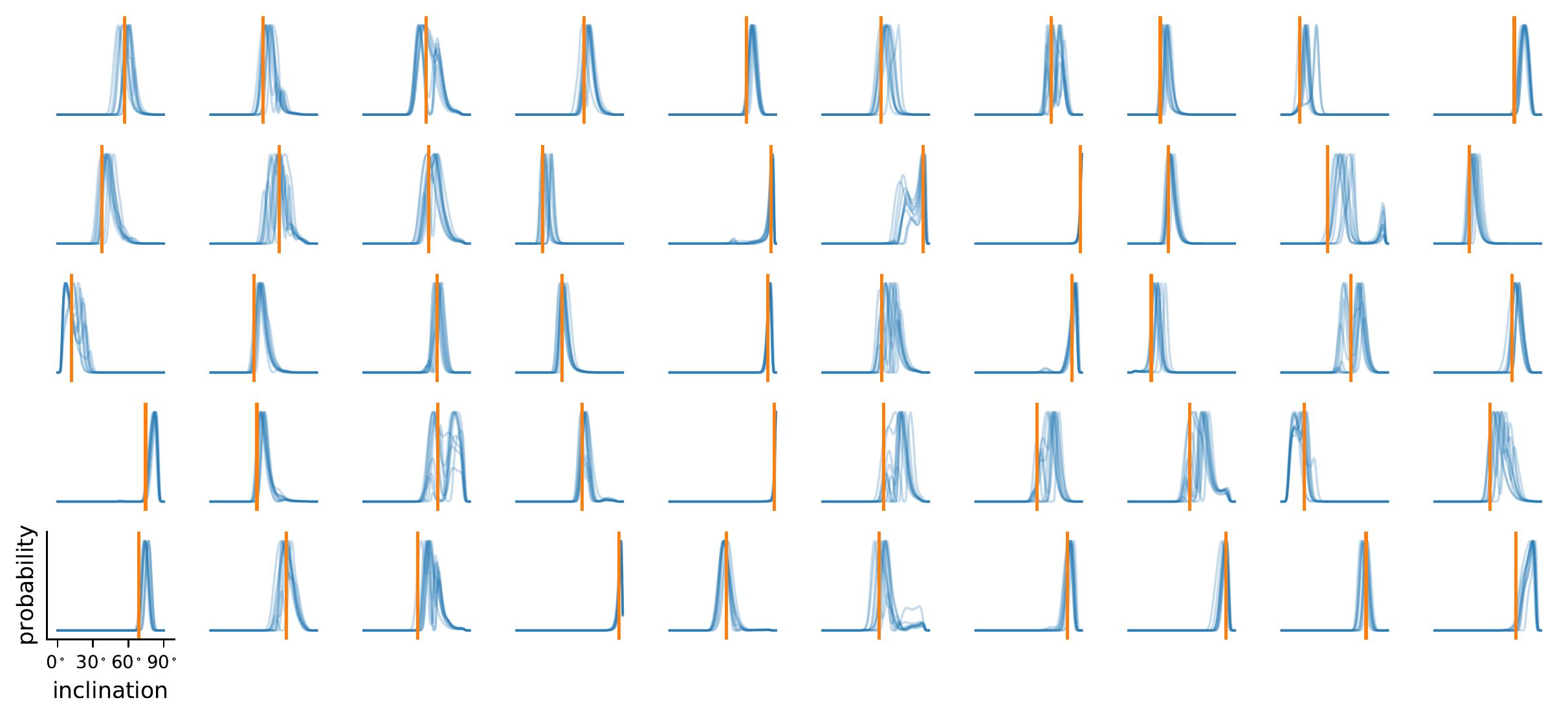}
        \oscaption{calibration_default}{%
            Posterior distributions for the inclinations of individual stars; individual
            panels correspond directly to those in Figure~\ref{fig:calibration_default_data}.
            As in Figure~\ref{fig:calibration_default_latitude}, each blue curve is a sample
            from the hyperdistribution of inclination PDFs, conditioned on a specific value
            of $\pmb{\theta}_\bullet$ drawn from the posterior. The orange lines indicate the
            true inclination of each star.
            \label{fig:calibration_default_inclination}
        }
    \end{centering}
\end{figure}

The input parameters for the default run are shown in Table~\ref{tab:synthetic},
and the corresponding light curves in Figure~\ref{fig:calibration_default_data}.
We run the nested sampler as described in the previous section and transform
the posteriors in $a$ and $b$ to posteriors in $\mu_\phi$ and $\sigma_\phi$ via
Equations~(\ref{eq:beta2gauss}), (\ref{eq:muphi}), and (\ref{eq:sigmaphi}).
The results are shown in Figure~\ref{fig:calibration_default_corner}, where
we correctly infer all five parameters within $2-3$ standard deviations.
Posterior distributions for the spot radius $r$, the central spot latitude $\mu_\phi$,
and the spot contrast $c$ are fairly tight, while the distribution for
the latitudinal scatter $\sigma_\phi$ has wider tails and the distribution for
the number of spots $n$ is very poorly constrained. The latter, in particular, is
degenerate with the spot contrast $c$; we discuss this at length in
\citetalias{PaperI}.

Figure~\ref{fig:calibration_default_latitude} shows samples from the spot
latitude posterior (hyper)distribution. Since the parameters $\mu_\phi$
and $\sigma_\phi$ characterize a distribution over spot latitudes, uncertainty
in their values translates to uncertainty in the actual shape of the spot
latitude distribution. Thus, the collection of blue curves in
Figure~\ref{fig:calibration_default_latitude} quantifies our knowledge of
how spots are distributed on the surfaces of the stars in the dataset.
These distributions are again consistent with the true distribution used
to generate the spots (orange curve) within less than 2 standard deviations.%
\footnote{%
    If the results in Figure~\ref{fig:calibration_default_latitude}
    seem biased, recall from Figure~\ref{fig:calibration_default_corner}
    that the mean of the latitude distribution is
    consistent with the truth at $2-3\sigma$. That is roughly the difference
    between the orange curve and the average of the blue curves. As we
    will see, inference with a larger ensemble
    (Figure~\ref{fig:calibration_default_1000})
    allows us to infer the mean latitude to within about two degrees.
}

Even though we explicitly marginalized over inclination, we can still
derive posterior constraints on the inclinations of the individual stars in
our ensemble by computing the log-likelihood as a function of $I$
conditioned on the value of $\pmb{\theta}_\bullet$ from a particular draw from
the posterior. We do this in Figure~\ref{fig:calibration_default_inclination},
where blue curves again correspond to samples from the posterior hyperdistribution,
and orange lines indicate the true inclination; the panels are arranged in the
same order as those in Figure~\ref{fig:calibration_default_data}.
In almost all cases, we are able to constrain the inclinations of individual
stars to within about $10^\circ$, consistent with the truth at less than
$2{-}3\sigma$.

\begin{figure}[p!]
    \begin{centering}
        \includegraphics[width=0.85\linewidth]{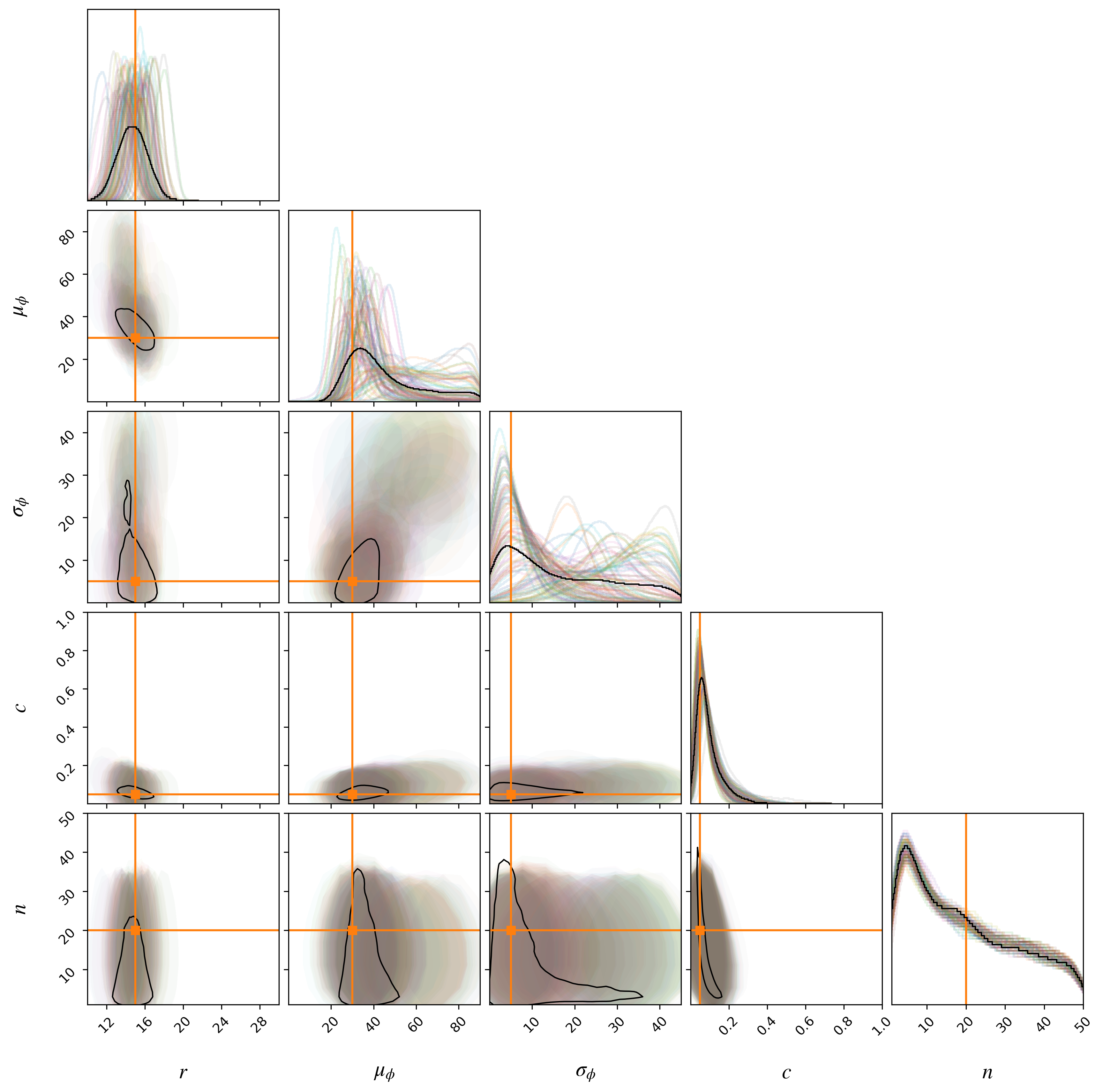}
        \oscaption{calibration_batch_default}{%
            Similar to Figure~\ref{fig:calibration_default_corner}, but
            showing the posterior distribution for 100 different synthetic
            datasets, all generated from the same default input parameters
            (Table~\ref{tab:synthetic}). Each colored histogram corresponds to a single
            run, with the corresponding $1\sigma$ contours shown as shaded ellipses for each pair
            of parameters. The black histograms correspond to the distributions of all
            samples from each of the 100 runs, and the black curves again indicate
            $1\sigma$ contours in the joint posterior.
            \label{fig:calibration_batch_default_corner}
        }
    \end{centering}
\end{figure}

\begin{figure}[p!]
    \begin{centering}
        \includegraphics[width=0.5\linewidth]{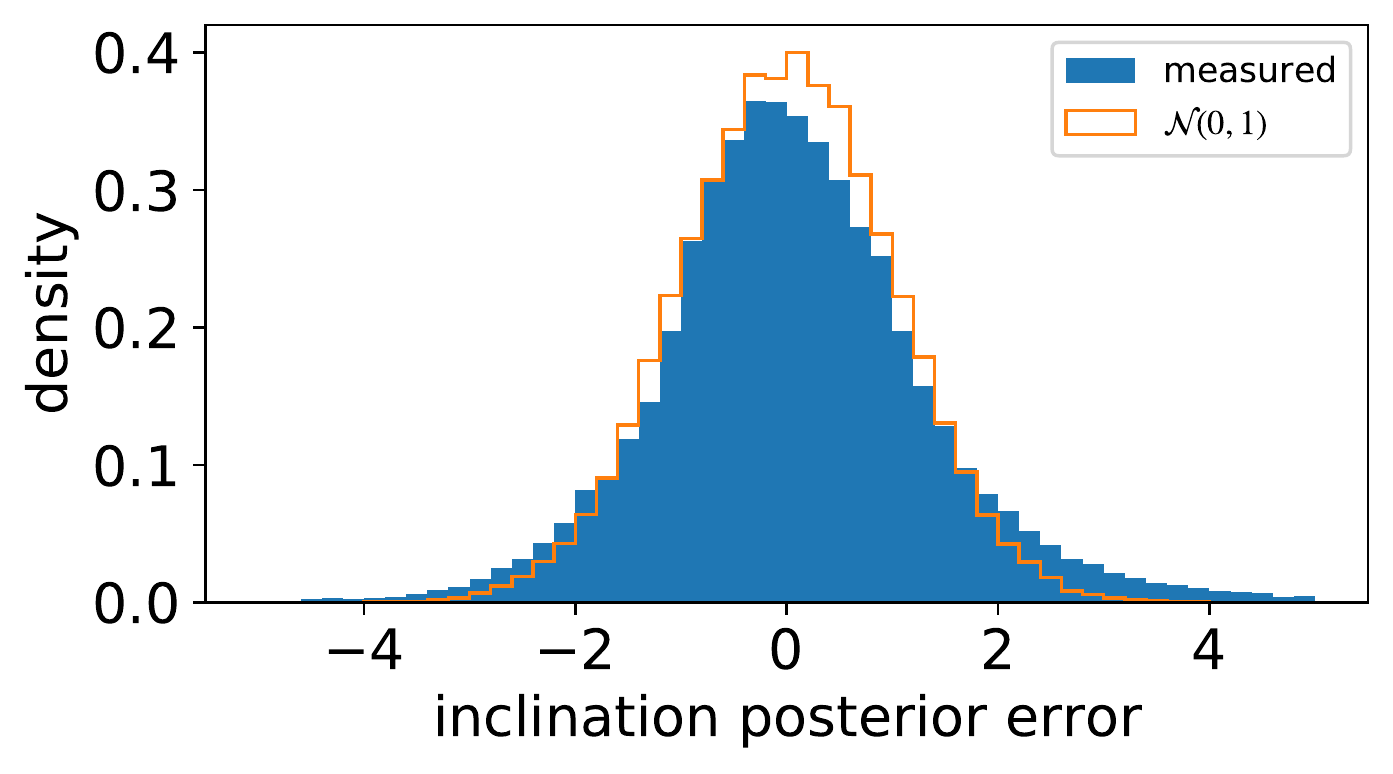}
        \oscaption{calibration_batch_default}{%
            Distribution of stellar inclination residuals normalized to the posterior
            standard deviation for all $5{,}000$ stars across the 100 trials in
            Figure~\ref{fig:calibration_batch_default_corner} (blue). The
            standard normal distribution is shown in orange for comparison.
            The inclination posteriors inferred with our GP are largely unbiased and
            have the expected variance.
            \label{fig:calibration_batch_default_inclinations}
        }
    \end{centering}
\end{figure}

\begin{figure}[t!]
    \begin{centering}
        \includegraphics[width=0.85\linewidth]{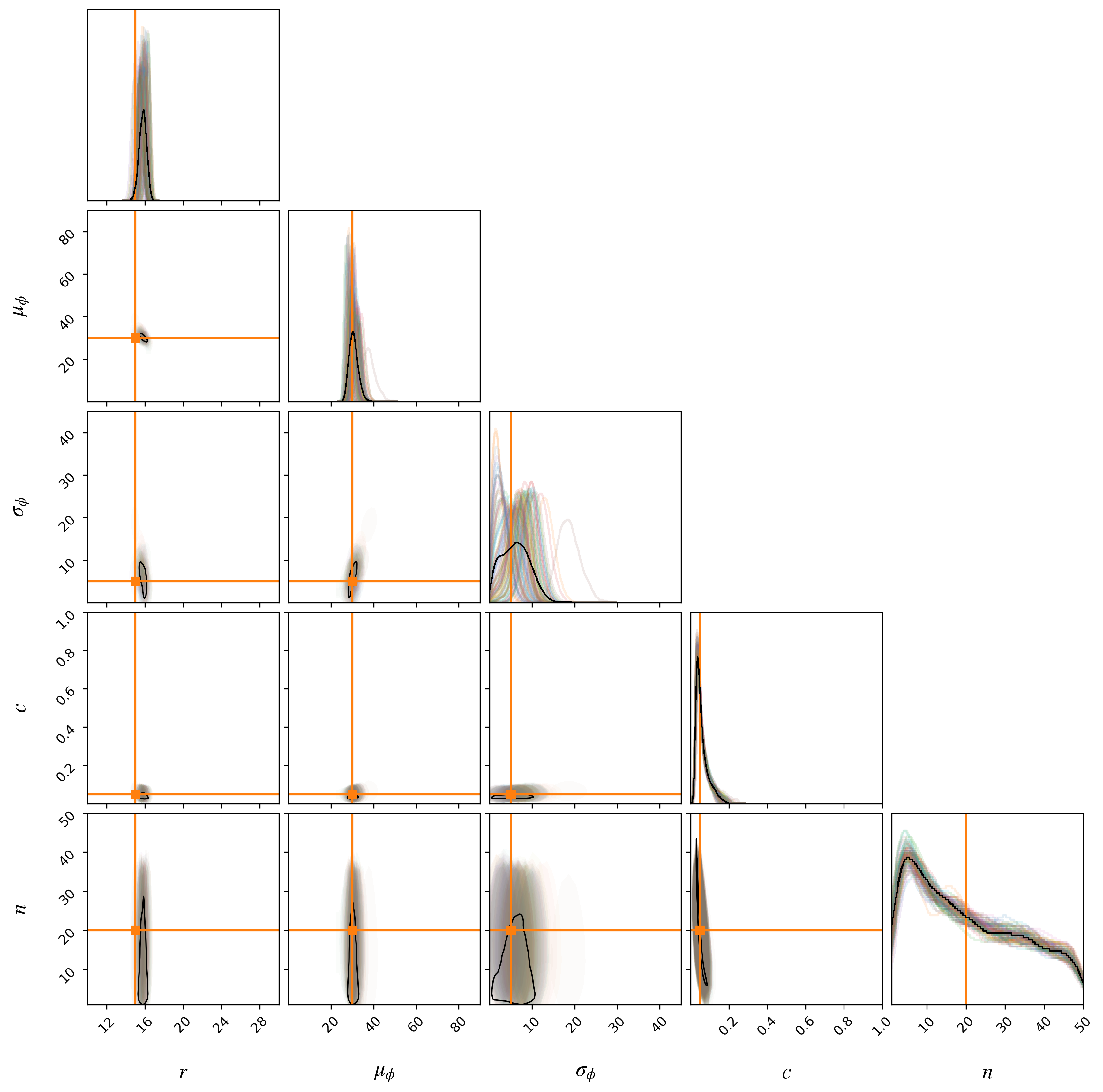}
        \oscaption{calibration_batch_default_1000}{%
        Same as Figure~\ref{fig:calibration_batch_default_corner}, but
        showing the posterior distributions of 100 realizations of ensembles
        with $M=1{,}000$ (instead of 50) light curves each. The constraints on
        mosts of the parameters are much tighter, and the polar spot degeneracy
        is gone.
        \label{fig:calibration_batch_default_1000}
        }
    \end{centering}
\end{figure}

The results in
Figures~\ref{fig:calibration_default_corner}---\ref{fig:calibration_default_inclination}
are based on a single run: i.e., a single realization of the light curve ensemble conditioned on
the properties of Table~\ref{tab:synthetic}. To properly gauge potential biases
in our model, it is useful to perform the run under many different realizations
of the dataset. We therefore generate 100 ensembles of light curves in
exactly the same way as above and perform inference on each of them.
Figure~\ref{fig:calibration_batch_default_corner} shows the marginal and joint
posterior distributions for $\pmb{\theta}_\bullet$ for all 100 trials.
Posteriors for individual trials are shown as the transluscent colored curves
(in the marginal plots) and as ellipses bounding the $1\sigma$ posterior level
(in the joint posterior plots). The black curves show the marginal distributions of all
samples across all trials, and the black contours show the corresponding $1\sigma$
levels in the joint posterior plots.

If our model is truly unbiased, in the limit of an infinite number of realizations
of the data, the expectation value of the distribution
of samples across all ensembles (the mean of the black curves in the marginal
posterior plots) should coincide with the true values (orange lines).
This is approximately
the case with the spot size $r$ posterior: on average, the posterior
distributions are centered on the correct value.
However, this is not the case for
the spot latitude parameters $\mu_\phi$ and $\sigma_\phi$, for which our
posterior means are biased high. While the \emph{modes} of their
posteriors are very close to the true values, the distributions have long
tails extending to high latitudes and high variance, respectively.

The reason for this bias has to do with the normalization degeneracy discussed in
\citetalias{PaperI}:
the total spottiness of a stellar surface is not an observable in single-band
photometry. In particular, this means spots near the poles lie \emph{almost}
entirely in the null space. Applied to the problem at hand, this degeneracy
makes it difficult to distinguish between stars with spots concentrated
exclusively at mid-latitudes (in this case, the truth) and
stars with spots centered closer to the poles but with large latitudinal
variance. The latter configuration leads to many spots close to the poles,
whose effect on the (relative) light curve is negligible, and some spots
at mid-latitudes, whose effect on the light curve is similar to that of
the former configuration.
Thus, the data alone cannot be used to discriminate between these two
scenarios, introducing the degeneracy we see in the posterior. In fact,
it is clear that in the tails of the distribution, the mean spot latitude
and the standard deviation of spot latitudes are positively correlated.
The bias we see is therefore not a shortcoming of the model, but of the
\emph{data} itself. To get around this, we either need to impose stronger
priors on $\mu_\phi$ and $\sigma_\phi$ (\S\ref{sec:caveats}),
observe in multiple wavelength bands
\citepalias{PaperI}, or simply collect more data.
As we will see below, the particular degeneracy described above is not perfect:
for very large ($M \sim 1{,}000$) ensembles, high-variance polar spots
can confidently be ruled out.

The posteriors for the contrast $c$ and the number of spots $n$
are mostly unbiased.
The contrast distribution has a bit of
a tail; inspection of Figure~\ref{fig:calibration_batch_default_corner}
reveals that it too is positively correlated with the mean spot latitude
and therefore suffers from the same degeneracy as above. And while the
mean of the spot number distribution is roughly correct, the posterior
is nearly unchanged across all runs, and equally uninformative in all
of them. This is yet another manifestation of the normalization degeneracy:
the total number of spots is not an observable in single-band photometry
\citepalias{PaperI}.

There is one final distribution that is instructive to consider: the
distribution of errors on the inferred stellar inclination.
Figure~\ref{fig:calibration_batch_default_inclinations} shows a histogram
of stellar inclination residuals (posterior mean minus true value)
normalized to the posterior standard deviation for all stars across the
100 trials described above. For a correctly calibrated model, this distribution
should equal the standard normal $\mathcal{N}(0, 1^2)$ in the limit of
infinite trials. This, in fact, is roughly what we find (compare to the
orange histogram in the figure). Our posterior has marginally
heavier tails, meaning we tend to \emph{slightly} underestimate the
posterior variance, but in general it is an excellent estimator of
individual stellar inclinations.

Finally, Figure~\ref{fig:calibration_batch_default_1000} shows
the same posterior distributions as in Figure~\ref{fig:calibration_batch_default_corner},
but for 100 runs each with $M=1{,}000$ light curves. In addition to the
constraints on all parameters (except the number of spots) being much
tighter, the larger amount of data breaks the polar spot degeneracy discussed
above. Given enough light curves, the model is capable of differentiating between
concentrated mid-latitude spots and high-latitude spots with large variance.
Interestingly, the inferred radius appears to be biased high by a small amount.
This is likely due to the fact that our prescription for
generating the spots (\S\ref{sec:calibration-setup}) is different from how
we actually model these spots. While we generate the spots as compact
circular disks expanded at high spherical harmonic degree, we model them
as sigmoids (\S\ref{sec:size}) expanded at significantly lower spherical
harmonic degree. Some minor disagreement is therefore to be expected in the inferred
radii.

\subsection{Other runs}
\label{sec:inference-other}
In this section we test the robustness of our model by changing one or
more of the fiducial values listed in Table~\ref{tab:synthetic}.
Each of the runs below corresponds to a single realization of the ensemble
dataset, and the corresponding figures are presented at the end of the Appendix.

Figures~\ref{fig:calibration_midlat}---\ref{fig:calibration_isotropic}
show the results for different latitudinal
distributions, keeping all other values in Table~\ref{tab:synthetic} the same.
Specifically, Figure~\ref{fig:calibration_midlat} corresponds to a run
with mid-latitude ($\mu_\phi = 45^\circ$ and $\sigma_\phi = 5^\circ$) spots,
Figure~\ref{fig:calibration_hilat} to a run with high-latitude
($\mu_\phi = 60^\circ$ and $\sigma_\phi = 5^\circ$) spots,
Figure~\ref{fig:calibration_equatorial} to a run with equatorial
($\mu_\phi = 0^\circ$ and $\sigma_\phi = 5^\circ$) spots, and
Figure~\ref{fig:calibration_isotropic} to a run with isotropically-distributed
spots ($\bbphi \sim \cos$).
The results are largely consistent with those of the default run: in all
cases we infer the correct spot radius, the mean and standard deviation of
the spot latitude, and the spot contrast within $2-3\sigma$; the number of
spots is equally unconstrained in all runs.
In Figure~\ref{fig:calibration_hilat} and to a lesser extent in
Figure~\ref{fig:calibration_midlat}, the polar spot degeneracy discussed
above is evident, particularly in the lower panels showing the latitudinal
distribution of spots. Nevertheless, the distribution peaks near the correct
latitude in both cases.
Figures~\ref{fig:calibration_equatorial} and \ref{fig:calibration_isotropic}
are interesting because, while the true latitude distribution is unimodal,
most of the posterior samples are not. In the equatorial case, the posterior
peaks at very low (but nonzero) latitudes and $\sigma_\phi$ appears to be
inconsistent with the true value at many standard deviations; however, recall
that $\sigma_\phi$ is a \emph{local} approximation the standard deviation of the
PDF at the mode (Appendix~\ref{sec:lat}), which deviates from the true
standard deviation (i.e., the
square root of the variance, computed from the expectation of the second
moment of the distribution) when the two modes are very closely spaced.
In fact, the latitude PDF samples (lower panel in the figure) nearly span
the true distribution, to the extent that our parametrization of the
latitude distribution can approximate a zero-mean Gaussian. While the Beta
distribution in $\cos\phi$ \emph{can} be unimodal in $\phi$ (see
Equation~\ref{eq:phi-pdf} and the first
column of Figure~\ref{fig:latitude_pdf}), this happens \emph{only} when
$\beta = 0$, which occupies an infinitesimally thin hyperplane in
parameter space. In practice, the majority of the posterior mass will be close
to but not exactly at $\beta = 0$, leading to the bimodality in the figure.
The same argument applies to Figure~\ref{fig:calibration_isotropic}.
In both cases, the posterior approximates the true distribution as best it
can given the constraints of the adopted PDF.

Figure~\ref{fig:calibration_tinyspots} tests the performance of the model
on light curves of stars with spots much smaller than the effective resolution
of the GP. Our expansion to $l_\mathrm{max} = 15$ only allows us to model spots
with radii $r \gtrsim 10^\circ$ (see Figure~\ref{fig:spot_profile}), so we
place zero prior mass below this value. The figure shows the results of inference
on a dataset generated from spots with $r = 3^\circ$ (and an increased
contrast $c = 1$ to enforce a comparable signal-to-noise to the other trials).
On the Sun, these would correspond to spots with diameters of about
$70{,}000$ km---typical of the larger spots during solar maximum.
While the radius posterior is biased (as it \emph{must} be, given our prior),
the fact that it peaks at the lower bound of the prior suggests the presence
of spots smaller than the model can capture. More importantly, however, the
latitudinal parameters are inferred correctly and at fairly high precision:
even though or model is biased against small spots, this does not affect
inference about their latitudes. On the other hand, the spot contrast is wrong
by many standard deviations, since the model must compensate for the fact
that the radii are biased high with a lower contrast to match the variability
amplitude of the light curves.

Figures~\ref{fig:calibration_default_1} and \ref{fig:calibration_default_1000}
show results for the default run but with extreme values of the number of
light curves in the ensemble: $M = 1$ and $M = 1{,}000$, respectively.
These two figures underscore the power of ensemble analyses: a single light
curve (Figure~\ref{fig:calibration_default_1}) is simply not informative
enough about the properties of its spots. On the other hand, a very large
ensemble can be \emph{extremely} informative: the radius, latitude, and even
the contrast are inferred correctly at high precision.

Figure~\ref{fig:calibration_ld}---\ref{fig:calibration_ld_500_no_ld}
show results for the default run but with limb darkening. In all cases
we assume quadratic limb darkening with fiducial values $u_1 = 0.5$
and $u_2 = 0.25$ for all stars. From Figure~\ref{fig:calibration_ld}, in which we
assume we know the limb darkening coefficients exactly, it is clear that
the presence of limb darkening significantly degrades our ability to
infer both the radii and latitudes of the spots. Limb darkening has a complicated
effect on the mapping between surface features and disk-integrated flux, as it
reveals information about odd harmonics at
the expense of introducing strong degeneracies with the even harmonics
\citepalias{PaperI}. In practice, this leads to higher uncertainty in
the spot radii and latitudes relative to the same dataset without limb
darkening (Figure~\ref{fig:calibration_default_corner}). Fortunately,
this uncertainty can be dramatically reduced with more data, as evident
in Figure~\ref{fig:calibration_ld_1000}, which shows the results of the
same run but with $M=1{,}000$ light curves in the ensemble. The constraints
on $r$, $\mu_\phi$, and $\sigma_\phi$ are now much tighter and in
good agreement with the truth. Finally, Figure~\ref{fig:calibration_ld_500_no_ld}
shows the results of inference on limb-darkened light curves under the
(wrong) assumption that limb darkening is not present ($\mathbf{u} = \mathbf{0}$).
Neglecting the effect of limb darkening can lead to biases in the spot radius
and latitude parameters. While the model still favors mid-latitude spots
(at $\sim 45^\circ$ instead of $30^\circ$), the constraints are deceptively tight
and discrepant by many standard deviations. We discuss these points in
more detail in \S\ref{sec:caveats}.

The runs so far correspond to stars with many ($n = 20$) spots, for which the
resulting light curves are smooth due to the fact that many spots are in
view at any given time. Figures~\ref{fig:calibration_hicontrast} and
\ref{fig:calibration_bigspots} show what happens when the model is applied to
stars with $n=2$ and $n=1$ spots, respectively. Despite large portions of the
light curves being flat (and therefore extremely non-stationary)
in these scenarios, the GP does surprisingly well, recovering the radii
and latitude parameters within $2-3\sigma$ in both cases.
Note that in order to preserve the same signal-to-noise ratio relative to
the other runs, we gave the spots in Figure~\ref{fig:calibration_hicontrast}
a much higher contrast ($c = 0.5$). Even though the contrast is degenerate
with the number of spots (which is very poorly constrained), the $c$ posterior
has a much heavier tail than in the other runs.
Thus, in spite of the arguments in \citet{PaperI} about the difficulty in
constraining $c$ and $n$ from single-band photometry, it is evident that
the full covariance structure of the data encodes \emph{some} information
about the contrast and---to a much lesser extent---the number of spots.
In Figure~\ref{fig:calibration_bigspots}, we compensate for the smaller number
of spots by increasing the spot radius to $r = 45^\circ \pm 5^\circ$ instead,
showing that the model can accurately model large spots, even in the
presence of some (unmodeled) scatter in their sizes.

In Figure~\ref{fig:calibration_variance} we add variance to all the spot
properties when generating the light curves: we add $n = 20 \pm 3$ spots
to each star with radii $r = 15^\circ \pm 3^\circ$, contrasts
$c = 0.05 \pm 0.01$, and at latitudes $\phi = 30^\circ \pm 5^\circ$.
As before, we only explicitly account for the variance of the latitude distribution in our
model. We correctly infer the latitude parameters and the contrast, but
our radii appear to be biased high. This is likely due to the fact that
larger spots have a bigger impact on the signal, so our inferred radius
is a weighted average of all spot radii. In Appendix~\ref{sec:size} we
derive an expression for the moment integrals of the spot size distribution
assuming a uniform distribution between $r - \Delta r$ and $r + \Delta r$
(instead of a delta function at $r$), which can be used to compute the
GP if one wishes to explicitly account for scatter in the spot sizes.
We find that repeating the run shown in Figure~\ref{fig:calibration_variance}
while explicitly sampling over the distribution in $\Delta r$ shifts the
posterior mass to lower radii, mitigating the bias described above.

Our final run is shown in Figure~\ref{fig:calibration_unnorm}, in which
we assume we know the true normalization of each light curve. That is, we
assume that we can measure all light curves in units of the flux we would
measure if the stars had no spots on them, and we \emph{do not}
normalize them (see \S\ref{sec:gp-norm}). In practice, this would require
knowledge of the brightness (or temperature) of the unspotted
photosphere, which is not an observable in single-band photometry.
This value can in principle be probed, however, in multi-band photometry
\citep[e.g.,][]{Gully2017,Guo2018}, for which this run is
extremely relevant. We again recover the radii and latitude parameters
to within $2-3\sigma$, but most importantly, we also infer the correct
spot contrast \emph{and} the correct number of spots with fairly high
precision. In particular, knowledge of the correct normalization breaks the $c-n$ degeneracy.
Photometric measurements in multiple bands (even just two!)
are therefore extremely useful when inferring spot properties.
We discussed this point in \citetalias{PaperI}.

\section{Discussion}
\label{sec:discussion}

\subsection{Small spots}
\label{sec:tinyspots}

One of the biggest downsides of adopting a spherical harmonic representation
of the stellar surface as the foundation of our GP is the inherent limitation
it imposes on the resolution of surface features. In order to maximize
computational efficiency and numerical stability, our default approach is to
model the surface using an expansion up to degree $l_\mathrm{max} = 15$,
which can model features only as small as
$\sim \nicefrac{180^\circ}{15} = 12^\circ$ across. Even on scales
slightly larger than this, the presence of ringing can be seen
(see panel (b) in Figure~\ref{fig:samples}, where ringing is just barely
noticeable in the equatorial region of the maps). The spherical harmonic basis consists
of global modes, all of which contribute to the intensity everywhere on the surface.
Localized features require constructive interference of modes inside
and destructive interference of modes outside, often leading to a wave-like ringing
pattern that gets worse as the size of the features gets smaller. Taken at
face value, this might suggest that a different basis---such as the common
choice of pixels on a grid, or perhaps a localized wavelet basis---would be better
at modeling small spots. While this is probably true, it may be quite
difficult to find closed-form expressions for the expectation integrals
(\S\ref{sec:integrals}) that make the GP covariance evaluation tractable.

One option is to bypass the computation of the covariance on the surface of
the star and to write down an expression for the flux directly in terms of
the properties of a starspot. Circular spots of uniform intensity can be
modeled as spherical caps, which are simply segments of ellipses when
projeted onto the sky. It is possible to express in closed form the projected area
they cover---and hence their contribution to the flux $\mathbbb{f}$---even in the case
where part of the spot is on the hemisphere facing away from the observer.
This was done in \citet{Luger2017}, who solved this problem in closed form
(see \S3.3.2 and Appendix A.3 in that paper). Computing the GP is then a matter of
integrating $\mathbbb{f}$ and $\mathbbb{f}\,\mathbbb{f}^\top$, weighted by
the hyperparameter PDFs, as we did in \S\ref{sec:gp-gp}.
However, the expression for
$\mathbbb{f}$ involves square roots and arctangents of functions of the spot
latitude and longitude
\citep[see Equation~40 in][]{Luger2017},
so computing these integrals in closed form is
likely to be very difficult. Even if a closed form solution can be found,
incorporating limb darkening (which we argued can be extremely important)
poses an even greater challenge. It is likely that several simplifications
must be made in order to make this approach tractable. This was done to some
extent in \citet{Morris2020}, who derived a closed form expression for the flux by
ignoring certain projection effects, such
as the self-occultation of large spots by the limb of the star, and neglected
variations in limb darkening within spots. Such a model could admit a closed-form
solution to the GP covariance and may be better at capturing the effects
of small spots, at the expense of the ability to model larger spots.

In principle, small spots can be modeled under our current
approach with negligible ringing by simply
increasing the degree of the spherical harmonic expansion. As we discuss in
\citep{JOSSPaper}, however, the algorithm presented here becomes
unstable for $l \gtrsim 15$, so doing so would require a reparametrization
of the equations in the Appendix. Importantly, however, as we showed in
\S\ref{sec:calibration}, the current implementation of the GP is suitable
to modeling light curves of stars with spots smaller than the limiting
resolution of $\sim 10^\circ$. Consider Figure~\ref{fig:calibration_tinyspots},
which shows the results of doing inference on an ensemble of light curves
of stars with small ($r = 3^\circ$) spots; these are comparable to some of
the larger spots seen on the Sun with diameters of about $70,000$ km.
Even though our inferred radius and contrast are wrong, the fact that the radius
posterior peaks
at the lower prior bound of $10^\circ$ is strongly suggestive of the presence
of spots smaller than the resolution of the model. Moreover, the
latitude mode and standard deviation posteriors are \emph{unbiased}, and we
correctly infer the presence of low-variance, mid-latitude spots.
With the above caveats in mind, our GP can therefore still be used to model
stars with small spots.

\subsection{Bright spots}
\label{sec:brightspots}

\begin{figure}[t!]
    \begin{centering}
        \includegraphics[width=\linewidth]{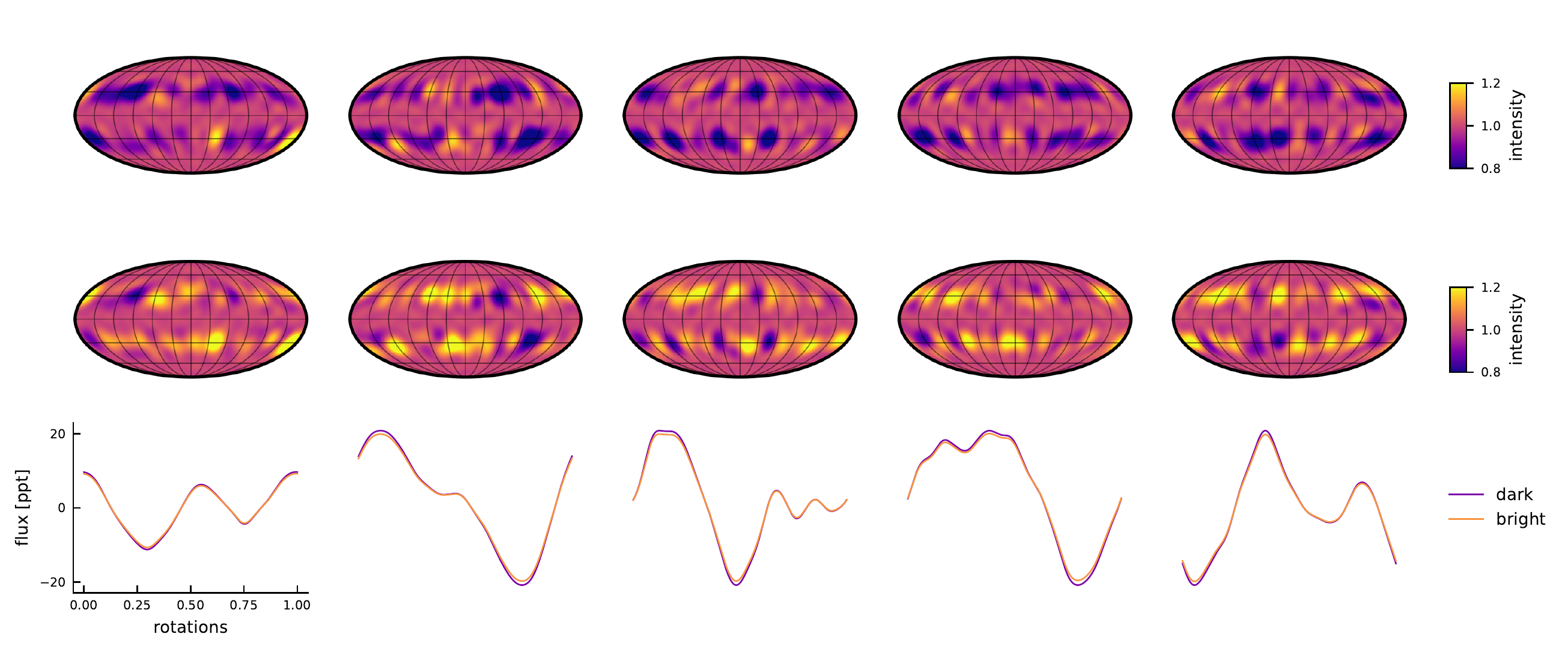}
        \oscaption{brightspots}{%
            Comparison of GP samples with dark spots (top) and bright
            spots (center) alongside their corresponding light curves
            viewed at $I = 75^\circ$ (bottom). Without knowledge of
            the correct normalization (\S\ref{sec:gp-norm}), it is very
            difficult to differentiate between the two from stellar
            light curves.
            \label{fig:brightspots}
        }
    \end{centering}
\end{figure}

All the calibration tests performed in the previous section assumed the
stellar surface was dominated by dark spots. We can easily model the
effect of bright spots by choosing negative values for the contrast
parameter, i.e. $c < 0$. The top two panels of Figure~\ref{fig:brightspots}
show five random samples from the GP with dark spots ($c = 0.1$) and
bright spots ($c = -0.1$), respectively; the random seed is the same for
both panels, so the maps are identical in all other respects.
While the surface maps can be easily distinguished by eye, the same is not
true for the corresponding light curve samples (bottom panel), which
are almost identical. There is a slight difference in amplitude between
the two cases: surfaces with dark spots have slightly higher light curve
amplitudes than surfaces with bright spots of the same contrast magnitude.
However, the magnitude of the bright spots can be increased slightly to get
a near-perfect match to the dark spot light curves, meaning it may be
difficult (if not impossible) to tell the difference between dark and bright
spots via the GP approach.

The reason for this degeneracy is rooted (once again) in the
fundamental issue with photometry: we lack any information about
the correct normalization of the light curve.
Consider the dependence of the (unnormalized) GP covariance on the
contrast: it enters in a single place, via Equation~(\ref{eq:bigEc}),
as $c^2$, meaning dark and bright spot models have exactly the same
covariance. These models differ only in the mean of the unnormalized
process, since that is proportional to $c$ (via Equation~\ref{eq:ec}).
However, as we argued in \citetalias{PaperI}, the mean is not a direct
observable. Instead, in single-band photometry, we are only sensitive to
the \emph{ratio} of the covariance to the square of the mean
(see Equation~\ref{eq:SigmaTilde}). From that equation, we can deduce
that stars with dark spots (for which
the light curve mean $\mu < 1$) will therefore have larger variance
than stars with bright spots ($\mu > 1$), leading to the slight difference
seen in the figure. However, since this is strictly a multiplicative
factor affecting the covariance, it is degenerate with the two other
properties that scale the covariance: the magnitude of the contrast and
the total number of spots.%
\footnote{There is also a small additive term in
    Equation~(\ref{eq:SigmaTilde}), but this, too, depends only on the ratio
    of entries in the covariance matrix to the mean, so it is of little help
    in breaking the degeneracy.}

We therefore conclude that \textbf{single-band photometry is largely insensitive to
    the difference between bright spots and dark spots}. However, it is
important to bear in mind that the
degeneracy described above exists only for the \emph{Gaussian approximation}
to the likelihood function. As we argued earlier, the true likelihood function
is not a Gaussian; in particular, the true probability distribution
has higher-order moments that we do not model here. These moments should
in principle encode information about the sign of the spot contrast, but they
may be very difficult to infer in practice. It may be possible to distinguish
between dark spots and bright spots with traditional forward models of stellar
surfaces, but (as we argued earlier) a statistically rigorous ensemble analysis
of stellar light curves using such forward models is probably computationally
intractable.

We can, however, skirt this degeneracy with observations in multiple bands,
which can provide limited information about the correct normalization. Recently,
\citet{Morris2018} used approximately coeval \emph{Kepler} and \emph{Spitzer}
light curves of TRAPPIST-1 to argue that a bright spot model for the star
is more consistent with the data. A detailed exploration of the
effect of multi-band photometry on the degeneracies of the mapping problem
is deferred to a future paper in this series.

\subsection{Comparison to other work}
\label{sec:other-work}

\subsubsection{Synthetic likelihoods, random fields, and approximate inference}

The core idea behind the methodology presented in this paper---to compute the
Gaussian approximation to an intractable multidimensional distribution in order
to obtain a likelihood function for inference---is not new. Although the
method likely goes by different names in different fields, it is a popular
technique particularly in the field of ecological population dynamics,
where it is referred to as the synthetic likelihood \citep[SL;][]{Wood2010}
or Bayesian synthetic likelihood \citep[BSL;][]{Price2018} method. In many
ecological systems, population growth is a chaotic process; observations of the
size of a population over time can be dominated by steep spikes and drops
in the population that occur due to sudden, random environmental pressure.
While population growth can be forward modeled with ease, it is very
difficult to use forward models to constrain basic growth parameters in an
inference setting, since that requires marginalization over the extremely
nonlinear noice processes. As a way around this, \citet{Wood2010} introduced
the SL method, in which, conditioned on a set of parameters of interest
$\pmb{\theta}$, one computes the forward model $f(\pmb{\theta})$ many times
under different realizations of the noise, and adopts the sample mean
and sample covariance (usually of a summary statistic of the data)
as the mean and covariance of the Gaussian likelihood
function $p(f \, | \, \pmb{\theta})$. \citet{Wood2010} showed that,
provided the number of forward model samples is large enough, this
``synthetic'' likelihood allows one to infer the population growth parameters
efficiently and without bias.

The method presented in this paper may be thought of as a synthetic likelihood
method in the limit of an infinite number of forward model samples. Unlike
\citet{Wood2010}, whose method determines the mean and covariance of the
distribution of some function of $f$ conditioned on $\pmb{\theta}$ by sampling, we are able
to actually compute the mean and covariance of $f$ directly \emph{in closed form}.
While traditional SL methods are inherently noisy, our method employs
the \emph{exact} Gaussian approximation to the likelihood function.

Our GP is also closely related to techniques commonly employed in models
of the cosmic microwave background (CMB). In particular, it is a type of
Gaussian random field (GRF) on the sphere, which is frequently used to
model perturbations in the CMB \citep{Wandelt2012}. In general, however,
GRFs used in cosmology are isotropic: when expressed in the spherical
harmonic basis, their covariance matrix is diagonal and admits a
representation as a (one-dimensional) power spectrum. Our GP, in contrast,
is anisotropic in
the polar coordinate (i.e., the latitude) by construction.

Our method is also related to various families of approximate inference, such as
variational inference (VI), in which a multivariate Gaussian is used to
approximate the \emph{posterior} distribution \citep[e.g,][]{Blei2016},
or to approximate Bayesian
computation (ABC), in which an (often intractable) likelihood function is
replaced with an approximation computed from simulations from the prior
\citep[e.g.,][]{Beaumont2019}.

\subsubsection{Starspots and stellar variability}

The methodology developed in this paper is closely related to that in
\citet{Perger2020}, who studied the effect of different starspot
configurations on the autocorrelation and covariance of stellar
radial velocity (RV) measurements. The authors of that study compared the
performance of various commonly used quasi-periodic kernels when applied
to synthetic RV datasets, arguing that a new four-parameter quasi-periodic
cosine kernel (QPC) can better capture the variability due to starspots.
However, their study was empirical and related spot configurations to their
effect on the covariance structure of the data primarily in a qualitative
fashion. Their QPC kernel is a function of two interpretable hyperparameters
(the rotation period and a spot timescale) as well as two amplitudes, which
are not explicitly related to physical spot properties. Our GP, in contrast,
is built from the ground up such that all of its hyperparameters directly
correspond to physical spot properties, allowing one to use it in starspot
inference (not just marginalization) problems. While the methodology presented
here applies to photometry, it is possible to extend it to model RV
datasets as well; we discuss this in \S\ref{sec:extensions}.

Recently, \citet{Morris2020} used \emph{Kepler}, \emph{K2}, and \emph{TESS}
light curves to derive a relationship between stellar age and spot
coverage using an ensemble analysis similar to that proposed here.
Because of the intractability of the marginal likelihood function,
that study used an approximate Bayesian computation (ABC) method to
infer spot properties from a large ensemble of stars. \citet{Morris2020}
developed a fast, approximate forward model for light curves of spotted stars
\citep[\textsf{fleck};][]{Morris2020b}, which they used to generate a large
number of prior samples for different values of the spot radii, contrasts,
and latitude distributions. For each collection of samples generated from a
given set of hyperparameters, \citet{Morris2020} computed the distribution of
the ``smoothed amplitude,''
the peak-to-trough difference of the (normalized, de-trended) light curve.
This distribution was then compared to the distribution of observed
smoothed amplitude values among stellar clusters of different ages
within an ABC algorithm, yielding approximate
posterior distributions for the hyperparameters as a function of
stellar age.
While we believe the spot coverage results of that paper are predominantly driven
by the prior \citep[due to the strong degeneracy between the spot contrast
    and the number of spots; see \S4.2 in][]{PaperI},
the ensemble analysis employed in that paper is nevertheless a powerful technique
to infer spot properties.
Our work builds on that of \citet{Morris2020} by deriving a closed form
solution to the likelihood function (as opposed to a sample-based
likelihood-free inference algorithm) and by harnessing the covariance
structure of the data when doing inference (as opposed to relying solely
on the amplitude of the data).

Finally, \citet{Basri2020} recently presented a large suite of forward models of light curves of
spotted stars, which they used to discuss the (complicated) dependence of
various light curve metrics on
the physical spot parameters used to generate the data. They concluded that it
is not possible to uniquely relate these metrics to the underlying starspot
configuration. While we agree this is the case for individual stars, our work
stresses that it is possible to circumvent many of these degeneracies with
ensemble analyses.
\citet{Basri2020} also concluded it is not in general possible to
uniquely disentangle differential rotation from spot evolution when their
timescales are comparable; nor is it possible to confidently measure a rotation
period when the evolution timescale is very short. However, their study
relied on the effect these processes have on simple light curve metrics,
which are almost certainly not sufficient statistics of the data. Inference
that takes into account the full covariance structure of the data, while
considering large ensembles of light curves, could in principle break
this degeneracy. While we do not explicitly model differential rotation in
this paper, it will be the subject of a future paper in this series.

\vfill
\pagebreak

\subsection{Caveats}
\label{sec:caveats}

We conclude our discussion with a list of several notes and caveats that should be kept
in mind when using our algorithm and its \Python implementation.

\begin{enumerate}
    \item \emph{The assumed latitude distribution is not Gaussian}.
          Because we require the first and second moment integrals
          (Equations~\ref{eq:ephi} and \ref{eq:Ephi}) to have closed
          form solutions, there are restrictions on the probability
          density function we can assume for the spot latitude.
          We find that a Beta distribution in the cosine of the latitude
          is integrable in closed form and can be evaluated efficiently
          in terms of recursion relations. In many cases, particularly
          when $\mu_\phi \lesssim 75^\circ$ and $\sigma_\phi \lesssim 10^\circ$,
          the distribution in the spot latitude
          is close to a bi-modal Gaussian with mean $\pm\mu_\phi$ and
          standard deviation $\sigma_\phi$ (see Figure~\ref{fig:latitude_pdf}).
          In general, however, $\mu_\phi$ is formally equal to the \emph{mode}
          (as opposed to the mean) of the distribution, and $\sigma_\phi$ is
          the Laplace approximation to the local standard deviation at the mode.
    \item \emph{The number of spots $n$ does not have to be an integer}.
          Samples from our GP prior will not in general have exactly $n$
          spots (see, e.g., Figure~\ref{fig:samples}). This is due to the fact
          that our model is only an approximation to the true distribution
          of stellar surfaces conditioned on the spot properties. A corollary
          of this point is that $n$ need not be an integer, which makes it
          easier in practice to sample over using modern inference
          techniques such as MCMC, HMC, ADVI, and nested sampling.
    \item \emph{Care should be taken when modeling large-amplitude light curves}.
          We discussed this point at length in \S\ref{sec:gp-norm}. Modeling
          a light curve that has been normalized to its mean (or median) as a
          Gaussian process is conceptually a bad idea when the amplitude
          of variability is large compared to the mean. As a rule of thumb, if
          the amplitude of variability exceeds $\sim 10\%$, we recommend
          not normalizing the light curve in this way, and instead modeling
          the normalization amplitude as a latent variable.
    \item \emph{Keep in mind the polar spot degeneracy}. Even when modeling
          ensembles of light curves, there are still strong degeneracies at
          play \citepalias{PaperI}. In particular, spots centered on the poles are always in
          the null space, so it can be difficult in practice to rule out
          their presence. This can be seen in Figure~\ref{fig:calibration_hilat},
          in which the model cannot distinguish between spots localized
          at $60^\circ$ and polar spots with high latitude variance.
          It may thus be advisable to adopt a prior that favors small values
          of $\sigma_\phi$, such as the common inverse gamma prior for the
          variance. Alternatively, one could place an isotropic prior on
          the latitude (with density proportional to $\cos\mu_\phi$)
          to downweight very high latitude spots.
    \item \emph{Limb darkening matters!} The null space is extremely sensitive
          to limb darkening \citepalias{PaperI}. It is therefore
          extremely important to model it correctly; otherwise, there may
          be substantial bias in the inferred spot parameters. For stars with
          transiting exoplanets, it may be possible to infer the limb darkening
          coefficients empirically, but in general we recommend modeling them
          as latent variables with priors informed by theoretical models.
    \item \emph{Careful with the data}. In general, covariances can be very
          hard to estimate from noisy data. This makes it especially important
          to ensure one is correctly modeling the noise. When applying our GP
          to model real data, we recommend the usual
          inference practices of clipping outliers,
          modeling a latent white noise (jitter) term, and modeling
          a small latent additive offset term to minimize the risk of
          bias in the posteriors of interest.
    \item \emph{Careful with the sample selection}. When performing an ensemble
          analysis of stellar variability, it is tempting to only analyze light curves
          that show variability in the first place. \emph{This is extremely
              dangerous}, since the lack of variability could simply be due to low
          inclinations. It is extremely important to ensure the sample selection step does
          not introduce bias. If there is reason to believe there are two
          distinct populations within an ensemble---say, a population of active
          stars and a population of quiet (spotless) stars---we strongly recommend
          the use of a Gaussian mixture model.
\end{enumerate}

\section{Extensions}
\label{sec:extensions}

\subsection{Composite GPs}
\label{sec:mixture}
Thus far we have assumed that spots are concentrated at a single latitude
(above and below the equator). We baked this assumption directly into our choice of
distribution function for the latitude (\S\ref{sec:lat}), which has exactly
two modes at $\pm \mu_\sigma$. However, it is possible (at least in principle)
that certain stars could have two or more active latitudes, in which case
our GP is not an appropriate description of the stellar surface.

\begin{figure}[t!]
    \begin{centering}
        \includegraphics[width=\linewidth]{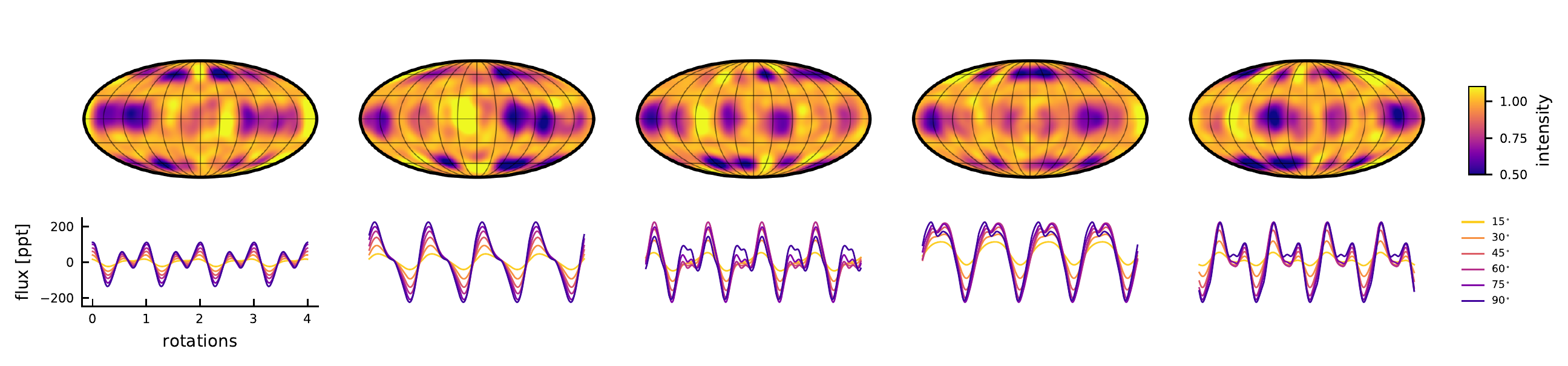}
        \oscaption{gpsum}{%
            Prior samples from a sum of two \starryprocess models. The
            first model consists of small ($r=10^\circ$)
            circumpolar ($\phi = 60^\circ \pm 3^\circ$) spots and the
            second model consists of larger ($r=20^\circ$)
            equatorial ($\phi = 0^\circ \pm 3^\circ$) spots.
            The sum of two \starryprocess models is also a
            \starryprocess, making it easy to model more complex
            distributions of spots.
            \label{fig:gpsum}
        }
    \end{centering}
\end{figure}

Fortunately, Gaussian distributions (and thus also Gaussian processes) are
closed under addition, meaning that the sum of two GPs is also a GP. We can thus
construct more complex models for stellar variability by summing GPs with different
spot hyperparameter vectors $\pmb{\theta}_\bullet$. The composite GP will then
have a mean equal to the sum of the means of each GP, and a covariance matrix equal
to the sum of the covariance matrices of each GP. One possible application of this
is to model stars with multiple active latitudes, as described above; an example
of this is shown in Figure~\ref{fig:gpsum}, where samples are drawn from a GP with
small circumpolar spots and large equatorial spots (see caption for details).
Since the composite GP inherits all of the properties of the standard GP, it can
be used to do inference under more complex priors than those presented here.

It is also worth noting that this technique may be employed to model arbitrary
distributions for parameters like the spot radius and the number of spots. In
the Appendix we present a formulation of our GP that admits a radius distribution
half-width parameter $\Delta r$; this generalizes our delta function distribution to
a uniform distribution between $r - \Delta r$ and $r + \Delta r$. One may then
compute the weighted sum of several GPs
with half-widths $\Delta \ne 0$ and central radii
$r = r_0$,
$r = r_0 + 2\Delta r$,
$r = r_0 + 4\Delta r$,
etc., to approximate \emph{any} distribution of spot radii. Similarly, one
may compute the weighted sum of several GPs with different values of the
number of spots $n$ to enforce any discrete distribution for that quantity.
The reader should keep in mind that the cost of computing the GP
covariance matrix will scale linearly with the number of GP components.
In many cases, however, the computational bottleneck is the covariance factorization
step \citep{JOSSPaper}, in which case adding components to the
model will result in negligible overhead.

\subsection{Time evolution}
\label{sec:temporal}

Another big limitation of the base algorithm is the implicit assumption that
stellar surfaces are static. Our GP hyperparameters $\pmb{\theta}_\bullet$
describe the \emph{spatial} configuration of starspots, but they say nothing
about their evolution in time. We know from observations of the Sun and of
\emph{Kepler} stars that temporal variability is extremely common: spots appear,
disappear, and even migrate in latitude over time. While it may be possible to
parametrize their evolution in a way that is general enough to capture all the
ways in which they may change over time, such an approach is beyond the scope
of the present paper. It is, however, straightforward to implement an
\emph{uninformative} temporal prior within the framework of our GP. To do this,
we will make two simplifying assumptions:

\begin{enumerate}
    \item \emph{The temporal process is stationary.} This implies that there is no
          preferred time (or phase) and that the spatial covariance is the
          same at all points in time. Stars may still have active longitudes
          under this assumption, but there is no preferred longitude
          \emph{across all stars}.
    \item \emph{The temporal and spatial covariances are independent.} This implies
          that the evolution of each spherical harmonic mode in time is independent of the
          evolution of any of the other modes in time.
\end{enumerate}

There is some tension between these assumptions and our knowledge of how stellar
surfaces evolve. Assumption (1) excludes surfaces whose total spottiness changes
significantly
or whose spots migrate in latitude, as both processes change the spatial covariance
over time. Assumption (2) ignores the correlation between spherical harmonic modes
due to the migration of spots, which requires the coherent evolution of many
modes at once. These assumptions likely limit the ability of our GP to model
light curves on very long baselines (i.e., on timescales of years) over which
stellar activity cycles take place. However, given that the use case of our
algorithm is likely to be the analysis of individual quarters of \emph{Kepler}
and individual sectors of \emph{TESS} data,
our assumptions are likely valid in most cases. Analyses of (say) all quarters of
\emph{Kepler} data could process each quarter at a time in a hierarchical
framework in which hyperparameters like the mean spot latitude in each quarter
are treated as functions of time.

The two assumptions listed above suggest a fairly straightforward form for the
GP covariance in spherical harmonics and time:
\begin{align}
    \label{eq:Sigmaty}
    \pmb{\Sigma}_\mathbf{y}^\mathbf{(t)} = \mathbf{K} \otimes \pmb{\Sigma}_\mathbf{y}
    \quad,
\end{align}
where $\mathbf{K}$ is a $(K \times K)$ matrix describing the
covariance among the $K$ points in time,
$\pmb{\Sigma}_\mathbf{y}$ is the $(N \times N)$ matrix describing describing
the covariance among the $N \equiv (l_\mathrm{max} + 1)^2$ spherical
harmonic coefficients (Equation~\ref{eq:cov_y}),
and $\otimes$ denotes the Kronecker product.
The quantity $\pmb{\Sigma}_\mathbf{y}^\mathbf{(t)}$
is the $(NK \times NK)$ temporal-spatial covariance, whose coefficient at index
$(Nk + n, Nk' + n')$ is the covariance between the spherical harmonic coefficient
$y_{n}(t_k)$ and the spherical harmonic coefficient $y_{n'}(t_{k'})$, where
the index $n$ is related to the spherical harmonic indices $l$ and $m$ via
Equation~(\ref{eq:n}). Finally, because of our assumption of stationarity,
the mean of the GP is still constant and equal to the mean of the static
process.

\begin{figure}[t!]
    \begin{centering}
        \includegraphics[width=\linewidth]{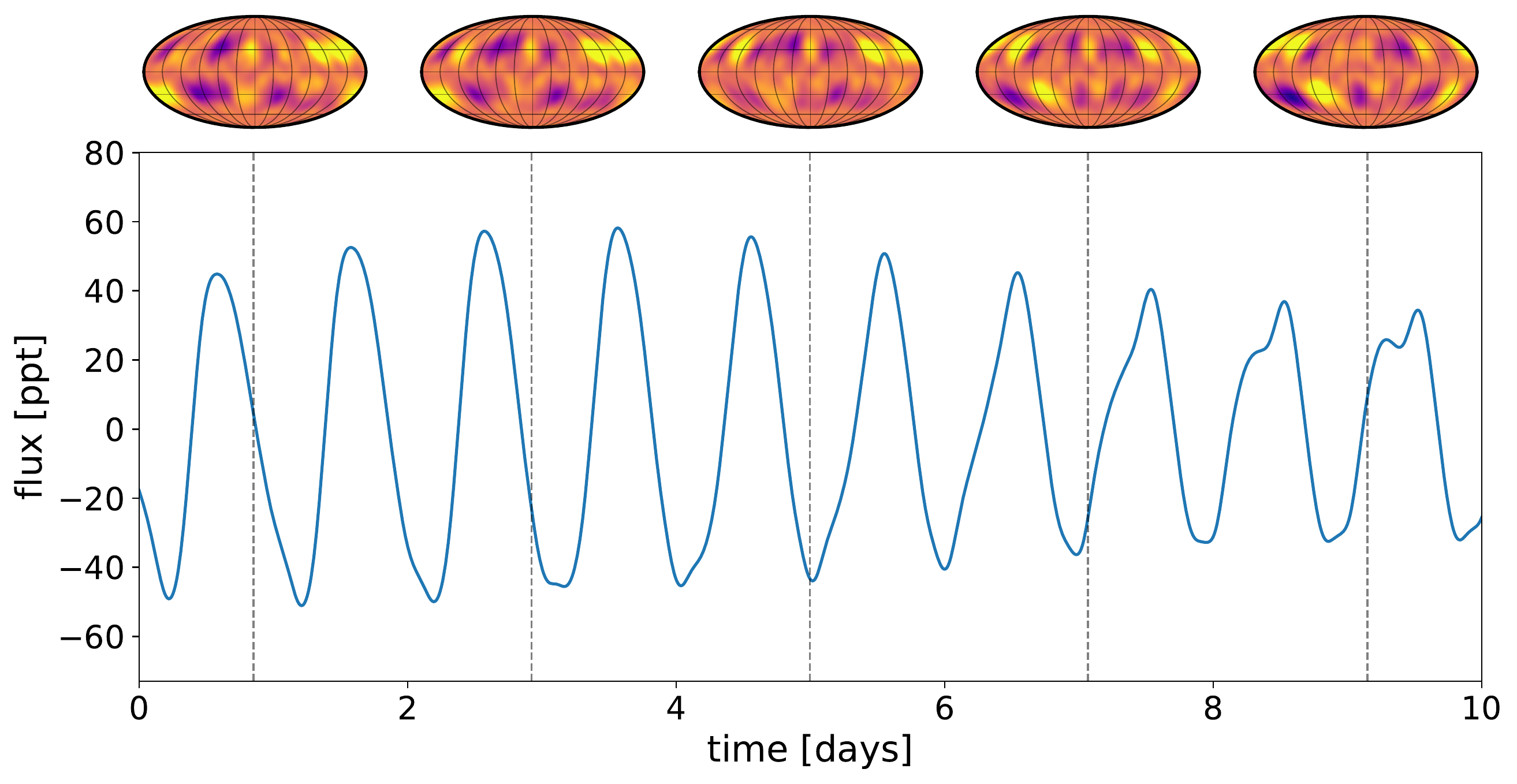}
        \oscaption{temporal}{%
            Prior sample from a time-variable \starryprocess, assuming a rotation
            period $P = 1$ day and a temporal evolution timescale $\tau = 20$ days
            modeled with an exponential squared kernel. Evolving map samples are
            shown at the top, with the corresponding light curve viewed at an
            inclination $I = 60^\circ$ at the bottom. The GP hyperparameters are
            set to their default values (Table~\ref{tab:synthetic}).
            \label{fig:temporal}
        }
    \end{centering}
\end{figure}

The covariance matrix in Equation~(\ref{eq:Sigmaty}) can be sampled from to
yield time-variable surface maps, or it can be transformed into flux space
for sampling light curves or computing likelihoods. The covariance in
flux space is given by
\begin{align}
    \label{eq:SigmatSlow}
    \pmb{\Sigma}^\mathbf{(t)} & =
    \pmb{\mathcal{A}}^\ddagger
    \,
    \pmb{\Sigma}_\mathbf{y}^\mathbf{(t)}
    \,
    {\pmb{\mathcal{A}}^\ddagger}^\top
\end{align}
where
\begin{align}
    \pmb{\mathcal{A}}^\ddagger
                      & \equiv
    \setstackgap{L}{1.25\baselineskip}
    \fixTABwidth{T}
    \parenMatrixstack{
    \mathbf{a}_0^\top &                   &        &                              \\
                      & \mathbf{a}_1^\top &        &                              \\
                      &                   & \ddots &                              \\
                      &                   &        & \,\,\, \mathbf{a}_{K-1}^\top
    }
\end{align}
is a design matrix constructed by staggering the rows of the standard design
matrix $\pmb{\mathcal{A}}$ (see Appendix \ref{sec:starry:basic}). The diagonal
structure of $\pmb{\mathcal{A}}^\ddagger$ is due to the fact that each snapshot of the surface is only observed
at a single phase.

In the temporally-variable version of our GP, the covariance matrix in
Equation~(\ref{eq:SigmatSlow}) replaces the standard covariance
$\pmb{\Sigma}$; it can similarly be modified (\S\ref{sec:gp-norm})
to obtain the covariance of the normalized process. While $\pmb{\Sigma}^\mathbf{(t)}$
is $(K \times K)$, it is computed from the contraction of a much larger
$(NK \times NK)$ matrix, which is extremely inefficient to instantiate and
operate on. Fortunately, it can be shown that
\begin{proof}{test_temporal}
    \pmb{\Sigma}^\mathbf{(t)} & =
    \pmb{\mathcal{A}}^\ddagger
    \,
    \pmb{\Sigma}_\mathbf{y}^\mathbf{(t)}
    \,
    {\pmb{\mathcal{A}}^\ddagger}^\top
    \nonumber \\
    &=
    \pmb{\mathcal{A}}^\ddagger
    \left(\mathbf{K} \otimes \pmb{\Sigma}_\mathbf{y}\right)
    {\pmb{\mathcal{A}}^\ddagger}^\top
    \nonumber \\
    &=
    \pmb{\Sigma} \odot \mathbf{K}
    \quad,
\end{proof}
that is, the flux covariance is just the elementwise $\odot$ product of the
GP covariance $\pmb{\Sigma}$ and the temporal covariance $\mathbf{K}$.
This fact makes the temporal GP \emph{just as efficient} to
evaluate as the standard GP!

Armed with this algorithm for computing $\pmb{\Sigma}^\mathbf{(t)}$, it only
remains for us to decide on a structure for $\mathbf{K}$. While this can
in principle be any covariance matrix constructed from a stationary kernel,
we recommend one of the common radial kernels such as the exponential
squared kernel
\begin{align}
    \label{eq:expsq}
    k_\mathrm{E^2}(\Delta t) & = \sigma^2 \exp\left(-\frac{\Delta t^2}{2\tau}\right)
\end{align}
or the Mat\'ern-3/2 kernel
\begin{align}
    \label{eq:mat32}
    k_\mathrm{M^\frac{3}{2}}(\Delta t) & = \sigma^2 \left( 1 + \frac{\sqrt{3}\Delta t}{\tau} \right) \exp\left(-\frac{\sqrt{3}\Delta t}{\tau}\right)
\end{align}
with variance parameter $\sigma^2$ set to
unity (since the variance is already specified within $\pmb{\Sigma}$). In this
case, the temporal covariance $\mathbf{K}$ is a function of a single
parameter: the timescale of the variability, $\tau$. Similar to the other
hyperparameters of our GP, this parameter can be estimated in an inference
setting. However, a detailed investigation of the ability of our temporal
GP to accurately capture spot variability is beyond the scope of this paper,
and will be revisited in the future, along with an algorithm to explicitly
model the effects of differential rotation on the covariance structure.

Figure~\ref{fig:temporal} shows a single prior sample from the temporal
GP described above, assuming a squared exponential covariance with $\tau = 20$ days
and a rotation period $P = 1$ day. We compute both the map samples at five
different times (top) and the light curve sample over ten rotations (bottom).
The flux variability is qualitatively similar to that seen in
temporally-variable \emph{Kepler} light curves.

\subsection{Marginalizing over period and limb darkening}
\label{sec:other-marg}

In \S\ref{sec:gp-inc} we discussed the value of marginalizing over the
stellar inclination when computing the GP covariance (and thus the likelihood).
When jointly analyzing the light curves of many stars (which we have repeatedly
argued is the best way to infer their spot properties), it can be extremely useful to
minimize the number of latent variables associated with individual stars by
analytically marginalizing over them. This can dramatically reduce the number
of free parameters, turning a difficult inference problem in (possibly) hundreds
or thousands of dimensions into a much easier problem in a handful of dimensions.
We showed how it is possible to marginalize away the dependence of our GP on
the inclinations of individual stars, which is a huge step in this direction.
However, our GP remains a function of two other quantities that will in general
be different for different stars: the stellar rotation period $P$ and the
limb darkening coefficient vector $\mathbf{u}$. We argued that in some cases
one may be able to fix the period of each star at an estimate obtained in a pre-processing
step (i.e., from a periodogram) and fix the limb darkening coefficients at
theoretical values, in which case the number of free parameters of the GP
is equal to the size of the spot hyperparameter vector $\pmb{\theta}_\bullet$
(five by default), and \emph{independent} of the number of light curves in
the ensemble.
However, this procedure ignores any uncertainty in the period and limb darkening
coefficients, which could be significant; it is also subject to bias due to
the fact that there may be systematic errors in theoretical models for the limb darkening
coefficients, particularly for low mass stars \citep[e.g.,][]{Kervella2017}. A much better
approach would be to analytically marginalize over these two quantities.

Consider Equation~(\ref{eq:akT}) in the Appendix, from which the rows of the
design matrix (which transforms vectors in the spherical harmonic basis to vectors
representing the flux at a point in time) are computed. Marginalization over the
inclination, which we demonstrated how to perform analytically in
Appendix~\ref{sec:inc}, entails integrating over the term
$\mathbf{R}_{\hat{\mathbf{x}}}\left(-I\right)$, the Wigner matrix that rotates
the star by the inclination angle $I$ into the observer's frame.
Similarly, marginalization over the
rotation period would entail integration over the term
$\mathbf{R}_{\hat{\mathbf{z}}}\left(\frac{2\pi}{P} t_k\right)$, another
Wigner matrix that rotates the star to the correct rotational phase at
time $t = t_k$. Given an appropriate prior on the rotational angular
frequency $\frac{2\pi}{P}$, it should be possible to follow the same
procedure outlined in Appendix~\ref{sec:inc} to analytically
compute the first two moments of the distribution of light curves marginalized over the
rotation period.

Equation~(\ref{eq:akT}) also makes clear the dependence of the GP on
the limb darkening cofficients, which enter via the limb darkening
operator $\mathbf{L}(\mathbf{u})$. The coefficients
of this matrix are linear in the limb darkening coefficients $u_1$ and $u_2$
(see Equation~\ref{eq:ld:L}), so it should be possible to derive closed
form solutions to the relevant integrals over $\mathbf{L}(\mathbf{u})$ to
yield the GP covariance marginalized over $\mathbf{u}$. It may be
possible to compute the marginalized covariance even when parametrizing
the quadratic limb darkening
coefficients in terms of the uncorrelated $q_1$ and $q_2$ parameters from
\citet{Kipping2013}, which would allow us to incorporate hard constraints on positivity
and strict limb darkening (as opposed to brightening) directly into the prior.

Given the complexity of the operations involved,
performing these marginalizations is beyond the scope of the present paper.
However, given the significant computational benefits of this marginalization,
as well as the fact that the mapping problem is particularly sensitive to
the limb darkening coefficients \citepalias{PaperI}, marginalizing over
$P$ and $\mathbf{u}$ will be the subject of an upcoming paper.

\subsection{Modeling transits and radial velocity datasets}
\label{sec:transits-rvs}

Finally, we would like to note that the GP developed in this paper is not
limited to modeling rotational light curves of stars. We derived the
covariance structure of spotted stellar surfaces in the spherical harmonic
basis, which we then linearly transformed into flux space to obtain the
light curve GP presented above. However, our GP may be used to model
\emph{any} kind of observation that
is linearly related to the spherical harmonic representation. If $\mathbf{A}$
is the linear operator that transforms the spherical harmonic representation
$\mathbf{y}$ to the data vector $\mathbf{d}$ via
$\mathbf{d} = \mathbf{A} \mathbf{y}$, then the mean and covariance of the
GP model for $\mathbf{d}$ are given by
\begin{align}
    \pmb{\mu}_\mathbf{d}    & = \mathbf{A} \, \pmb{\mu}_\mathbf{y}
    \\
    \pmb{\Sigma}_\mathbf{d} & = \mathbf{A} \, \pmb{\Sigma}_\mathbf{y} \, \mathbf{A}^\top
    \quad,
\end{align}
respectively, where $\pmb{\mu}_\mathbf{y}$ and $\pmb{\Sigma}_\mathbf{y}$
are the mean and covariance in the spherical harmonic basis derived in this paper.
In \citet{Luger2019} we showed that
occultation light curves are also linearly related to the
spherical harmonic representation of the stellar surface, so it is
straightforward to use our GP to model transits of planets across spotted
stars, either to marginalize over the spot variability or to constrain the
surface map of the star. In this context, $\mathbf{A}$ may be computed via the
\textsf{design\_matrix()} method of a \textsf{Map} instance of the \starry
package.

The formalism developed here may also be extended to model radial
velocity (RV) datasets, although this requires a bit more work. The
instantaneous radial
velocity shift $v$ induced by a rotating spotted star may be approximated as
\begin{align}
    v & = \frac{\iint_S I \, V \, \mathrm{d} S}{\iint_S I \mathrm{d} S}
\end{align}
where $I$ is the stellar intensity at a point on the surface, $V$ is the
radial component of the rotational velocity vector at that point, and the
integral is taken over the projected disk of the star.
If we expand the surface intensity distribution in spherical harmonics,
the integral in the denominator is just a classical \starry integral, as it
is just the disk-integrated intensity (i.e., the flux).
The numerator may also be computed following the \starry formalism,
provided we weight the surface intensity representation by the velocity
field $V$. In the case of rigid body rotation, $V$ is exactly a dipole ($l = 1$)
field, so the quantity $I V$ can be expressed exactly as a product of
spherical harmonics.%
\footnote{
    Higher order effects such as differential rotation and convective
    blueshift can be easily modeled with a higher degree expansion of $u$.
}
Since spherical harmonics are closed under multiplication,
we may write $I V$ as a linear combination of spherical harmonics, meaning
the integral in the numerator is \emph{also} a \starry integral.
For some linear operators $\mathbf{A}$ and $\mathbf{B}$, we may therefore
write
\begin{align}
    \mathbbb{v} & = \frac{\mathbf{B} \, \mathbbb{y}}{\mathbf{A} \, \mathbbb{y}}
\end{align}
where $\mathbbb{v}$ is the random variable representing the observed radial
velocity time series,
$\mathbbb{y}$ is the Gaussian random variable representing the spherical harmonic
coefficients describing the stellar surface, and the division is performed
elementwise.
Since $\mathbbb{v}$ is the ratio of two Gaussian random variables, its
distribution is not Gaussian. However, we can still compute the first two moments
of the distribution of $\mathbbb{v}$ to derive a Gaussian approximation to it
(similar to what we did in \S\ref{sec:gp-norm}), which will yield the mean
and covariance of the Gaussian process representation of the radial velocity
time series.
Given the complexity of the operations described above,
we defer this calculation (and the calibration of the resulting GP model)
to future work.

\section{Conclusions}
\label{sec:conclusions}

This paper is the second in a series devoted to the development of
statistically rigorous techniques to model stellar surfaces based on
unresolved photometric and spectroscopic measurements.
Here, we presented a new Gaussian process (GP) model for stellar variability
whose hyperparameters explicitly correspond to physical properties of the stellar
surface. Our GP allows one to efficiently compute the likelihood function
for stellar light curves
marginalized over nuisance parameters such as the specific sizes,
positions, and contrasts of individual spots,
which are generally unknowable due to the extreme degeneracies
involved in the light curve mapping problem. Our GP therefore
makes it easy to
do posterior inference on the real quantities of interest: parameters
controlling the \emph{distribution} of spot sizes, latitudes, and
contrasts within a star and/or across many stars in an ensemble.

Because our expression for the GP covariance has an exact, closed-form
solution as a function of the spot parameters, it can be computed
efficiently: a typical likelihood evaluation on a dataset
consisting of $K \sim 1{,}000$ points takes about 20ms on a modern
laptop. Our algorithm is implemented in the open-source, user-friendly
\Python package \starryprocess, which is \textsf{pip}-installable, available on
\href{https://github.com/rodluger/starry_process}{GitHub}, and
described in \citet{JOSSPaper}.
The algorithm is implemented in a combination of \cpp and \Python,
linked using the \theano package. Because our GP covariance has an exact
representation, so too do its derivatives. We therefore implement
backpropagated derivatives with respect to all input parameters for
out-of-the-box usage with gradient-based inference and optimization
tools such as Hamiltionain Monte Carlo (HMC), autodifferentation
variational inference (ADVI), and gradient-based nested sampling.

We devoted a large portion of this paper to testing the algorithm on a variety
of synthetic datasets, showing that it is a well-calibrated and in most
cases unbiased estimator for starspot properties. Below we list our main
results:

\begin{enumerate}
    \item \textbf{Our GP works best for ensemble analyses.} The light curve mapping
          problem is extremely degenerate, as light curves contain a vanishingly small
          fraction of the total information about a stellar surface. However, the
          degenerate surface modes are a strong function of the observer's viewing
          angle, so light curves of stars seen at different inclinations constrain
          different components of the surface. We have shown that if we jointly
          analyze the light curves of many stars,
          we can break many of the degeneracies at play and uniquely infer the
          statistical properties of the spots across the ensemble. This type
          of analysis works best if the stars in the ensemble are statistically
          similar: i.e., the properties of their spots are all drawn from the
          same parent distribution, whose parameters we can constrain.
    \item \textbf{Typically, an ensemble of at least $M \sim 50$ light curves
              is needed to place meaningful constraints on starspot properties}.
          This estimate is based on ensemble analyses of light curves with
          $K = 1{,}000$ cadence each and per-cadence precision of one part
          per thousand. Lower quality, shorter baseline observations, or
          datasets contaminated by outliers will in general require larger values
          of $M$ for the same constraining power. The presence of strong limb
          darkening also degrades the information content of light curves, in
          which case an ensemble of hundreds or even a thousand light curves
          is recommended.
    \item \textbf{Our GP is in most cases an unbiased estimator for the spot radius and
              latitude distributions.} We showed that our GP can accurately infer the
          angular size of spots and the mode and standard deviation of their
          distribution in latitude from stellar light curves. For the fiducial
          ensemble of $M = 50$ light curves described above, we are able to
          constrain the average spot radii to within a couple degrees and the
          average spot latitudes to within $5^\circ$.
    \item \textbf{Our GP can accurately infer stellar inclinations.} We presented
          two versions of our GP: one conditioned on a specific value of the stellar
          inclination, and one marginalized over inclination under an isotropic
          prior. In both cases, we find that we can accurately infer the inclinations
          of individual stars in an ensemble analysis (in the latter case, a
          simple post-processing step can yield the inclination posterior distribution).
          While the inclination is not an observable for an individual stellar light
          curve, the population-level constraints on the spot properties achieved
          by the GP can break the degeneracies involving the inclination, allowing
          us to usually infer it to within about $10^\circ$ and without bias.
    \item \textbf{Our GP can be used to model small, Sun-like spots.} The algorithm
          presented here is limited to a surface resolution of about $10^\circ$,
          corresponding to spots about an order of magnitude larger than typical
          sunspots. However, we have showed that when we apply our model to
          light curves of stars with small ($r \sim 3^\circ$) spots, we can
          still infer their latitudinal distribution without bias, as well as
          the presence of spots below our resolution limit.
    \item \textbf{Our GP can be extended to model time-variable surfaces.} The
          algorithm presented here was derived for static stellar surfaces,
          corresponding to perfectly periodic light curves. However, time
          variability can easily be modeled as the product of the \starryprocess
          kernel and a kernel describing the covariance of the process in
          time, such as a simple exponential squared or Mat\'ern-3/2 kernel.
          The hyperparameters of the temporal covariance are then strictly
          tied to the timescale on which the surface evolves. More
          complex temporal variability, such as that induced by differential
          rotation, will be the subject of a future paper in this series.
    \item \textbf{Our GP can be used in exoplanet transit modeling and
              radial velocity datasets.} At its core, the \starryprocess GP
          defines a distribution over spherical harmonic representations
          of stellar surfaces. It can therefore be used as a physically
          interpretable prior when modeling transits of exoplanets across
          spotted stars, either to marginalize over the stellar inhomogeneity
          or to explicitly infer spot properties. When combined with the
          \starry package, it can also be used as a
          prior on the stellar surface in Doppler imaging, Doppler tomography,
          or even radial velocity searches for exoplanets. The latter will
          be the subject of a future paper in this series.
\end{enumerate}

The GP presented here has far-ranging applications for stellar light
curve studies. It serves as a drop-in replacement for commonly used
GP kernels for stellar variability, which currently do not have
physically interpretable parameters other than the rotation
period and, in some cases, a spot evolution timescale.
As such, it can be used to marginalize over stellar rotational variability
signals in (say) transiting exoplanet searches, asteroseismic characterization
of stars, radial velocity searches, etc. It can also be used to learn
about stellar surfaces directly: to infer spot properties of main sequence
stars as a function of spectral type and age, to differentiate between
spot-dominated and plage-dominated stellar surfaces, to better understand
chemically peculiar massive stars, and to better understand the spot
properties of transiting exoplanet hosts for unbiased spectroscopic
characterization of their atmospheres (to name a few).

The next papers in this series will focus on (in no particular order)
\begin{itemize}
    \item \emph{A more rigorous treatment of time variability}.
          This paper will focus
          on modeling differential rotation, whose effect on the covariance of the
          process can be derived in a similar fashion to what we did here. This
          will enable direct inference about the differential rotation rates and
          spot evolution timescales of stars, processes whose effects on light curves
          are too similar for current methodology to reliably discern between them.
    \item \emph{Explicit marginalization over the remaining stellar parameters}.
          These include the stellar rotation period
          and limb darkening coefficients, which can be marginalized over analytically
          under certain choices of prior. This will eliminate all per-star
          hyperparameters in the expression for the GP covariance, greatly
          speeding up inference for large ensembles of stellar light curves.
    \item \emph{Extension of this formalism to radial velocity datasets}.
          As we discussed in \S\ref{sec:transits-rvs}, it is possible to extend
          the methodology presented here to model the contribution of stellar
          surface variability to radial velocity measurements, which can be used
          (for instance) to mitigate systematics in extreme precision radial velocity
          (EPRV) searches for exoplanets.
    \item \emph{Extension of this formalism to Doppler imaging}.
          As we show in upcoming work, it is possible to derive an exact
          linear relationship between the wavelength-dependent spherical harmonic
          representation of a rotating star and its time-variable spectrum.
          This linearity makes it possible to adapt our GP formalism to the Doppler
          imaging problem, providing an efficient marginal likelihood function
          for rigorous inference studies.
\end{itemize}

In keeping with other papers in the \starry series, all figures in this
paper are generated automatically from open-source scripts linked to in
each of the captions \codeicon, and the principal equations link to associated
unit tests that ensure the accuracy and reproducibility of the algorithm
presented here \testpassicon/\testfailicon.

\vspace{2em}

We would like to thank David W. Hogg, Michael Gully-Santiago, Adam Jermyn,
Megan Bedell, Will Farr, Dylan Simon, and the
Astronomical Data Group at the Center for Computational Astrophysics for
their help and for
many thought-provoking discussions that made this paper possible.

\vfill
\pagebreak
\bibliography{bib}

\pagebreak

\input{table}
\vfill
\clearpage

\pagebreak

\appendix

\section{Notation}
\label{sec:notation}
Unless otherwise noted, we adopt
the following conventions throughout this paper:
integers are represented by italic uppercase letters (i.e., $K$),
scalars are represented by italic lowercase
letters (i.e., $x$), column vectors are
represented by boldface lowercase letters
($\mathbf{x}$), and matrices are represented
by boldface capital letters ($\mathbf{X}$). In general, the elements of a vector
$\mathbf{x}$ are denoted $x_i$ and the elements of a matrix $\mathbf{X}$
are denoted $X_{i,j}$. Importantly, we make a distinction between
quantities like $X_{i,j}$ and $\mathbf{X}_{i,j}$: the former is a scalar
element of a matrix, while the latter is a \emph{matrix}, which is itself
a component of a higher-dimensional (in this case, 4-dimensional) linear
operator. Thus, lowercase bold symbols \emph{always} represent vectors, and
uppercase bold symbols \emph{always} represent matrices.

We also make an explicit distinction
between numerical quantities and random variables. The former are typeset
in serif font (as above), while the latter are typeset in blackboard font.
For example, the quantity $\mathbb{x}$ denotes a scalar random variable,
while $x$ denotes a particular realization of that variable. The same
applies to vector-valued ($\mathbf{x}$ is a realization of $\mathbbb{x}$)
and matrix-valued random variables
($\mathbf{X}$ is a realization of $\mathbbb{X}$).

Much of the math in this paper involves vectors representing
coefficients in the spherical harmonic basis, which are
customarily indexed by two integers $l$ and $m$.
We therefore make an exception to our indexing notation for
quantities in the spherical harmonic basis: we use
\emph{two} indices to represent a scalar vector element, $x^l_m$, and \emph{four}
indices to represent a scalar matrix element, $X^{l,l'}_{m,m'}$.
The upper indices corresponds to the spherical harmonic degree,
$l \in [0, l_{\mathrm{max}}]$ and $l' \in [0, l_{\mathrm{max}}]$,
while the lower indices correspond to the
spherical harmonic order, $m \in [-l, l]$ and $m' \in [-l', l']$.
Vector elements are arranged in order of increasing $l$ and,
within each $l$, in order of increasing $m$.
For example, a vector $\mathbf{x}$
representing a quantity in the spherical harmonic basis up to degree
$l_\mathrm{max}$ has components given by
\begin{align}
    \mathbf{x}
     & =
    \left(
    x^0_0 \,\,\,
    \,\,\,\,\,\,
    x^1_{-1} \,\,\,
    x^1_{0} \,\,\,
    x^1_{1} \,\,\,
    \,\,\,\,\,\,
    \cdots \,\,\,
    \,\,\,\,\,\,
    x^{l_\mathrm{max}}_{-l_\mathrm{max}}
    \cdots \,\,
    x^{l_\mathrm{max}}_{l_\mathrm{max}} \,\,\,
    \right)^\top
    \quad,
\end{align}
while a matrix $\mathbf{X}$ in the same basis has components given by
\begin{align}
    \setstackgap{L}{1.25\baselineskip}
    \fixTABwidth{T}
    \mathbf{X} =
    \parenMatrixstack{
    X^{0,0}_{0,0}  &  & X^{0,1}_{0,-1}  & X^{0,1}_{0,0}  & X^{0,1}_{0,1}  &        \\
                   &  &                 &                &                &        \\
    X^{1,0}_{-1,0} &  & X^{1,1}_{-1,-1} & X^{1,1}_{-1,0} & X^{1,1}_{-1,1} &        \\
    X^{1,0}_{0,0}  &  & X^{1,1}_{0,-1}  & X^{1,1}_{0,0}  & X^{1,1}_{0,1}  &        \\
    X^{1,0}_{1,0}  &  & X^{1,1}_{1,-1}  & X^{1,1}_{1,0}  & X^{1,1}_{1,1}  &        \\
                   &  &                 &                &                & \ddots
    }\quad.
\end{align}
For completeness, the element of a spherical harmonic vector $\mathbbb{x}$ with
degree $l$ and order $m$ is at (flattened) index
\begin{align}
    \label{eq:n}
    n = l^2 + l + m
    \quad.
\end{align}
Conversely, the element at (flattened) index $n$ has degree and order
\begin{align}
    \label{eq:lm}
    \begin{split}
        l & = \floor{\sqrt{n}}
        \\
        m & = n - l^2 - l
        \quad,
    \end{split}
\end{align}
respectively.
Note, finally, that our use of upper and lower indices is purely a
notational convenience,
and should not be confused with
exponentiation or a distinction between covariant and contravariant
tensors. It should also not be confused with the notation used for the complex
spherical harmonics, which also uses upper and lower indexing.

For reference, Table~\ref{tab:variables} lists the principal symbols,
operators, and variables used throughout the paper, with links to
the equations and/or section in which they are presented.

\section{Computing the flux}
\label{sec:starry}

\subsection{Basic expression}
\label{sec:starry:basic}
As we mentioned in \S\ref{sec:gp-gp}, the flux $\mathbf{f}$ is a purely
linear function of the spherical harmonic coefficient vector $\mathbf{y}$:
\begin{align}
    \mathbf{f} = \mathbf{1} + \pmb{\mathcal{A}} \, \mathbf{y}
    \quad.
\end{align}
Even though this is derived in detail in \citet{Luger2019}, it is useful to
expand on the computation of the design matrix $\pmb{\mathcal{A}}$ in more
detail here. Let $\mathbf{a}_k^\top$ denote the $k^{th}$ row of $\pmb{\mathcal{A}}$,
such that
\begin{align}
    \label{eq:Arows}
    \pmb{\mathcal{A}}
     & =
    \setstackgap{L}{1.25\baselineskip}
    \fixTABwidth{T}
    \parenMatrixstack{
        \mathbf{a}_0^\top \\
        \mathbf{a}_1^\top \\
        \vdots            \\
        \mathbf{a}_{K-1}^\top
    }\quad.
\end{align}
The row vector $\mathbf{a}_k^\top$ encodes how the spherical harmonic
coefficient vector projects onto the $k^\mathrm{th}$ cadence in the flux time series, and
may be computed from
\begin{align}
    \label{eq:akT}
    \mathbf{a}_k^\top = \mathbf{r}^\top \,
    \mathbf{A_1} \,
    \mathbf{R}_{\hat{\mathbf{x}}}\left(-I\right) \,
    \mathbf{R}_{\hat{\mathbf{z}}}\left(\frac{2\pi}{P}t_k\right) \,
    \mathbf{R}_{\hat{\mathbf{x}}}\left(\frac{\pi}{2}\right)
    \quad.
\end{align}
To understand the expression above, let us proceed from right to left,
starting with the spherical harmonic vector $\mathbbb{y}$, which we assume
describes the surface intensity of the star at time $t = 0$
in a frame where $\hat{\mathbf{x}}$
points to the right, $\hat{\mathbf{y}}$ points up, and $\hat{\mathbf{z}}$
points out of the page. The quantity $\mathbf{R}_{\hat{\mathbf{x}}}$
is a Wigner rotation matrix
(described in detail in \S\ref{sec:wigner}),
which in this case rotates the spherical harmonic representation
of the star by an angle $\nicefrac{\pi}{2}$ counter-clockwise about $\hat{\mathbf{x}}$
such that the north pole of the star points along $\hat{\mathbf{z}}$. In this
frame, we apply a second Wigner rotation matrix, $\mathbf{R}_{\hat{\mathbf{z}}}$,
to rotate the star about $\hat{\mathbf{z}}$ counter-clockwise (i.e., eastward) by an angle
$\nicefrac{2\pi t_k}{P}$, where $P$ is the rotation period and $t_k$ is the
time at cadence $t$.
Next, we rotate the star by a \emph{clockwise} angle of $I$ about $\hat{\mathbf{x}}$,
where $I$ is the stellar inclination ($I = 0$ corresponding to a pole-on view and
$I = \nicefrac{\pi}{2}$ corresponding to an edge-on view). With this last rotation,
we are now in the observer's frame.%
\footnote{In principle, one last rotation could be performed about $\hat{\mathbf{z}}$
    to orient the projected disk of the star on the plane of the sky; however, the disk-integrated
    flux is independent of the rotation angle along the plane of the sky
    (which we refer to as the \emph{obliquity}), so this step is unnecessary.}

Following \citet{Luger2019}, the next step is to project the representation of
the star into a more convenient basis for performing the integration over the stellar
disk. The change-of-basis matrix $\mathbf{A}_1$ \citep[see Appendix~B in][]{Luger2019}
projects the stellar map into the \emph{polynomial basis}
\citep[Equation~7 in][]{Luger2019}, comprised of the sequence of monomials in Cartesian
coordinates
$\left( 1 \,\, x \,\, z \,\, y \,\, x^2 \,\, xz \,\, xy \,\, yz \,\, y^2 \cdots \right)$
where $z = \sqrt{1 - x^2 - y^2}$ on the surface of the unit sphere. We
can now compute the disk-integrated flux by integrating each of the terms in the basis
over the unit disk, which is straightforward in the polynomial basis; the
individual terms integrate to simple ratios of Gamma functions. These are then
assembled into the row vector $\mathbf{r}^\top$, given by Equation~(20) in
\citet{Luger2019}, which we dot into our expression (and add one) to obtain the
flux at the $k^\mathrm{th}$ cadence.

\subsection{With limb darkening}
%
\label{sec:ld}
We must adjust our expression for the flux in the presence of limb darkening.
For any polynomial limb darkening law of the form
\begin{align}
    \label{eq:ld:I}
    \frac{I(\upmu)}{I(\upmu = 1)} = 1 - \sum_{n=1}^{n_\mathrm{max}} u_n(1 - \upmu)^n
    \quad,
\end{align}
where $I$ is the intensity on the stellar surface,
$\upmu = z = \sqrt{1 - x^2 - y^2}$ is the radial coordinate on the
projected disk, and $u_n$ is a limb darkening coefficient, the effect of
limb darkening on the stellar map can be expressed exactly as a linear
operation on the spherical harmonic coefficient vector
\citep{Luger2019}.
This includes the popular linear and quadractic limb darkening laws
and generalizes to \emph{any} limb darkening law in the limit
$n_\mathrm{max} \rightarrow \infty$. The linearity of the problem
can be understood by noting that all terms
in Equation~(\ref{eq:ld:I}) are strictly polynomials in $x$, $y$, and $z$,
all of which can be expressed exactly as sums of spherical harmonics
\citep{Luger2019}. When weighting the surface intensity by the limb darkening
profile, the resulting intensity is simply a product of spherical harmonics,
which is itself a linear combination of spherical harmonics.
Thus, given a limb darkening law
of degree $n_\mathrm{max}$ with coefficients $\mathbf{u}$,
we can construct a matrix $\mathbf{L}(\mathbf{u})$
that transforms a spherical harmonic vector $\mathbbb{y}$ of degree
$l_\mathrm{max}$ to a limb-darkened spherical harmonic vector $\mathbf{y'}$
of degree $l_\mathrm{max} + n_\mathrm{max}$. As an example, consider
a map of degree $l_\mathrm{max} = 1$ and the linear limb darkening law
($n_\mathrm{max} = 1$) with coefficient vector $\mathbf{u} = ( u_1 )$.
The transformation matrix from $\mathbbb{y}$ to $\mathbf{y'}$ is
\begin{proof}{test_ld}
    \label{eq:ld:L}
    \setstackgap{L}{1.5\baselineskip}
    \mathbf{L}(\mathbf{u}) =
    \frac{1}{1 - \frac{u_1}{3}}
    \begin{pmatrix}
        \quadquad1-u_1\quadquad & 0                       & \frac{u_1}{\sqrt{3}}    & 0                       \\
        0                       & \quadquad1-u_1\quadquad & 0                       & 0                       \\
        \frac{u_1}{\sqrt{3}}    & 0                       & \quadquad1-u_1\quadquad & 0                       \\
        0                       & 0                       & 0                       & \quadquad1-u_1\quadquad \\
        0                       & 0                       & 0                       & 0                       \\
        0                       & \frac{u_1}{\sqrt{5}}    & 0                       & 0                       \\
        0                       & 0                       & \frac{2u_1}{\sqrt{15}}  & 0                       \\
        0                       & 0                       & 0                       & \frac{u_1}{\sqrt{5}}    \\
        0                       & 0                       & 0                       & 0
    \end{pmatrix}
    \quad.
\end{proof}
The columns of $\mathbf{L}$ are constructed from the coefficient vectors of
each transformed spherical harmonic, which are in turn computed by
multiplying each spherical harmonic by the spherical harmonic representation
of the particular limb darkening law.

In the presence of limb darkening, we may therefore replace our expression for the
$k^\mathrm{th}$ row of the flux design matrix $\mathcal{A}$ (Equation~\ref{eq:akT})
with
\begin{align}
    \label{eq:akTld}
    \mathbf{a}_k^\top = \mathbf{r}^\top \,
    \mathbf{A_1} \,
    \mathbf{L}(\mathbf{u})
    \mathbf{R}_{\hat{\mathbf{x}}}\left(-I\right) \,
    \mathbf{R}_{\hat{\mathbf{z}}}\left(\frac{2\pi}{P}t_k\right) \,
    \mathbf{R}_{\hat{\mathbf{x}}}\left(\frac{\pi}{2}\right)
\end{align}
for a given limb darkening coefficient vector $\mathbf{u}$.

\section{The Expectation Integrals}
\label{sec:integrals}

Our goal in this section is to find closed-form solutions to the
first and second moments of the spherical harmonic representation of the
stellar surface $\mathbbb{y}$,
\begin{align}
    \label{eq:exp_y_app}
    \mathrm{E} \Big[ \mathbbb{y} \, \Big| \, \pmb{\theta}_\bullet \Big]
     & =
    \int \mathbbb{y}(\mathbbb{x} ) \, p(\mathbbb{x} \, \big| \, \pmb{\theta}_\bullet)\mathrm{d}\mathbbb{x}
    \\
    \label{eq:exp_yy_app}
    \mathrm{E} \Big[ \mathbbb{y} \, \mathbbb{y}^\top \, \Big| \, \pmb{\theta}_\bullet \Big]
     & =
    \int \mathbbb{y}(\mathbbb{x} ) \mathbbb{y}^\top(\mathbbb{x} ) \, p(\mathbbb{x} \, \big| \, \pmb{\theta}_\bullet)\mathrm{d}\mathbbb{x}
    \quad,
\end{align}
which are linearly related to the mean and the covariance of our GP
(\S\ref{sec:gp-gp}).
Recall that $\mathbbb{x}$ is a vector of parameters describing the exact
configuration of features on the surface of a star, and
$p(\mathbbb{x} \, \big| \, \pmb{\theta}_\bullet)$ is its probability
density function conditioned on hyperparameters $\pmb{\theta}_\bullet$,
which describe the \emph{distribution} of the features on the surface
of one or many stars.
As we are specifically interested in modeling the effect of starspots
on stellar light curves, we let
\begin{align}
    \label{eq:x}
    \mathbbb{x} =
    \left(
    \mathbb{n} \,\,\,\,
    \mathbb{c}_0 \,\, \cdots \,\, \mathbb{c}_{\mathbb{n}-1} \,\,\,\,
    \bblambda_0 \,\, \cdots \,\, \bblambda_{\mathbb{n}-1} \,\,\,\,
    \bbphi_0 \,\, \cdots \,\, \bbphi_{\mathbb{n}-1} \,\,\,\,
    \mathbb{r}_0 \,\, \cdots \,\, \mathbb{r}_{\mathbb{n}-1}
    \right)^\top
\end{align}
and
\begin{align}
    \label{eq:RRs}
    \mathbbb{y}(\mathbbb{x}) =
    -
    \sum_{i=0}^{\mathbb{n}-1}
    \mathbb{c}_i
    \,
    \mathbf{R}_{\hat{\mathbf{y}}}(\bblambda_i)
    \,
    \mathbf{R}_{\hat{\mathbf{x}}}(\bbphi_i)
    \,
    \mathbf{s}(\mathbb{r}_i)
    \quad,
\end{align}
where $\mathbb{n}$ is the total number of spots,
$\mathbb{c}_i$ is the contrast of the $i^\mathrm{th}$ spot,
$\bblambda_i$ is its longitude, $\bbphi_i$ is its latitude,
and $\mathbb{r}_i$ is its radius.
The vector function $\mathbf{s}(\mathbb{r}_i)$
returns the spherical harmonic expansion of a negative unit brightness
circular spot of radius $\mathbb{r}_i$ at $\bblambda = \bbphi = 0$,
$\mathbf{R}_{\hat{\mathbf{x}}}(\bbphi_i)$ is the Wigner matrix that rotates the
expansion about $\hat{\mathbf{x}}$ such that the spot is centered at a
latitude $\bbphi_i$, and $\mathbf{R}_{\hat{\mathbf{y}}}(\bblambda_i)$ is the Wigner
matrix that then rotates the
expansion about $\hat{\mathbf{y}}$ such that the spot is centered at a
longitude $\bblambda_i$; these three functions are detailed in the sections below.
Equation~(\ref{eq:RRs}) thus provides a way of converting a random variable
$\mathbbb{x}$ describing the size, brightness, and position of spots to the
corresponding representation in terms of spherical harmonics.
Regarding this equation,
two things should be noted. First, we define $\mathbbb{y}$ relative to
a baseline of zero: i.e., a star with no spots on it will have
$\mathbf{y} = \mathbf{0}$ (which is why we add unity in the expression for the flux
in Equation~\ref{eq:fAy}). Second, and more importantly,
we are not interested in any specific value of
$\mathbbb{y}$; rather, we would like to know its expectation value under
the probability distribution governing the different spot properties $\mathbbb{x}$,
i.e., $p(\mathbbb{x} \, \big| \, \pmb{\theta}_\bullet)$.

For simplicity, we assume that
the total number of spots is fixed to a value $n$, i.e.,
\begin{align}
    p(\mathbb{n} \, \big| \, \pmb{\theta}_\bullet)
     & =
    \delta(\mathbb{n} - n)
    \quad,
\end{align}
where $\delta$ is the delta function.%
\footnote{%
    When modeling a single star using the GP, this assumption is justified by
    definition. It is less justified when the GP is used to model an ensemble of
    stars, where each star may have a different total number of spots $\mathbb{n}$.
    However, as we argue in the text, $\mathbb{n}$ is extremely difficult to constrain
    from light curves, in particular because of how degenerate it is with
    the spot contrast. In practice, we find that assuming that all stars in the
    ensemble have the same number of spots $n$ leads to higher variance in the
    estimate of $n$, but it does not lead to noticeable bias in $n$ or in any of
    the other hyperparameters.
}
We further asume that
$p(\mathbbb{x} \, \big| \, \pmb{\theta}_\bullet)$
is separable in each of the four other spot properties, and that all of the spots
are drawn from the same distribution:
\begin{align}
    p(\mathbbb{x} \, \big| \, \pmb{\theta}_\bullet)
    =
    \prod_{i=0}^{n-1}
    p(\mathbb{c}_i \, \big| \, \pmb{\theta}_{c}) \,
    p(\bblambda_i \, \big| \, \pmb{\theta}_{\lambda}) \,
    p(\bbphi_i \, \big| \, \pmb{\theta}_{\phi})\,
    p(\mathbb{r}_i \, \big| \, \pmb{\theta}_{r})
    \quad,
\end{align}
where
\begin{align}
    \pmb{\theta}_\bullet = \left(
    n \, \,
    \pmb{\theta}_{c} \, \,
    \pmb{\theta}_{\lambda} \, \,
    \pmb{\theta}_{\phi} \, \,
    \pmb{\theta}_{r} \right)^\top
\end{align}
is the vector of hyperparameters describing the process and
$\pmb{\theta}_{c}$,
$\pmb{\theta}_{\lambda}$,
$\pmb{\theta}_{\phi}$, and
$\pmb{\theta}_{r}$ are yet to be specified.
This allows us to rewrite the expectation integrals (\ref{eq:exp_y_app})
and (\ref{eq:exp_yy_app}) as
\begin{align}
    \label{eq:exp_y_sep}
    \mathrm{E} \Big[ \mathbbb{y} \, \Big| \, \pmb{\theta}_\bullet \Big]
     & =
    n \, \mathbf{e}_c
    \\[1em]
    \label{eq:exp_yy_sep}
    \mathrm{E} \Big[ \mathbbb{y} \, \mathbbb{y}^\top \, \Big| \, \pmb{\theta}_\bullet \Big]
     & =
    n \, \mathbf{E}_c
\end{align}
where we define the first moment integrals
\begin{align}
    \label{eq:e1}
    \mathbf{e}_r
     & \equiv
    \int
    \mathbf{s}(\mathbb{r}) \,
    p(\mathbb{r} \, \big| \, \pmb{\theta}_{r}) \,
    \mathrm{d}\mathbb{r}
    \\[1em]
    \label{eq:e2}
    \mathbf{e}_\phi
     & \equiv
    \int
    \mathbf{R}_{\hat{\mathbf{x}}}(\bbphi) \,
    \mathbf{e}_r \,
    p(\bbphi \, \big| \, \pmb{\theta}_{\phi}) \,
    \mathrm{d}\bbphi
    \\[1em]
    \label{eq:e3}
    \mathbf{e}_\lambda
     & \equiv
    \int
    \mathbf{R}_{\hat{\mathbf{y}}}(\bblambda) \,
    \mathbf{e}_\phi \,
    p(\bblambda \, \big| \, \pmb{\theta}_{\lambda}) \,
    \mathrm{d}\bblambda
    \\[1em]
    \label{eq:e4}
    \mathbf{e}_c
     & \equiv
    -
    \int
    \mathbb{c} \,
    \mathbf{e}_\lambda \,
    p(\mathbb{c} \, \big| \, \pmb{\theta}_{c}) \,
    \mathrm{d}\mathbb{c}
\end{align}
and the second moment integrals
\begin{align}
    \label{eq:E1}
    \mathbf{E}_r
     & \equiv
    \int
    \mathbf{s}(\mathbb{r}) \, \mathbf{s}^\top(\mathbb{r}) \,
    p(\mathbb{r} \, \big| \, \pmb{\theta}_{r}) \,
    \mathrm{d}\mathbb{r}
    \\[1em]
    \label{eq:E2}
    \mathbf{E}_\phi
     & \equiv
    \int
    \mathbf{R}_{\hat{\mathbf{x}}}(\bbphi) \,
    \mathbf{E}_r \,
    \mathbf{R}_{\hat{\mathbf{x}}}^\top(\bbphi) \,
    p(\bbphi \, \big| \, \pmb{\theta}_{\phi})
    \mathrm{d}\bbphi
    \\[1em]
    \label{eq:E3}
    \mathbf{E}_\lambda
     & \equiv
    \int
    \mathbf{R}_{\hat{\mathbf{y}}}(\bblambda) \,
    \mathbf{E}_\phi \,
    \mathbf{R}_{\hat{\mathbf{y}}}^\top(\bblambda) \,
    p(\bblambda \, \big| \, \pmb{\theta}_{\lambda})
    \mathrm{d}\bbphi
    \\[1em]
    \label{eq:E4}
    \mathbf{E}_c
     & \equiv
    \int
    \mathbb{c}^2 \,
    \mathbf{E}_\lambda \,
    p(\mathbb{c} \, \big| \, \pmb{\theta}_c)
    \mathrm{d}\mathbb{c}
    \quad.
\end{align}
In Equations~(\ref{eq:exp_y_sep}) and (\ref{eq:exp_yy_sep}), we used the fact
that both the mean and the variance of the sum of $n$
independent, identically-distributed random variables are equal to $n$
times the individual mean and the variance, respectively.

We devote the remainder of this section to the computation of these eight
integrals.

\subsection{The Radius Integrals}
\label{sec:size}
Below we compute the first and second moments
of the radius distribution ($\mathbf{e}_r$, $\mathbf{E}_r$) under
a suitable spherical harmonic expansion $\mathbf{s}(\mathbb{r})$ of the spot profile
and a suitable probability distribution function for the spot radius,
$p(\mathbb{r} \, \big| \, \pmb{\theta}_{r})$.

\subsubsection{Spot profile}
We model the brightness $b$ an angle $\vartheta$ away from the
center of a spot of negative unit intensity and radius $\mathbb{r}$ as
\begin{align}
    \label{eq:brvartheta}
    b(r; \vartheta) & = \frac{1}{1 + \exp\left(\dfrac{r-\vartheta}{s}\right)} - 1
\end{align}
for some (constant) shape parameter $s$. In the limit $s \rightarrow 0$, $b$ approaches an
inverted top-hat function with half-width equal to $r$,
corresponding to a circular spot of uniform intensity. For $s > 0$, each half
of $b$ is a sigmoid with half-width at half-minimum equal to $r$.
In our implementation of the algorithm we choose $s = 0.2^\circ$,
which is small compared to features of interest but not too small as to
create numerical issues when computing model gradients (which would be
undefined at the spot boundary if the spot profile were truly an inverted top-hat).

Our goal now is to expand the function above in spherical harmonics. To that end,
we note that in a frame where the spot is centered on $\hat{\mathbf{z}}$
(i.e., at polar angle $\vartheta = 0$), the brightness profile is azimuthally
symmetric, so the only nonzero coefficients in the spherical harmonic
expansion are those with order $m = 0$. The corresponding spherical harmonics are
simply proportional to the Legendre polynomials in $\cos\vartheta$, so our task is
simplified to finding the Legendre polynomial expansion of $b$.
Define a vector $\pmb{\vartheta}$ of $K$ equally-spaced points
between $0$ and $\pi$, with coefficients given by
\begin{align}
    \vartheta_k = \frac{k\pi}{K-1}
    \quad.
\end{align}
We wish to model the brightness evaluated at each $\vartheta_k$
as a weighted combination of Legendre polynomials,
\begin{align}
    \mathbf{B} \, \mathbf{s_0}(r) = \mathbf{b}(r)
\end{align}
where $\mathbf{b}(r)$ is computed by evaluating Equation~(\ref{eq:brvartheta})
at each of the $\vartheta_k$,
$\mathbf{B}$ is a design matrix whose columns are
the weighted Legendre polynomials,
\begin{align}
    B_{k,l} & = \sqrt{2l + 1} \, P_l\left( \cos \vartheta_k \right)
    \quad,
\end{align}
and $\mathbf{s_0}(r)$ are the coefficients of the expansion.
These are related to the full vector of spherical harmonic coefficients describing
the spot, $\mathbf{s}(r)$, by
\begin{align}
    s^l_m(r) = s^l_0 \delta_{m,0}
    \quad,
\end{align}
or, in vector form,
\begin{align}
    \label{eq:sofr}
    \mathbf{s}(r) = \pmb{\mathcal{I}} \, \mathbf{s_0}(r)
\end{align}
where $\pmb{\mathcal{I}}$ is a rectangular
$\left( \left(l_\mathrm{max} + 1\right)^2 \times \left(l_\mathrm{max} + 1\right) \right)$
identity-like matrix with components
\begin{align}
    \mathcal{I}_{n, l} = \delta_{n, l^2 + l}
    \quad,
\end{align}
and $\delta$ is the Kronecker delta function.
To find the coefficients $\mathbf{s_0}(r)$
(and hence $\mathbf{s}(r)$), we solve the (linear) inverse problem,
\begin{align}
    \mathbf{s_0}(r) & = \mathbf{B}^+ \, \mathbf{b}(r)
\end{align}
where
\begin{align}
    \mathbf{B}^+ & = \mathbf{S} \Big(\mathbf{B}^\top \mathbf{B} +
    \epsilon \mathbf{I}\Big)^{-1} \mathbf{B}^\top
\end{align}
is the smoothed pseudo-inverse of $\mathbf{B}$ with small regularization
parameter $\epsilon$, $\mathbf{I}$ is the identity matrix, and
$\mathbf{S}$ is a diagonal smoothing matrix with coefficients
\begin{align}
    S_{k,l} = \exp\left[-\frac{l(l + 1)}{2\xi^2}\right] \delta_{k,l}
\end{align}
for smoothing strength $\xi$. For $\epsilon \rightarrow 0$ and
$\xi \rightarrow \infty$,
$\mathbf{B}^+$ is the exact pseudo-inverse of
$\mathbf{B}$. However,
$\epsilon > 0$ is chosen for improved numerical stability and
$\xi > 0$ is chosen to mitigate the effect of ringing in the solution.
In practice, we obtain good results with $\epsilon \approx 10^{-9}$
and $\xi \approx 15$.

\begin{figure}[t!]
    \begin{centering}
        \includegraphics[width=\linewidth]{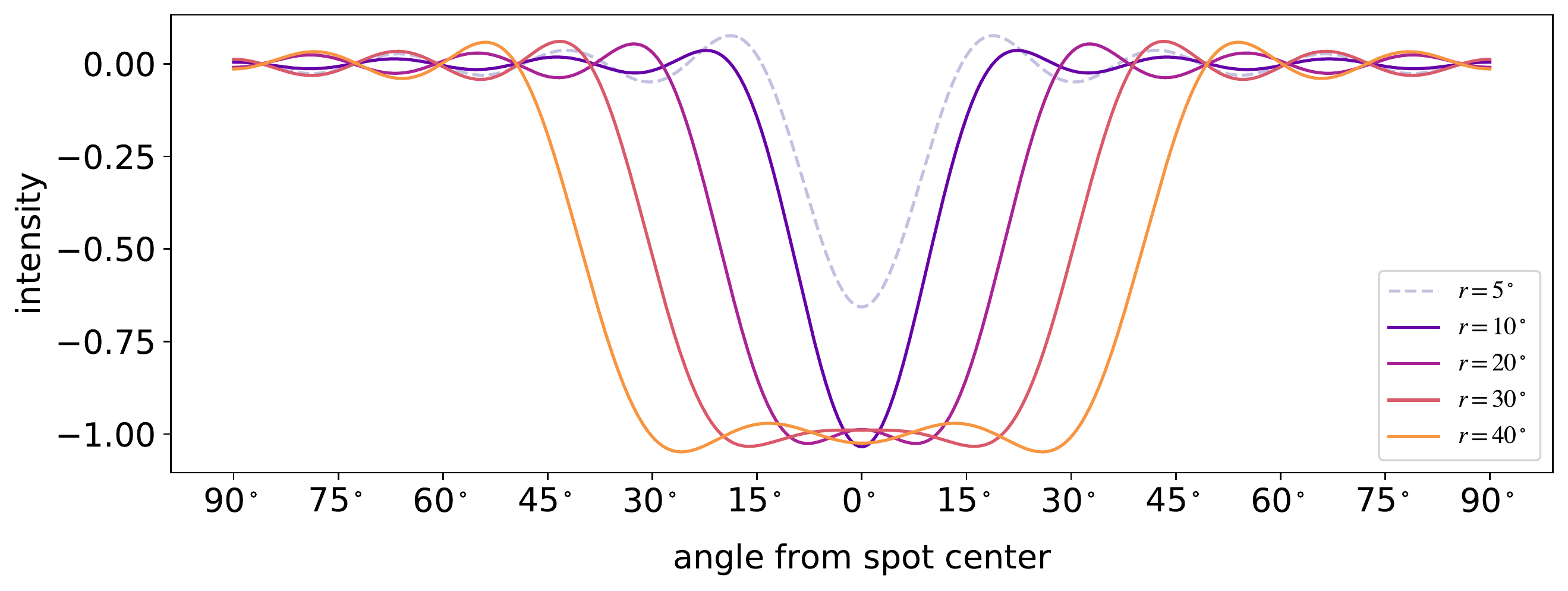}
        \oscaption{spot_profile}{%
            Intensity profiles for spots with different
            radii $r$ computed at spherical harmonic
            degree $l_{\mathrm{max}} = \LMAX$.
            For $r \gtrsim 10^\circ$, the spherical harmonic expansion
            captures the spot shape and intensity reasonably well,
            albeit with some ringing due to the truncated expansion.
            \label{fig:spot_profile}
        }
    \end{centering}
\end{figure}

Figure~\ref{fig:spot_profile} shows the intensity profile for spots of
different radii expanded to spherical harmonic degree $l_{\mathrm{max}} = \LMAX$.
The average intensity
within the spots is close to $-1$ and the half-widths at half-minimum are
equal to the spot radii, as expected.
The effect of ringing due to the truncated spherical harmonic expansion is
evident, although it is strongly suppressed compared to an expansion
without the smoothing term (i.e., $\xi = \infty$). However, for
$r \lesssim 10^\circ$, an expansion to $l_{\mathrm{max}} = \LMAX$ is
insufficient to correctly model the spot, as can be seen from the
$r = 5^\circ$ profile (dashed curve). Expansions to higher spherical harmonic
degree allow one to model spots with radii smaller than $10^\circ$, although
at increased computational cost and potential numerical stability issues;
we discuss this point at length in \S\ref{sec:tinyspots}.

\subsubsection{Probability density function}
For simplicity, we will adopt a uniform probability distribution for the
spot radius, characterized by a mean radius $r$ and a half-width $\Delta r$:
\begin{align}
    p(\mathbb{r} \, \big| \, \pmb{\theta}_{r})
     & =
    \begin{cases}
        \frac{1}{2\Delta r} & r - \Delta r \leq r \leq r + \Delta r
        \\
        0                   & \mathrm{otherwise}
        \quad,
    \end{cases}
\end{align}
where the hyperparameters of the distribution are
\begin{align}
    \pmb{\theta}_r = \left(
    r \, \, \, \,
    \Delta r \right)^\top
    \quad.
\end{align}
As we argue in the text, in practice it is often difficult to constrain the
moments of the radius distribution above the first (the mean). It is
therefore useful to also consider the limiting case of the radius distribution
as $\Delta r \rightarrow 0$, in which case the PDF becomes
\begin{align}
    p(\mathbb{r} \, \big| \, \pmb{\theta}_{r}, \Delta r = 0)
     & =
    \delta(\mathbb{r} - r)
    \quad,
\end{align}
where $\delta$ is the delta function.

\subsubsection{First moment}
The first moment of the radius distribution is (Equation~\ref{eq:e1})
\begin{align}
    \mathbf{e}_r
     & \equiv
    \int
    \mathbf{s}(\mathbb{r}) \,
    p(\mathbb{r} \, \big| \, \pmb{\theta}_{r}) \,
    \mathrm{d}\mathbb{r}
    \nonumber \\
     & =
    \frac{1}{2\Delta r}
    \int_{r_0 - \Delta r}^{r_0 + \Delta r}
    \mathbf{s}(\mathbb{r}) \,
    \mathrm{d}\mathbb{r}
    \quad.
\end{align}
Using the equations from the previous section, its components may be written
\begin{proof}{test_er}
    (e_r)^l_m
    & =
    \frac{\delta_{m,0}}{2\Delta r}
    \int_{r - \Delta r}^{r+ \Delta r}
    \tilde{s}_{l}(\mathbb{r}) \,
    \mathrm{d}\mathbb{r}
    \nonumber \\
    & =
    \frac{\delta_{m,0}}{2\Delta r}
    \sum_{k=0}^{K-1} B^+_{l,k}
    \int_{r - \Delta r}^{r + \Delta r}
    b_{k}(\mathbb{r})
    \mathrm{d}\mathbb{r}
    \nonumber \\
    & =
    \delta_{m,0}
    \sum_{k=0}^{K-1} B^+_{l,k}
    c_k(r, \Delta r)
\end{proof}
where
\begin{proof}{test_er}
    c_k(r, \Delta r) & =
    \frac{s}{2\Delta r}
    \ln
    \left(
    \frac{
            1+\chi_k^-(r, \Delta r)
        }
        {
            1+\chi_k^+(r, \Delta r)
        }
    \right)
\end{proof}
and
\begin{proof}{test_er}
    \chi_k^-(r, \Delta r) & \equiv \exp\left(\frac{r -\Delta r -\vartheta_k}{s}\right)
    \nonumber                                                              \\
    \chi_k^+(r, \Delta r) & \equiv \exp\left(\frac{r +\Delta r -\vartheta_k}{s}\right)
    \quad.
\end{proof}
In vector form, we may simply write
\begin{proof}{test_er}
    \mathbf{e}_r
    & =
    \pmb{\mathcal{I}} \,
    \mathbf{B}^+ \, \mathbf{c}(r, \Delta r)
    \quad.
\end{proof}
Note, finally, that in the limit $\Delta r \rightarrow 0$,
\begin{proof}{test_er}
    \lim_{\Delta r \rightarrow 0}
    \mathbf{e}_r
    & =
    \pmb{\mathcal{I}} \,
    \mathbf{B}^+ \mathbf{b}(r)
    \quad.
\end{proof}

\subsubsection{Second moment}
The second moment of the radius distribution is (Equation~\ref{eq:E1})
\begin{align}
    \mathbf{E}_r
     & \equiv
    \int
    \mathbf{s}(\mathbb{r}) \,
    \mathbf{s}^\top(\mathbb{r}) \,
    p(\mathbb{r} \, \big| \, \pmb{\theta}_{r}) \,
    \mathrm{d}\mathbb{r}
    \nonumber \\
     & =
    \frac{1}{2\Delta r}
    \int_{r - \Delta r}^{r + \Delta r}
    \mathbf{s}(\mathbb{r}) \,
    \mathbf{s}^\top(\mathbb{r}) \,
    \mathrm{d}\mathbb{r}
    \quad.
\end{align}
As before, its components may be written
\begin{proof}{test_bigEr}
    (E_\mathbb{r})^{l,l'}_{m,m'}
    & =
    \frac{\delta_{m,0}\delta_{m',0}}{2\Delta r}
    \int_{r - \Delta r}^{r + \Delta r}
    \tilde{s}^l_{m}(\mathbb{r}) \,
    \tilde{s}^{l'}_{m'}(\mathbb{r}) \,
    \mathrm{d}\mathbb{r}
    \nonumber \\
    & =
    \frac{\delta_{m,0}\delta_{m',0}}{2\Delta r}
    \sum_{k=0}^{K-1} B^+_{l,k}
    \sum_{k'=0}^{K-1} B^+_{l',k'}
    \int_{r - \Delta r}^{r + \Delta r}
    b_{k}(\mathbb{r})
    b_{k'}(\mathbb{r})
    \mathrm{d}\mathbb{r}
    \nonumber \\
    & =
    \delta_{m,0}\delta_{m',0}
    \sum_{k=0}^{K-1} B^+_{l,k}
    \sum_{k'=0}^{K-1} B^+_{l',k'}
    C_{k,k'}(r,\Delta r)
    \quad,
\end{proof}
where
\begin{proof}{test_bigEr}
    C_{k,k'}(r,\Delta r)
    & \equiv
    \frac{s}{2\Delta r}
    \begin{cases}
        \displaystyle
        \frac{
            \exp\left(\frac{\vartheta_{k} - \vartheta_{k'}}{s}\right)
            \ln
            \left(
            \frac{
                1+\chi_k^-
            }
            {
                1+\chi_k^+
            }
            \right)
            -
            \ln
            \left(
            \frac{
                1+\chi_{k'}^-
            }
            {
                1+\chi_{k'}^+
            }
            \right)
        }{
            1
            -
            \exp\left(\frac{\vartheta_{k} - \vartheta_{k'}}{s}\right)
        }
         & \hspace{-2em} k \ne k'
        \\[0.5em]
        \frac{1}{1 + \chi_k^+}
        +
        \frac{\chi_k^-}{1 + \chi_k^-}
        -
        \ln\left( \frac{
            1 + \chi_k^-
        }{
            1 + \chi_k^+
        } \right)
        - 1
         & \hspace{-2em} k = k'
        \quad.
    \end{cases}
\end{proof}
In vector form, this may be written
\begin{proof}{test_bigEr}
    \mathbf{E}_r
    &=
    \pmb{\mathcal{I}} \,
    \mathbf{B}^+ \mathbf{C} (r, \Delta r) {\mathbf{B}^+}^\top
    \pmb{\mathcal{I}}^\top
    \quad.
\end{proof}
Finally, in the limit $\Delta r \rightarrow 0$,
\begin{proof}{test_bigEr}
    \lim_{\Delta r \rightarrow 0}
    \mathbf{E}_r
    & =
    \pmb{\mathcal{I}} \,
    \mathbf{B}^+ \mathbf{b}(r) \mathbf{b}^\top(r) {\mathbf{B}^+}^\top
    \pmb{\mathcal{I}}^\top
    \quad.
\end{proof}

\subsection{The Latitude Integrals}
\label{sec:lat}
Our goal in this section is to compute the first and second moments
of the latitude distribution ($\mathbf{e}_\phi$ and $\mathbf{E}_\phi$,
given by Equations \ref{eq:e2} and \ref{eq:E2}, respectively).
These involve integrals over the terms in the Wigner
rotation matrix for spherical harmonics, which we discuss below.

\subsubsection{Rotation matrices}
\label{sec:wigner}
The Wigner rotation matrix for real spherical harmonics up to degree $l_{\mathrm{max}}$
may be written as the block-diagonal matrix
\begin{align}
    \setstackgap{L}{1.25\baselineskip}
    \fixTABwidth{T}
    \mathbf{R} =
    \parenMatrixstack{
    \quad\quad \, \mathbf{R}^0 \, \quad\quad
     &              &              &        &                               \\
     & \mathbf{R}^1 &              &        &                               \\
     &              & \mathbf{R}^2 &        &                               \\
     &              &              & \ddots &                               \\
     &              &              &        & \mathbf{R}^{l_{\mathrm{max}}}
    }\quad,
\end{align}
where
\begin{align}
    \mathbf{R}^l = {\mathbf{U}^l}^{-1} \mathbf{D}^l \mathbf{U}^l
\end{align}
is the Wigner rotation matrix for a single spherical harmonic degree,
\begin{align}
    \label{eq:U}
    \setstackgap{L}{1.25\baselineskip}
    \fixTABwidth{T}
    \mathbf{U}^l =
    \frac{1}{\sqrt{2}}
    \parenMatrixstack{
        \quad\quad\, \ddots \, \quad\quad\quad\quad
           &       &        &       &          &    &   &    & \Ddots \\
           & \imag &        &       &          &    &   & 1  &        \\
           &       & \imag  &       &          &    & 1 &    &        \\
           &       &        & \imag &          & 1  &   &    &        \\
           &       &        &       & \sqrt{2} &    &   &    &        \\
           &       &        & \imag &          & -1 &   &    &        \\
           &       & -\imag &       &          &    & 1 &    &        \\
           & \imag &        &       &          &    &   & -1 &        \\
    \Ddots &       &        &       &          &    &   &    & \ddots
    }
\end{align}
describes the transformation from complex to real spherical harmonics,
and $\mathbf{D}$ is the Wigner matrix for complex spherical harmonics,
whose terms are given by the expression
\begin{align}
    D_{m,m'}^l(\upalpha, \upbeta, \upgamma) = \exp\left({-\imag m' \upalpha}\right)
    d_{m,m'}^l(\upbeta) \exp\left({-\imag m \upgamma}\right)
\end{align}
where $\upalpha$, $\upbeta$, and $\upgamma$ are the Euler angles describing
the rotation in the
$\hat{\mathbf{z}}{-}\hat{\mathbf{y}}{-}\hat{\mathbf{z}}$ convention
and $\imag$ is the imaginary unit
\citep{AlvarezCollado1989}. The terms of the $d$-matrix
depend on powers of $\sin\left(\nicefrac{\upbeta}{2}\right)$
and $\cos\left(\nicefrac{\upbeta}{2}\right)$
\citep[see Equation C15 in][]{Luger2019}, but it is convenient to use
the half-angle formula to express these terms instead as
\begin{proof}{test_dlmmp}
    d_{m,m'}^l(\upbeta) =
    \sum\limits_{i=0}^{2l} c_{m,m',i}^{l}
    \mathrm{sgn}(\sin\upbeta)^{i}
    (1 - \cos\upbeta)^{\frac{2l - i}{2}}
    (1 + \cos\upbeta)^\frac{i}{2}
\end{proof}
where
\begin{proof}{test_dlmmp}
    c_{m,m',i}^{l} & =
    \resizebox{.75\hsize}{!}{$
            \begin{cases}
                \dfrac{
                \left(-1\right)^{\frac{2l - m + m' - i}{2}}
                \sqrt{\left(l - m\right)!
                    \left(l + m\right)!
                    \left(l - m'\right)!
                    \left(l + m'\right)!}
                }{
                2^l
                \left(\frac{i - m - m'}{2}\right)!
                \left(\frac{i + m + m'}{2}\right)!
                \left(\frac{2l - i - m + m'}{2}\right)!
                \left(\frac{2l - i + m - m'}{2}\right)!
                }
                 & \hspace{-2em} m - m' - i \,\, \mathrm{even}
                \\[0.5em]
                0
                 & \hspace{-2em} m - m' - i \,\, \mathrm{odd}
            \end{cases}
        $}
\end{proof}

\subsubsection{Probability density function}
The latitude integrals (Equations~\ref{eq:e2} and \ref{eq:E2}) involve
rotations by an angle $\bbphi$ about $\hat{\mathbf{x}}$, which
may be accomplished by choosing
Euler angles $\upalpha = \nicefrac{\pi}{2}$, $\upbeta = \bbphi$, and
$\upgamma = -\nicefrac{\pi}{2}$, such that
\begin{align}
    \mathbf{R}^l_{\hat{\mathbf{x}}}(\bbphi)
     & =
    {\mathbf{U}^l}^{-1} \mathbf{D}^l_{\hat{\mathbf{x}}}(\bbphi) \mathbf{U}^l
\end{align}
with
\begin{align}
    \mathbf{D}^l_{\hat{\mathbf{x}}}(\bbphi)
     & =
    \mathbf{D}^l\left(\frac{\pi}{2}, \bbphi, -\frac{\pi}{2}\right)
    \quad.
\end{align}
From the expressions above, it is clear that all terms in
$\mathbf{R}_{\hat{\mathbf{x}}}(\bbphi)$ are equal to (weighted) sums of powers
of $(1 \pm \cos\bbphi)$.
Since our goal is to compute integrals of these terms multiplied by
a probability density function, it is convenient to model
$\cos\bbphi$ as a Beta-distributed variable. As we will see, this
choice will allow us to
analytically compute the first two moments of the distribution of
$\bbphi$ conditioned on $\pmb{\theta}_\phi$.

The Beta distribution in $\cos\bbphi$ has hyperparameters $\alpha$ and $\beta$
(not to be confused with the Euler angles $\upalpha$ and $\upbeta$) and PDF given by
\begin{align}
    \label{eq:cosphi-pdf}
    p \big(\cos\bbphi \, \big| \, \alpha, \beta \big)
     & =
    \dfrac{\Gamma(\alpha + \beta)}{\Gamma(\alpha)\Gamma(\beta)}
    (\cos\bbphi)^{\alpha - 1}
    (1 - \cos\bbphi)^{\beta - 1}
    \quad,
\end{align}
where $\Gamma$ is the Gamma function. The implied distribution for $\bbphi$
may be computed by a straightforward change of variable:
\begin{proof}{test_beta_transform}
    \label{eq:phi-pdf}
    p \big(\bbphi \, \big| \, \alpha, \beta \big)
    & =
    \dfrac{\Gamma(\alpha + \beta)}{2\Gamma(\alpha)\Gamma(\beta)}
    \big|
    \sin\bbphi
    \big|
    (\cos\bbphi)^{\alpha - 1}
    (1 - \cos\bbphi)^{\beta - 1}
    \quad,
\end{proof}
for $\bbphi \in \left[ -\frac{\pi}{2}, \frac{\pi}{2} \right]$.
Both $\alpha$ and $\beta$ are restricted to $(0, \infty)$.
However, in practice it is necessary to limit the values of these parameters
to a finite range to ensure the numerical stability of the algorithm.
It is also convenient
to work with the log of these quantities because of their large dynamic range.
We therefore introduce the modified parameters
\begin{align}
    \label{eq:gauss2beta}
    a & \equiv \frac{\ln\alpha - K_{00}}{K_{10} - K_{00}}
    \nonumber                                             \\[0.5em]
    b & \equiv \frac{\ln\beta - K_{10}}{K_{11} - K_{10}}
\end{align}
with inverse transform
\begin{align}
    \label{eq:beta2gauss}
    \alpha & = \exp\left({K_{00} + (K_{10} - K_{00})a}\right)
    \nonumber                                                 \\
    \beta  & = \exp\left({K_{10} + (K_{11} - K_{10})b}\right)
    \quad,
\end{align}
where the matrix
\begin{align}
    \setstackgap{L}{1.25\baselineskip}
    \fixTABwidth{T}
    \mathbf{K}
    =
    \parenMatrixstack{
    0              & 10 \\
    \ln\frac{1}{2} & 10
    }
\end{align}
defines the minimum and maximum values of $\ln\alpha$ (top row) and $\ln\beta$
(bottom row) we adopt in our implementation of the algorithm. The lower limits
correspond to $\alpha > 1$ and $\beta > \frac{1}{2}$, which excludes
distributions with unphysically sharp peaks at $\phi = 90^\circ$. Both $a$
and $b$ are restricted to the domain $(0, 1)$, and together comprise the
hyperparameter vector
\begin{align}
    \label{eq:ab}
    \pmb{\theta}_\phi = \left(
    a \, \, \, \,
    b \right)^\top
    \quad.
\end{align}
Because of their trivial domain, these parameters are convenient to sample
in when doing inference (provided we account for their implied prior on
the spot latitudes; see below). However, $a$ and $b$ do not intuitively
relate to physical quantities of interest. In many cases it is more
desirable to parametrize the latitude distribution in terms of a parameter
$\mu_\phi$ controlling the central latitude and a parameter $\sigma_\phi$
controlling the dispersion in latitude among the spots.
In this case, we may choose instead
\begin{align*}
    \tag{\ref*{eq:ab}*}
    \pmb{\theta}_\phi = \left(
    \mu_\phi \, \, \, \,
    \sigma_\phi \right)^\top
    \quad.
\end{align*}
Over most of the parameter space in $a$ and $b$,
the spot latitude distribution defined above is well approximated
by a bimodal Gaussian. In particular, there exists
a one-to-one relationship between $a$ and $b$ and the mean $\mu_\phi$
and standard deviation $\sigma_\phi$ of a normal approximation to the
distribution. Moreover, we find that
if we let $\mu_\phi$ be the \emph{mode} of the PDF and
$\sigma_\phi^2$ be a \emph{local} approximation to the variance of the PDF,
the relationship has a convenient closed form.

To compute $\mu_\phi$, we differentiate Equation~(\ref{eq:phi-pdf}) with respect
to $\phi$, set the expression equal to zero, and solve
for $\phi$ to obtain
\begin{proof}{test_lat_transforms}
    \label{eq:muphi}
    \mu_\phi &= 2 \tan^{-1}
    \left(
    \sqrt{
        2 \alpha + \beta - 2 -
        \sqrt{
            4 \alpha^2
            - 8 \alpha
            - 6 \beta
            + 4 \alpha \beta
            + \beta^2
            + 5
        }
    }
    \right)
    \quad.
\end{proof}
To compute $\sigma_\phi$, we note that the variance of a Gaussian distribution
$\varphi(\phi; \mu, \sigma^2)$
is the negative reciprocal of its curvature in log space:
\begin{align}
    \sigma^2 = -\left(\frac{\mathrm{d}^2 \ln \varphi(\phi; \mu, \sigma^2)}{\mathrm{d}\phi^2}\right)^{-1}
\end{align}
We therefore twice differentiate the log of Equation~(\ref{eq:phi-pdf}), negate it,
take the reciprocal, and evaluate it at $\phi = \mu_\phi$ to obtain a local
approximation to the standard deviation of the distribution at the mode:
\begin{proof}{test_lat_transforms}
    \label{eq:sigmaphi}
    \sigma_\phi &= \frac{\sin\mu_\phi}{
        \sqrt{1
            - \alpha
            + \beta
            + \left(\beta - 1\right) \cos\mu_\phi
            + \dfrac{\alpha - 1}{\cos\mu_\phi^2}}}
    \quad.
\end{proof}
For completeness, the inverse transform also has a closed form:
\begin{proof}{test_lat_transforms}
    \alpha &= \frac{2 + 4 \sigma_\phi^2 + (3 + 8 \sigma_\phi^2) \cos\mu_\phi + 2 \cos\left( 2\mu_\phi \right) + \cos\left( 3\mu_\phi \right)}{16 \sigma_\phi^2 \cos\left( \frac{\mu_\phi}{2} \right)^4} \\
    \beta &= \frac{\cos\mu_\phi + 2 \sigma_\phi^2 (3 + \cos\left( 2\mu_\phi \right)) - \cos\left( 3\mu_\phi \right)}{16 \sigma_\phi^2 \cos\left( \frac{\mu_\phi}{2} \right)^4}
    \quad.
\end{proof}

\begin{figure}[t!]
    \begin{centering}
        \includegraphics[width=\linewidth]{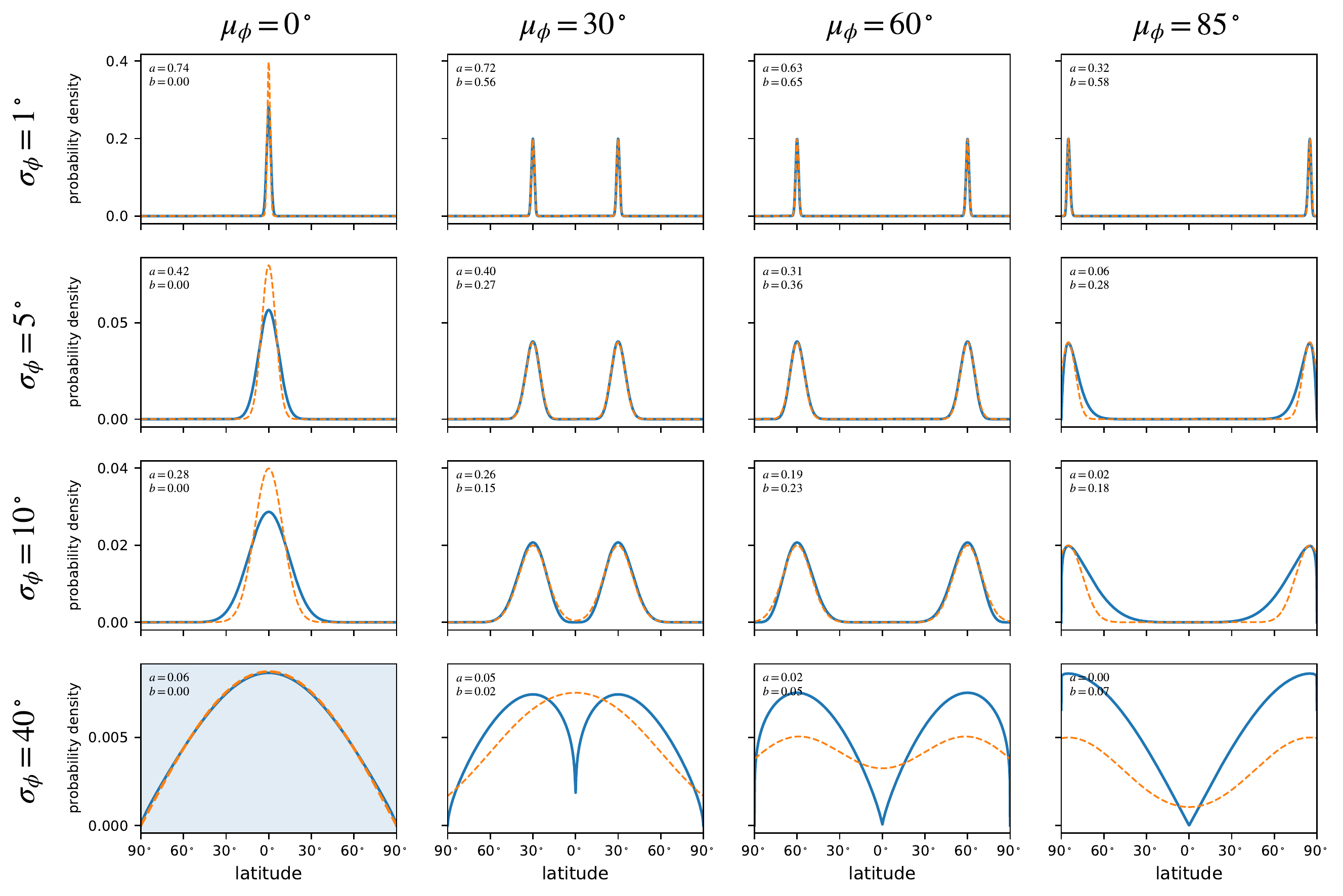}
        \oscaption{latitude_pdf}{%
            Probability density function for the spot latitude (blue curves)
            for different values of the mode $\mu_\phi$ (columns)
            and local standard deviation $\sigma_\phi$ (rows).
            The corresponding values of $a$ and $b$ are
            indicated within each panel. The bimodal
            normal distribution with mean $\mu_\phi$ and standard deviation
            $\sigma_\phi$ is shown as the dashed orange curves; for mid-latitude
            modes and low standard deviations, the Gaussian approximation is
            quite good.
            The shaded panel in the lower right
            ($\mu_\phi = 0^\circ$, $\sigma_\phi = 0^\circ$)
            corresponds to an approximately isotropic distribution of
            spots over the surface of the star; in this panel only, the
            dashed orange curve corresponds to a cosine distribution in
            $\phi$ (i.e., the exact isotropic distribution).
            \label{fig:latitude_pdf}
        }
    \end{centering}
\end{figure}

For reference, Figure~\ref{fig:latitude_pdf} shows the latitude PDF
and the corresponding Gaussian approximation for different values of
$\mu_\phi$ and $\sigma_\phi$. The corresponding values of $a$ and $b$ are
indicated in the top left of each panel. For $\mu_\phi$ at intermediate
latitudes and moderate values of $\sigma_\phi$, the approximation is quite good.
However, for $\mu_\phi$ very close to the equator or to the poles, the curvature of
the distribution changes significantly as a function of $\phi$, so the variance is somewhat
underestimated by the approximation; and for $\sigma_\phi$ large, the
distribution becomes noticeably non-Gaussian.

The shaded panel at the lower left is a special case of the distribution
($\mu_\phi = 0^\circ$, $\sigma_\phi \approx 40^\circ$; or equivalently,
$a \approx 0.06$, $b = 0$), which is approximately isotropic in latitude. In
this panel, the orange curve instead corresponds to an isotropic (cosine)
distribution in $\phi$; note the excellent agreement.
Thus, in addition to having closed-form moments, the Beta distribution is
quite flexible and, via the transforms outlined above, intuitive in how
it affects the distribution of spots on the surface of a star.

In the following sections we derive expressions
for the moments of the distribution in terms
of $\alpha$ and $\beta$, as this is somewhat more convenient; these
can easily be transformed into expressions involving either $\mu_\phi$ and $\sigma_\phi$
or $a$ and $b$ via the equations above. The former parametrization is convenient
when these properties are known and can be fixed; i.e., when using the GP as
a restrictive prior for the light curve of a star whose spot distribution is
already understood.
However, for the purposes of posterior inference---that is, when trying to
constrain the hyperparameters of the GP---we recommend sampling in the
parameters $a$ and $b$, since their domains are trivial,
with uncorrelated boundaries. Posterior constraints on these quantities may
easily be transformed into constraints on $\mu_\phi$ and $\sigma_\phi$
via the equations above. Note, importantly, that this
requires us to explicitly add a Jacobian term to the likelihood
to account for the prior implied by sampling uniformly in $a$ and $b$
in the range $(0, 1)$.
The Jacobian is given by
\begin{proof}{test_lat_jac}
    \label{eq:J}
    J &=
    \frac{\partial{\mu_\phi}}{\partial{a}}
    \frac{\partial{\sigma_\phi}}{\partial{b}} -
    \frac{\partial{\mu_\phi}}{\partial{b}}
    \frac{\partial{\sigma_\phi}}{\partial{a}}
    \\
    &=\resizebox{.925\hsize}{!}{$
            \frac{
                \left(K_{10} - K_{00}\right) \left(K_{11} - K_{10}\right)
                \alpha \beta \left(1 + \cos\mu_\phi\right)^3
                \sin\left(2\mu_\phi\right)^3
            }{
                \sigma_\phi
                \left(
                2 \alpha + \beta - 3 +
                \left(2 \alpha + \beta - 1\right) \cos\mu_\phi
                \right)
                \left(
                2 \left( \alpha + \beta - 1\right) +
                3 (\beta - 1) \cos\mu_\phi
                - 2 \left( \alpha - \beta - 1 \right) \cos\left( 2\mu_\phi \right)
                + \left( \beta - 1\right) \cos\left( 3\mu_\phi \right)
                \right)^2
            }
        $}
    \nonumber
\end{proof}
Adding the log of the absolute value of $J$ to the log likelihood
corrects for the \emph{ad hoc}
prior on the latitude parameters introduced by our particular choice of
parametrization, enforcing instead a uniform prior on the quantities $\mu_\phi$
and $\sigma_\phi$.

\subsubsection{First moment}
Since the Wigner matrices are block diagonal, we may evaluate the moments of the
distribution one spherical harmonic degree at a time. To that end, let us
write the first moment integral as
\begin{proof}{test_ephi}
    \label{eq:ephi}
    \mathbf{e}_\phi
    & \equiv
    \int
    \mathbf{R}_{\hat{\mathbf{x}}}(\bbphi) \,
    \mathbf{e}_r \,
    p(\bbphi \, \big| \, \pmb{\theta}_{\phi}) \,
    \mathrm{d}\bbphi
    \nonumber
    \\
    & =
    \left(
    \mathbf{e}_\phi^0
    \,\,\,
    \mathbf{e}_\phi^1
    \,\,\,
    \mathbf{e}_\phi^2
    \,\,\,
    \cdots
    \,\,\,
    \mathbf{e}_\phi^{l_{\mathrm{max}}}
    \right)^\top
    \quad,
\end{proof}
where
\begin{proof}{test_ephi}
    \mathbf{e}_\phi^l
    & =
    \int
    \mathbf{R}^l_{\hat{\mathbf{x}}}(\bbphi) \,
    \mathbf{e}^l_r \,
    p(\bbphi \, \big| \, \pmb{\theta}_{\phi}) \,
    \mathrm{d}\bbphi
    \nonumber \\
    & =
    {\mathbf{U}^l}^{-1}
    \mathbf{p}^l_\phi
    \quad,
\end{proof}
and we define
\begin{align}
    \label{eq:plphi}
    \mathbf{p}^l_\phi
     & \equiv
    \int
    \mathbf{D}^l_{\hat{\mathbf{x}}}(\bbphi) \,
    \bar{\mathbf{e}}^l_r \,
    p(\phi \, \big| \, \pmb{\theta}_{\phi}) \,
    \mathrm{d}\bbphi
    \\
    \bar{\mathbf{e}}^l_r
     & \equiv
    \mathbf{U}^l
    \mathbf{e}^l_r
    \quad.
\end{align}
The integral $\mathbf{p}_\phi^l$ defined above has a closed-form solution.
To show this, we write the terms of $\mathbf{p}^l_\phi$ as
\begin{align}
    {({p^l_\phi})_{}}_m
     & =
    \int
    \sum\limits_{\mu=-l}^l
    {({D^l_{\hat{\mathbf{x}}}})_{}}_{m,\mu}(\bbphi) \,
    {({\bar{e}}^l_r)_{}}_{\mu} \,
    p(\bbphi \, \big| \, \pmb{\theta}_{\phi}) \,
    \mathrm{d}\bbphi
    \nonumber \\[0.5em]
     & =
    \dfrac{\Gamma(\alpha + \beta)}{\Gamma(\alpha)\Gamma(\beta)}
    \sum\limits_{\mu=-l}^l
    {({\bar{e}}^l_r)_{}}_{\mu}
    \exp\left[{\frac{\imag \pi}{2}(m - \mu)}\right]
    \sum\limits_{i=0}^{2l} c_{m,\mu,i}^{l}
    \,
    {({q^l_\phi})_{}}_i(\pmb{\theta}_{\phi})
    \quad,
\end{align}
where
\begin{proof}{test_qlphii}
    {({q^l_\phi})_{}}_i(\pmb{\theta}_{\phi})
    & \equiv
    \frac{1}{2}
    \int_{-\frac{\pi}{2}}^{\frac{\pi}{2}}
    \mathrm{sgn}(\sin\bbphi)^{i}
    \big| \sin\bbphi \big|
    (\cos\bbphi)^{\alpha - 1}
    (1 - \cos\bbphi)^{l + \beta - \frac{i}{2} - 1}
    (1 + \cos\bbphi)^\frac{i}{2}
    \,
    \mathrm{d}\bbphi
    \nonumber \\[0.5em]
    & =
    \begin{cases}
        \displaystyle\int_{0}^{1}
        \mathbb{x}^{\alpha - 1}
        (1 - \mathbb{x})^{l + \beta - \frac{i}{2} - 1 }
        (1 + \mathbb{x})^\frac{i}{2}
        \,
        \mathrm{d}\mathbb{x}
         & i \,\, \mathrm{even}
        \\
        0
         & i \,\, \mathrm{odd} \quad,
    \end{cases}
\end{proof}
and in the last line we made use of the transformation $\mathbb{x} = \cos\bbphi$.
The integral in the expression above has a closed-form solution in terms
of the hypergeometric function $_2F_1$:
\begin{proof}{test_qlphii}
    \label{eq:qlphii}
    {({q^l_\phi})_{}}_i(\pmb{\theta}_{\phi})
    & =
    \resizebox{.72\hsize}{!}{$
            \begin{cases}
                \dfrac{\Gamma(\alpha)\Gamma(l + \beta - \frac{i}{2})}{\Gamma(l + \alpha + \beta - \frac{i}{2})}
                \,
                _2F_1\Big(
                -\frac{i}{2},
                \alpha;
                l + \alpha + \beta - \frac{i}{2};
                -1
                \Big)
                 & \hspace{-3em} i \,\, \mathrm{even}
                \\
                0
                 & \hspace{-3em} i \,\, \mathrm{odd} \quad.
            \end{cases}
        $}
\end{proof}
In order to compute the integral $\mathbf{e}_\phi$
(Equation~\ref{eq:e2}) we must evaluate Equation~(\ref{eq:qlphii})
for all $0 \leq l \leq l_{\mathrm{max}}$, $0 \leq i \leq 2l$,
which can be done efficiently via upward recursion relations for both
the gamma function and the hypergeometric function.

\subsubsection{Second moment}
Similarly as before, let us write the second moment integral as
\begin{proof}{test_bigEphi}
    \label{eq:Ephi}
    \mathbf{E}_\phi
    & \equiv
    \int
    \mathbf{R}_{\hat{\mathbf{x}}}(\bbphi) \,
    \mathbf{E}_r \,
    \mathbf{R}_{\hat{\mathbf{x}}}^\top(\bbphi) \,
    p(\bbphi \, \big| \, \pmb{\theta}_{\phi}) \,
    \mathrm{d}\bbphi
    \nonumber
    \\
    & =
    \setstackgap{L}{1.25\baselineskip}
    \fixTABwidth{T}
    \parenMatrixstack{
    \mathbf{E}_\phi^{0,0} & \mathbf{E}_\phi^{0,1} & \mathbf{E}_\phi^{0,2} & \cdots & \mathbf{E}_\phi^{0,l_{\mathrm{max}}} \\
    \mathbf{E}_\phi^{1,0} & \mathbf{E}_\phi^{1,1} & \mathbf{E}_\phi^{1,2} & \cdots & \mathbf{E}_\phi^{1,l_{\mathrm{max}}} \\
    \mathbf{E}_\phi^{2,0} & \mathbf{E}_\phi^{2,1} & \mathbf{E}_\phi^{2,2} & \cdots & \mathbf{E}_\phi^{2,l_{\mathrm{max}}} \\
    \vdots & \vdots & \vdots & \ddots & \vdots \\
    \mathbf{E}_\phi^{l_{\mathrm{max}},0} & \mathbf{E}_\phi^{l_{\mathrm{max}},1} & \mathbf{E}_\phi^{l_{\mathrm{max}},2} & \cdots & \mathbf{E}_\phi^{l_{\mathrm{max}},l_{\mathrm{max}}}
    }
    \quad,
\end{proof}
where
\begin{proof}{test_bigEphi}
    \mathbf{E}_\phi^{l,l'}
    & =
    \int
    \mathbf{R}^l_{\hat{\mathbf{x}}}(\bbphi) \,
    \mathbf{E}^{l,l'}_r \,
    {\mathbf{R}^{l'}_{\hat{\mathbf{x}}}}^\top(\bbphi) \,
    p(\bbphi \, \big| \, \pmb{\theta}_{\phi}) \,
    \mathrm{d}\bbphi
    \nonumber \\
    & =
    {\mathbf{U}^l}^{-1}
    \mathbf{P}^{l,l'}_\phi
    {{\mathbf{U}^{l'}}^{-1}}^\top
\end{proof}
and we define
\begin{align}
    \label{eq:Plphi}
    \mathbf{P}^{l,l'}_\phi
     & \equiv
    \int
    \mathbf{D}^l_{\hat{\mathbf{x}}}(\bbphi) \,
    \bar{\mathbf{E}}^{l,l'}_r \,
    {\mathbf{D}^{l'}_{\hat{\mathbf{x}}}}^\top(\bbphi) \,
    p(\bbphi \, \big| \, \pmb{\theta}_{\phi}) \,
    \mathrm{d}\bbphi
    \\
    \bar{\mathbf{E}}^{l,l'}_r
     & \equiv
    \mathbf{U}^l
    \mathbf{E}^{l,l'}_r
    {\mathbf{U}^{l'}}^\top
    \quad.
\end{align}
As before, we may express the solution to the integral $\mathbf{P}^{l,l'}_\phi$ in
closed form. Let us write the terms of $\mathbf{P}^{l,l'}_\phi$ as
\begin{align}
    {({P^{l,l'}_\phi})_{}}_{m,m'}
     & =
    \int
    \sum\limits_{\mu=-l}^l
    \sum\limits_{{\mu'}=-l'}^{l'}
    {({D^l_{\hat{\mathbf{x}}}})_{}}_{m,\mu}(\bbphi) \,
    {(\bar{E}^{l,l'}_r)_{}}_{\mu,{\mu'}} \,
    {({D^{l'}_{\hat{\mathbf{x}}}})_{}}_{m',{\mu'}}(\bbphi) \,
    p(\bbphi \, \big| \, \pmb{\theta}_{\phi}) \,
    \mathrm{d}\bbphi
    \\[0.5em]
     & =
    \resizebox{.75\hsize}{!}{$
            \dfrac{\Gamma(\alpha + \beta)}{\Gamma(\alpha)\Gamma(\beta)}
            \sum\limits_{\mu=-l}^l
            \sum\limits_{{\mu'}=-l'}^{l'}
            {(\bar{E}^{l,l'}_r)_{}}_{\mu,{\mu'}} \,
            \exp\left[{\frac{\imag \pi}{2}(m - \mu + m' - {\mu'})}\right]
            \sum\limits_{i=0}^{2l}
            \sum\limits_{i'=0}^{2l'}
            c_{m,\mu,i}^{l}
            c_{m',{\mu'},i'}^{l'}
            \,
            {({Q^{l,l'}_\phi})_{}}_{i,i'}(\pmb{\theta}_{\phi})
        $}
    \quad,
    \nonumber
\end{align}
where, similarly to before,
\begin{proof}{test_bigEphi}
    {({Q^{l,l'}_\phi})_{}}_{i,i'}(\pmb{\theta}_{\phi})
    & \equiv
    \resizebox{.8\hsize}{!}{$
            \frac{1}{2}
            \int_{-\frac{\pi}{2}}^{\frac{\pi}{2}}
            \mathrm{sgn}(\sin\bbphi)^{i+i'}
            \big| \sin\bbphi \big|
            (\cos\bbphi)^{\alpha - 1}
            (1 - \cos\bbphi)^{l + l' + \beta - \frac{i+i'}{2} - 1}
            (1 + \cos\bbphi)^\frac{i+i'}{2}
            \,
            \mathrm{d}\bbphi
        $}
    \nonumber \\[0.5em]
    & =
    \begin{cases}
        \displaystyle\int_{0}^{1}
        \mathbb{x}^{\alpha - 1}
        (1 - \mathbb{x})^{l + l' + \beta - \frac{i+i'}{2} - 1 }
        (1 + \mathbb{x})^\frac{i+i'}{2}
        \,
        \mathrm{d}\mathbb{x}
         & \hspace{-2em} i+i' \,\, \mathrm{even}
        \\
        0
         & \hspace{-2em} i+i' \,\, \mathrm{odd} \quad.
    \end{cases}
\end{proof}
We may again express this integral in closed form:
\begin{proof}{test_bigEphi}
    \label{eq:Qlphiij}
    {({Q^{l,l'}_\phi})_{}}_{i,i'}(\pmb{\theta}_{\phi})
    & =
    \resizebox{.68\hsize}{!}{$
            \begin{cases}
                \dfrac{\Gamma(\alpha)\Gamma(l {+} l' {+} \beta {-} \frac{i+i'}{2})}{\Gamma(l {+} l' {+} \alpha {+} \beta {-} \frac{i+i'}{2})}
                \,
                _2F_1\Big(
                -\frac{i+i'}{2},
                \alpha;
                l {+} l' {+} \alpha {+} \beta {-} \frac{i+i'}{2};
                -1
                \Big)
                 & \hspace{-3em} i+i' \,\, \mathrm{even}
                \\
                0
                 & \hspace{-3em} i+i' \,\, \mathrm{odd} \quad.
            \end{cases}
        $}
\end{proof}
As before, this integral may be evaluated recursively to efficiently compute all
of the terms in $\mathbf{E}_\phi$.

\subsection{The Longitude Integrals}
\label{sec:lon}

In this section we will compute the first and second moments
of the longitude distribution
($\mathbf{e}_\lambda$ and $\mathbf{E}_\lambda$, given by Equations
\ref{eq:e3} and \ref{eq:E3}, respectively). The math here is
very similar to that in the previous section, as we are again
dealing with integrals of Wigner matrices (\S\ref{sec:wigner}).

\subsubsection{Probability density function}
The longitude integrals (Equations~\ref{eq:e3} and \ref{eq:E3}) involve
rotations by an angle $\bblambda$ about $\hat{\mathbf{y}}$, which
may be accomplished by choosing
Euler angles $\upalpha = 0$, $\upbeta = \bblambda$, and
$\upgamma = 0$, such that
\begin{align}
    \mathbf{R}^l_{\hat{\mathbf{y}}}(\bblambda)
     & =
    {\mathbf{U}^l}^{-1} \mathbf{D}^l_{\hat{\mathbf{y}}}(\bblambda) \mathbf{U}^l
\end{align}
with
\begin{align}
    \mathbf{D}^l_{\hat{\mathbf{y}}}(\bblambda)
     & =
    \mathbf{D}^l\left(0, \bblambda, 0\right)
    \quad.
\end{align}
Since we expect the longitudinal distribution of features on the surfaces
of stars to be (on average) isotropic, we will place a uniform prior on
$\bblambda \in [-\pi, \pi)$:
\begin{align}
    p(\bblambda \big| \pmb{\theta}_\lambda)
     & =
    \begin{cases}
        \frac{1}{2\pi} & -\pi \leq \bblambda < \pi
        \\
        0              & \mathrm{otherwise} \quad.
    \end{cases}
\end{align}
We therefore have no hyperparameters
controlling the longitudinal distribution, i.e.,
\begin{align}
    \pmb{\theta}_\lambda = \left( \,\,\, \right)
    \quad.
\end{align}

\subsubsection{First moment}
As before, we will solve for the terms of the moment integrals one
spherical harmonic degree at a time:
\begin{proof}{test_elam}
    \mathbf{e}_\lambda
    & \equiv
    \int
    \mathbf{R}_{\hat{\mathbf{y}}}(\bblambda) \,
    \mathbf{e}_\phi \,
    p(\bblambda \, \big| \, \pmb{\theta}_{\lambda}) \,
    \mathrm{d}\bblambda
    \nonumber
    \\
    & =
    \left(
    \mathbf{e}_\lambda^0
    \,\,\,
    \mathbf{e}_\lambda^1
    \,\,\,
    \mathbf{e}_\lambda^2
    \,\,\,
    \cdots
    \,\,\,
    \mathbf{e}_\lambda^{l_{\mathrm{max}}}
    \right)^\top
    \quad,
\end{proof}
where
\begin{proof}{test_elam}
    \mathbf{e}_\lambda^l
    & =
    \int
    \mathbf{R}^l_{\hat{\mathbf{y}}}(\bblambda) \,
    \mathbf{e}^l_\phi \,
    p(\bblambda \, \big| \, \pmb{\theta}_{\lambda}) \,
    \mathrm{d}\bblambda
    \nonumber \\
    & =
    {\mathbf{U}^l}^{-1}
    \mathbf{p}^l_\lambda
    \quad,
\end{proof}
and we define
\begin{align}
    \label{eq:pllam}
    \mathbf{p}^l_\lambda
     & \equiv
    \int
    \mathbf{D}^l_{\hat{\mathbf{x}}}(\bblambda) \,
    \bar{\mathbf{e}}^l_\phi \,
    p(\bblambda \, \big| \, \pmb{\theta}_{\lambda}) \,
    \mathrm{d}\bblambda
    \\
    \bar{\mathbf{e}}^l_\phi
     & \equiv
    \mathbf{U}^l
    \mathbf{e}^l_\phi
    \quad.
\end{align}
The integral $\mathbf{p}^l_\lambda$ defined above has a closed-form solution.
To show this, we write the terms of $\mathbf{p}^l_\lambda$ as
\begin{align}
    {({p^l_\lambda})_{}}_m
     & =
    \int
    \sum\limits_{\mu=-l}^l
    {({D^l_{\hat{\mathbf{y}}}})_{}}_{m,\mu}(\bblambda) \,
    {({\bar{e}^l_\phi})_{}}_{\mu} \,
    p(\bblambda \, \big| \, \pmb{\theta}_{\lambda}) \,
    \mathrm{d}\bblambda
    \nonumber \\[0.5em]
     & =
    \sum\limits_{\mu=-l}^l
    {({\bar{e}^l_\phi})_{}}_{\mu} \,
    \sum\limits_{i=0}^{2l} c_{m,\mu,i}^{l}
    \,
    {({q^l_\lambda})_{}}_i
    \quad,
\end{align}
where
\begin{proof}{test_qllam}
    {({q^l_\lambda})_{}}_i
    & \equiv
    \frac{1}{2\pi}
    \int_{-\pi}^{\pi}
    \mathrm{sgn}(\sin\bblambda)^{i}
    (1 - \cos\bblambda)^{l - \frac{i}{2}}
    (1 + \cos\bblambda)^\frac{i}{2}
    \,
    \mathrm{d}\bblambda
    \nonumber \\[0.5em]
    & =
    \begin{cases}
        \dfrac{2^l \Gamma \left(\frac{1+i}{2}\right)
            \Gamma \left(l+\frac{1-i}{2}\right)}{\pi  \Gamma (l+1)}
         & i \,\, \mathrm{even}
        \\
        0
         & i \,\, \mathrm{odd} \quad,
    \end{cases}
\end{proof}
whose terms may easily be computed by upward recursion.
Since $\mathbf{q}^l_I$ does not depend on any user inputs,
it may be computed a single time as a pre-processing step
for efficiency.

\subsubsection{Second moment}
We write the second moment integral as
\begin{proof}{test_bigElam}
    \mathbf{E}_\lambda
    & \equiv
    \int
    \mathbf{R}_{\hat{\mathbf{y}}}(\bblambda) \,
    \mathbf{E}_\phi \,
    \mathbf{R}_{\hat{\mathbf{y}}}^\top(\bblambda) \,
    p(\bblambda \, \big| \, \pmb{\theta}_{\bblambda}) \,
    \mathrm{d}\bblambda
    \nonumber
    \\
    & =
    \setstackgap{L}{1.25\baselineskip}
    \fixTABwidth{T}
    \parenMatrixstack{
    \mathbf{E}_\lambda^{0,0} & \mathbf{E}_\lambda^{0,1} & \mathbf{E}_\lambda^{0,2} & \cdots & \mathbf{E}_\lambda^{0,l_{\mathrm{max}}} \\
    \mathbf{E}_\lambda^{1,0} & \mathbf{E}_\lambda^{1,1} & \mathbf{E}_\lambda^{1,2} & \cdots & \mathbf{E}_\lambda^{1,l_{\mathrm{max}}} \\
    \mathbf{E}_\lambda^{2,0} & \mathbf{E}_\lambda^{2,1} & \mathbf{E}_\lambda^{2,2} & \cdots & \mathbf{E}_\lambda^{2,l_{\mathrm{max}}} \\
    \vdots & \vdots & \vdots & \ddots & \vdots \\
    \mathbf{E}_\lambda^{l_{\mathrm{max}},0} & \mathbf{E}_\lambda^{l_{\mathrm{max}},1} & \mathbf{E}_\lambda^{l_{\mathrm{max}},2} & \cdots & \mathbf{E}_\lambda^{l_{\mathrm{max}},l_{\mathrm{max}}}
    }
    \quad,
\end{proof}
where
\begin{proof}{test_bigElam}
    \mathbf{E}_\lambda^{l,l'}
    & =
    \int
    \mathbf{R}^l_{\hat{\mathbf{y}}}(\bblambda) \,
    \mathbf{E}^{l,l'}_\phi \,
    {\mathbf{R}^{l'}_{\hat{\mathbf{y}}}}^\top(\bblambda) \,
    p(\bblambda \, \big| \, \pmb{\theta}_{\lambda}) \,
    \mathrm{d}\bblambda
    \nonumber \\
    & =
    {\mathbf{U}^l}^{-1}
    \mathbf{P}^{l,l'}_\lambda
    {{\mathbf{U}^{l'}}^{-1}}^\top
\end{proof}
and we define
\begin{align}
    \label{eq:Pllam}
    \mathbf{P}^{l,l'}_\lambda
     & \equiv
    \int
    \mathbf{D}^l_{\hat{\mathbf{x}}}(\bblambda) \,
    \bar{\mathbf{E}}^{l,l'}_\phi \,
    {\mathbf{D}^{l'}_{\hat{\mathbf{x}}}}^\top(\bblambda) \,
    p(\bblambda \, \big| \, \pmb{\theta}_{\lambda}) \,
    \mathrm{d}\bblambda
    \\
    \bar{\mathbf{E}}^{l,l'}_\phi
     & \equiv
    \mathbf{U}^l
    \mathbf{E}^{l,l'}_\phi
    {\mathbf{U}^{l'}}^\top
    \quad.
\end{align}
We then express the terms of $\mathbf{P}^{l,l'}_\lambda$ as
\begin{align}
    {({P^{l,l'}_\lambda})_{}}_{m,m'}
     & =
    \int
    \sum\limits_{\mu=-l}^l
    \sum\limits_{{\mu'}=-l'}^{l'}
    ({D^l_{\hat{\mathbf{x}}}})_{m,\mu}(\bblambda) \,
    (\bar{E}^{l,l'}_\phi)_{\mu,{\mu'}} \,
    ({D^{l'}_{\hat{\mathbf{x}}}})_{m',{\mu'}}(\bblambda) \,
    p(\bblambda \, \big| \, \pmb{\theta}_{\lambda}) \,
    \mathrm{d}\bblambda
    \nonumber \\[0.5em]
     & =
    \sum\limits_{\mu=-l}^l
    \sum\limits_{{\mu'}=-l'}^{l'}
    (\bar{E}^{l,l'}_\phi)_{\mu,{\mu'}}
    \sum\limits_{i=0}^{2l}
    \sum\limits_{i'=0}^{2l'}
    c_{m,\mu,i}^{l}
    c_{m',{\mu'},i'}^{l'}
    \,
    (Q^{l,l'}_\lambda)_{i,i'}
    \quad,
\end{align}
where
\begin{proof}{test_bigElam}
    (Q^{l,l'}_\lambda)_{i,i'}
    & \equiv
    \frac{1}{2\pi}
    \int_{-\pi}^{\pi}
    \mathrm{sgn}(\sin\bblambda)^{i+i'}
    (1 - \cos\bblambda)^{l + l' - \frac{i+i'}{2}}
    (1 + \cos\bblambda)^\frac{i+i'}{2}
    \,
    \mathrm{d}\bblambda
    \nonumber \\[0.5em]
    & =
    \begin{cases}
        \dfrac{2^{l+l'} \Gamma \left(\frac{1+i+i'}{2}\right)
            \Gamma \left(l+l'+\frac{1-i-i'}{2}\right)}{\pi  \Gamma (l+l'+1)}
         & i+i' \,\, \mathrm{even}
        \\
        0
         & i+i' \,\, \mathrm{odd} \quad,
    \end{cases}
\end{proof}
whose terms may again be computed by upward recursion in a single
pre-processing step.

\subsection{The Contrast Integrals}
\label{sec:contrast}
The final integrals we must take in our computation of
$\mathrm{E} \big[ \mathbbb{y} \, \big| \, \pmb{\theta}_\bullet \big]$
and $\mathrm{E} \big[ \mathbbb{y} \mathbbb{y}^\top \, \big| \, \pmb{\theta}_\bullet \big]$
are the integrals over
the spot contrast distribution, $\mathbf{e}_c$ and $\mathbf{E}_c$.
These are by far the easiest, since the spot contrast is a scalar
multiplier of the spherical harmonic coefficient vector, so we can
pull the terms $\mathbf{e}_\lambda$ and $\mathbf{E}_\lambda$ out
of the
integrals in Equations (\ref{eq:e4}) and (\ref{eq:E4}) to write
\begin{align}
    \mathbf{e}_c
     & \equiv
    -
    \mathbf{e}_\lambda \,
    \int
    \mathbb{c} \,
    p(\mathbb{c} \, \big| \, \pmb{\theta}_{c}) \,
    \mathrm{d}\mathbb{c}
    \\
    \mathbf{E}_c
     & \equiv
    \mathbf{E}_\lambda \,
    \int
    \mathbb{c}^2 \,
    p(\mathbb{c} \, \big| \, \pmb{\theta}_c)
    \mathrm{d}\mathbb{c}
    \quad.
\end{align}
These integrals may be computed analytically for any choice of
probability density function $p(\mathbb{c} \, \big| \, \pmb{\theta}_c)$
with closed-form moments.
However, in practice, it is quite difficult to constrain the
spot contrast from light curves, let alone higher moments of its
distribution; this is due largely to the fact that the contrast
is extremely degenerate with the total number of spots
\citepalias{PaperI}.
In our implementation of the algorithm,
we therefore choose the simplest possible probability distribution,
a delta function:
\begin{align}
    p(\mathbb{c} \, \big| \, \pmb{\theta}_{c}) = \delta(\mathbb{c} - c)
\end{align}
characterized by a single parameter, the contrast of the spots:
\begin{align}
    \pmb{\theta}_c = \left( \, c \, \right)^\top
    \quad.
\end{align}
The moment integrals are then trivial to evaluate:
\begin{align}
    \label{eq:ec}
    \mathbf{e}_c & = - c \, \mathbf{e}_\lambda
    \\
    \label{eq:bigEc}
    \mathbf{E}_c & = c^2 \mathbf{E}_\lambda
    \quad.
\end{align}

\section{Inclination}
\label{sec:inc}

In this section we will compute the first and second moment integrals of the
inclination distribution (Equations~\ref{eq:eI} and \ref{eq:EI}),
which allow us to compute the mean and covariance of the process
that describes the flux marginalized over all values of the inclination
(Equations~\ref{eq:mu_marg} and \ref{eq:cov_marg}).

\subsection{Probability density function}
Similar to the latitude integrals, the process of inclining a star relative to the
observer (see Equation~\ref{eq:akT}) involves
rotations by an angle $-\mathbb{I}$ about $\hat{\mathbf{x}}$, which
may be accomplished by choosing
Euler angles $\upalpha = \nicefrac{\pi}{2}$, $\upbeta = -\mathbb{I}$, and
$\upgamma = -\nicefrac{\pi}{2}$, such that
\begin{align}
    \mathbf{R}^l_{\hat{\mathbf{x}}}\left(-\mathbb{I}\right)
     & =
    {\mathbf{U}^l}^{-1} \mathbf{D}^l_{\hat{\mathbf{x}}}\left(-\mathbb{I}\right) \mathbf{U}^l
\end{align}
with
\begin{align}
    \mathbf{D}^l_{\hat{\mathbf{x}}}\left(-\mathbb{I}\right)
     & =
    \mathbf{D}^l\left(\frac{\pi}{2}, -\mathbb{I}, -\frac{\pi}{2}\right)
    \quad.
\end{align}
Since we expect an isotropic distribution of stellar rotation axes (absent
prior constraints on individual stars), the prior probability density for
the inclination $I$ is simply
\begin{align}
    p(\mathbb{I}) = \sin \mathbb{I}
\end{align}
for $\mathbb{I} \in [0, \frac{\pi}{2}]$.

\subsection{First moment}
\label{sec:inc-mom1}
The expression for the first moment is
\begin{align}
    \mathbf{e}_I
     & \equiv
    \int
    \pmb{\mathcal{A}}(\mathbb{I}, P, \mathbf{u}) \,
    \mathbf{e}_y \,
    p(\mathbb{I}) \,
    \mathrm{d}\mathbb{I}
\end{align}
where
\begin{align}
    \mathbf{e}_y
     & \equiv
    \mathrm{E} \Big[ \mathbbb{y} \, \Big| \, \pmb{\theta}_\bullet \Big]
\end{align}
is the first moment of the distribution over spherical harmonic coefficients
(Equation~\ref{eq:exp_y}). We can use Equations~(\ref{eq:Arows})
and (\ref{eq:akT})
to express the element at index $k$ (corresponding to the mean of the
GP at time $t = t_k$) as
\begin{align}
    \label{eq:eIk}
    \left(e_I\right)_k
     & =
    \int
    \mathbf{a}_k^\top(\mathbb{I})
    \mathbf{e}_y \,
    p(\mathbb{I}) \,
    \mathrm{d}\mathbb{I}
    \nonumber \\
     & =
    \mathbf{r}^\top \,
    \mathbf{A_1} \,
    \int
    \mathbf{R}_{\hat{\mathbf{x}}}\left(-\mathbb{I}\right) \,
    \,
    \left(\mathbf{e}_{y'}\right)_k \,
    p(\mathbb{I}) \,
    \mathrm{d}\mathbb{I}
    \quad,
\end{align}
where we define
\begin{align}
    \left(\mathbf{e}_{y'}\right)_k
     & \equiv
    \mathbf{R}_{\hat{\mathbf{z}}}\left(\frac{2\pi}{P}t_k\right) \,
    \mathbf{R}_{\hat{\mathbf{x}}}\left(\frac{\pi}{2}\right) \,
    \mathbf{e}_y
\end{align}
as the expectation of $\mathbbb{y}$ in the polar frame at time $t = t_k$.%
\footnote{In the presence of limb darkening, we must include the limb
    darkening operator $\mathbf{L}(\mathbf{u})$ in Equation~(\ref{eq:eIk});
    see \S\ref{sec:ld}.}
At this point, it is convenient to invoke the fact that our GP
is longitudinally isotropic: there is no preferred longitude on
the surface of the star, or, equivalently, no preferred phase
in the light curve. The rotation about $\hat{\mathbf{z}}$ (i.e., the
rotational axis of the star) therefore cannot change the expectation
of $\mathbbb{y}$, so
\begin{align}
    \left(\mathbf{e}_{y'}\right)_k
     & =
    \left(\mathbf{e}_{y'}\right)_0
    \nonumber                 \\
     & =
    \mathbf{R}_{\hat{\mathbf{x}}}\left(\frac{\pi}{2}\right) \,
    \mathbf{e}_y
    \nonumber                 \\
     & \equiv \mathbf{e}_{y'}
    \quad.
\end{align}
We therefore have
\begin{align}
    \mathbf{e}_I
     & \equiv
    e_I \, \mathbf{1}
\end{align}
where
\begin{align}
    e_I
     & =
    \mathbf{r}^\top \,
    \mathbf{A_1} \,
    \mathbf{e}_{y''}
\end{align}
and
\begin{align}
    \mathbf{e}_{y''}
     & =
    \int
    \mathbf{R}_{\hat{\mathbf{x}}}\left(-\mathbb{I}\right) \,
    \,
    \mathbf{e}_{y'} \,
    p(\mathbb{I}) \,
    \mathrm{d}\mathbb{I}
    \nonumber \\
     & =
    \left(
    \mathbf{e}_{y''}^0
    \,\,\,
    \mathbf{e}_{y''}^1
    \,\,\,
    \mathbf{e}_{y''}^2
    \,\,\,
    \cdots
    \,\,\,
    \mathbf{e}_{y''}^{l_{\mathrm{max}}}
    \right)^\top
\end{align}
is the expectation of $\mathbbb{y}$ in the observer's frame,
and as before we explicitly separate it out by spherical harmonic degree.
As in \S\ref{sec:lat}, we may write
\begin{proof}{test_eI}
    \mathbf{e}_{y''}^l
    & =
    \int
    \mathbf{R}^l_{\hat{\mathbf{x}}}(-\mathbb{I}) \,
    \mathbf{e}^l_{y'} \,
    p(\mathbb{I}) \,
    \mathrm{d}\mathbb{I}
    \nonumber \\
    & =
    {\mathbf{U}^l}^{-1}
    \mathbf{p}^l_I
    \quad,
\end{proof}
and we define
\begin{align}
    \mathbf{p}^l_I
     & \equiv
    \int
    \mathbf{D}^l_{\hat{\mathbf{x}}}(-\mathbb{I}) \,
    \bar{\mathbf{e}}^l_{y'} \,
    p(\mathbb{I}) \,
    \mathrm{d}\mathbb{I}
    \\
    \bar{\mathbf{e}}^l_{y'}
     & \equiv
    \mathbf{U}^l
    \mathbf{e}^l_{y'}
    \quad.
\end{align}
The integral $\mathbf{p}^l_I$ defined above has a closed-form solution.
To show this, we write the terms of $\mathbf{p}^l_I$ as
\begin{proof}{test_eI}
    {({p^l_I})_{}}_m
    & =
    \int
    \sum\limits_{\mu=-l}^l
    {({D^l_{\hat{\mathbf{x}}}})_{}}_{m,\mu}(-\mathbb{I}) \,
    {({\bar{e}}^l_{y'})_{}}_{\mu} \,
    p(\mathbb{I}) \,
    \mathrm{d}\mathbb{I}
    \nonumber \\[0.5em]
    & =
    \sum\limits_{\mu=-l}^l
    {({\bar{e}}^l_{y'})_{}}_{\mu}
    \exp\left[{\frac{\imag \pi}{2}(m - \mu)}\right]
    \sum\limits_{i=0}^{2l} c_{m,\mu,i}^{l}
    \,
    {({q^l_I})_{}}_i
    \quad,
\end{proof}
where
\begin{proof}{test_eI}
    {({q^l_I})_{}}_i
    & \equiv
    \int_{0}^{\frac{\pi}{2}}
    (-1)^{i}
    (1 - \cos \mathbb{I})^{\frac{2l - i}{2}}
    (1 + \cos \mathbb{I})^\frac{i}{2}
    \sin \mathbb{I}
    \,
    \mathrm{d}\mathbb{I}
    \nonumber \\[0.5em]
    & =
    (-1)^{i}
    \displaystyle\int_{0}^{1}
    (1 - \mathbb{x})^{\frac{2l - i}{2}}
    (1 + \mathbb{x})^\frac{i}{2}
    \,
    \mathrm{d}\mathbb{x}
    \nonumber \\[0.5em]
    &=
    \frac{(-1)^{i}}{l-\frac{i}{2}+1}
    \,
    {_2F_1}\left(
    1, -\frac{i}{2}; 2 + l - \frac{i}{2}; -1
    \right)
    \quad,
\end{proof}
which may easily be computed recursively.
As with the longitude integrals, the vector
$\mathbf{q}^l_I$ need only be computed a single
time as a pre-processing step, as it does not
depend on any user inputs.

\subsection{Second moment}
\label{sec:inc-mom2}
The expression for the second moment is
\begin{align}
    \mathbf{E}_I
     & \equiv
    \int
    \pmb{\mathcal{A}}(\mathbb{I}, P, \mathbf{u}) \,
    \mathbf{E}_y \,
    \pmb{\mathcal{A}}^\top(I, P, \mathbf{u}) \,
    p(\mathbb{I}) \,
    \mathrm{d}\mathbb{I}
\end{align}
where
\begin{align}
    \mathbf{E}_y
     & \equiv
    \mathrm{E} \Big[ \mathbbb{y} \, \mathbbb{y}^\top \, \Big| \, \pmb{\theta}_\bullet \Big]
\end{align}
is the second moment of the distribution over spherical harmonic coefficients
(Equation~\ref{eq:exp_yy}). We can use Equations~(\ref{eq:Arows})
and (\ref{eq:akT})
to express the element at index $k, k'$ (corresponding to the covariance of the
GP between times $t = t_k$ and $t' = t_{k'}$) as
\begin{align}
    \label{eq:Eikkp}
    \left(E_I\right)_{k, k'}
     & =
    \int
    \mathbf{a}_k^\top(\mathbb{I})
    \mathbf{E}_y \,
    \mathbf{a}_{k'}(\mathbb{I})
    p(\mathbb{I}) \,
    \mathrm{d}\mathbb{I}
    \nonumber \\
     & =
    \mathbf{r}^\top \,
    \mathbf{A_1} \,
    \left(\mathbf{E}_{y''}\right)_{k,k'}
    \mathbf{A_1}^\top \,
    \mathbf{r}
    \quad,
\end{align}
where
\begin{align}
    \label{eq:Eyppkkp}
    \left(\mathbf{E}_{y''}\right)_{k,k'}
     & =
    \int
    \mathbf{R}_{\hat{\mathbf{x}}}\left(-\mathbb{I}\right) \,
    \left(\mathbf{E}_{y'}\right)_{k,k'} \,
    \mathbf{R}_{\hat{\mathbf{x}}}^\top\left(-\mathbb{I}\right) \,
    p(\mathbb{I}) \,
    \mathrm{d}\mathbb{I}
\end{align}
is the expectation of $\mathbbb{y}\,\mathbbb{y}^\top$ in the
observer's frame at times $t = t_k$ and $t' = t_{k'}$
and
\begin{align}
    \left(\mathbf{E}_{y'}\right)_{k,k'}
     & \equiv
    \mathbf{R}_{\hat{\mathbf{z}}}\left(\frac{2\pi}{P}t_k\right) \,
    \mathbf{R}_{\hat{\mathbf{x}}}\left(\frac{\pi}{2}\right) \,
    \mathbf{E}_y \,
    \mathbf{R}_{\hat{\mathbf{x}}}^\top\left(\frac{\pi}{2}\right) \,
    \mathbf{R}_{\hat{\mathbf{z}}}^\top\left(\frac{2\pi}{P}t_{k'}\right)
\end{align}
is the expectation of $\mathbbb{y}\,\mathbbb{y}^\top$ in the
polar frame at times $t = t_k$ and $t' = t_{k'}$.
The rest of the computation follows what we did in \S\ref{sec:lat},
except that the number of operations required to compute $\mathbf{E}_I$
is a factor of $K^2$ larger than in the computation of expectations
like $\mathbf{E}_\phi$ (Equation~\ref{eq:Ephi}). That is because we
must compute the integral of all terms of a matrix
for \emph{each} of the $K^2$ elements of $\mathbf{E}_I$.
We discuss in \S\ref{sec:inc-speedup} below strategies that can drastically
improve the computational scaling of marginalizing over the inclination.

Let us write Equation~(\ref{eq:Eyppkkp}) in terms of its spherical harmonic
components:
\begin{proof}{test_bigEI}
    \resizebox{.83\hsize}{!}{$
            \left(\mathbf{E}_{y''}\right)_{k,k'}
            =
            \setstackgap{L}{1.25\baselineskip}
            \fixTABwidth{T}
            \parenMatrixstack{
                \left(\mathbf{E}_{y''}^{0,0}\right)_{k,k'} & \left(\mathbf{E}_{y''}^{0,1}\right)_{k,k'} & \left(\mathbf{E}_{y''}^{0,2}\right)_{k,k'} & \cdots & \left(\mathbf{E}_{y''}^{0,l_{\mathrm{max}}}\right)_{k,k'} \\
                \left(\mathbf{E}_{y''}^{1,0}\right)_{k,k'} & \left(\mathbf{E}_{y''}^{1,1}\right)_{k,k'} & \left(\mathbf{E}_{y''}^{1,2}\right)_{k,k'} & \cdots & \left(\mathbf{E}_{y''}^{1,l_{\mathrm{max}}}\right)_{k,k'} \\
                \left(\mathbf{E}_{y''}^{2,0}\right)_{k,k'} & \left(\mathbf{E}_{y''}^{2,1}\right)_{k,k'} & \left(\mathbf{E}_{y''}^{2,2}\right)_{k,k'} & \cdots & \left(\mathbf{E}_{y''}^{2,l_{\mathrm{max}}}\right)_{k,k'} \\
                \vdots & \vdots & \vdots & \ddots & \vdots \\
                \left(\mathbf{E}_{y''}^{l_{\mathrm{max}},0}\right)_{k,k'} & \left(\mathbf{E}_{y''}^{l_{\mathrm{max}},1}\right)_{k,k'} & \left(\mathbf{E}_{y''}^{l_{\mathrm{max}},2}\right)_{k,k'} & \cdots & \left(\mathbf{E}_{y''}^{l_{\mathrm{max}},l_{\mathrm{max}}}\right)_{k,k'}
            }
        $}
    \quad,
\end{proof}
where
\begin{proof}{test_bigEI}
    \left(\mathbf{E}_{y''}^{l,l'}\right)_{k,k'}
    & =
    \int
    \mathbf{R}^l_{\hat{\mathbf{x}}}(-\mathbb{I}) \,
    \left(\mathbf{E}_{y'}^{l,l'}\right)_{k,k'} \,
    {\mathbf{R}^{l'}_{\hat{\mathbf{x}}}}^\top(-\mathbb{I}) \,
    p(\mathbb{I}) \,
    \mathrm{d}\mathbb{I}
    \nonumber \\
    & =
    {\mathbf{U}^l}^{-1}
    \left(\mathbf{P}^{l,l'}_I\right)_{k,k'}
    {{\mathbf{U}^{l'}}^{-1}}^\top
\end{proof}
and we define
\begin{align}
    \label{eq:PllpIkkp}
    \left(\mathbf{P}^{l,l'}_I\right)_{k,k'}
     & \equiv
    \int
    \mathbf{D}^l_{\hat{\mathbf{x}}}(-\mathbb{I}) \,
    \left(\bar{\mathbf{E}}^{l,l'}_{y'}\right)_{k,k'} \,
    {\mathbf{D}^{l'}_{\hat{\mathbf{x}}}}^\top(-\mathbb{I}) \,
    p(\mathbb{I}) \,
    \mathrm{d}\mathbb{I}
    \\
    \left(\bar{\mathbf{E}}^{l,l'}_{y'}\right)_{k,k'}
     & \equiv
    \mathbf{U}^l
    \left(\mathbf{E}^{l,l'}_{y'}\right)_{k,k'}
    {\mathbf{U}^{l'}}^\top
    \quad.
\end{align}
As before, we may express the solution to the integral in Equation~(\ref{eq:PllpIkkp}) in
closed form. Let us write its terms as
\begin{align}
    \label{eq:nightmare}
    \left[{\left({P^{l,l'}_I}\right)_{}}_{k,k'}\right]_{m,m'}
     & =
    \resizebox{.65\hsize}{!}{$
        \int
        \sum\limits_{\mu=-l}^l
        \sum\limits_{{\mu'}=-l}^l
        {({D^l_{\hat{\mathbf{x}}}})_{}}_{m,\mu}(-\mathbb{I}) \,
        \left[{\left({\bar{E}^{l,l'}_{y'}}\right)_{}}_{k,k'}\right]_{\mu,\mu'}
        {({D^{l'}_{\hat{\mathbf{x}}}})_{}}_{{\mu'},m'}(-\mathbb{I}) \,
        p(\mathbb{I}) \,
        \mathrm{d}\mathbb{I}
    $}
    \\[0.5em]
     & =
    \resizebox{.75\hsize}{!}{$
            \sum\limits_{\mu=-l}^l
            \sum\limits_{{\mu'}=-l'}^{l'}
            \left[{\left({\bar{E}^{l,l'}_{y'}}\right)_{}}_{k,k'}\right]_{\mu,\mu'}
            \exp\left[{\frac{\imag \pi}{2}(m - \mu + m' - {\mu'})}\right]
            \sum\limits_{i=0}^{2l}
            \sum\limits_{i'=0}^{2l'}
            c_{m,\mu,i}^{l}
            c_{m',{\mu'},i'}^{l'}
            \,
            {({Q^{l,l'}_I})_{}}_{i,i'}
        $}
    \quad,
    \nonumber
\end{align}
where, similarly to before,
\begin{proof}{test_bigEI}
    \label{eq:QllpIiip}
    {({Q^{l,l'}_I})_{}}_{i,i'}
    & \equiv
    \int_{0}^{\frac{\pi}{2}}
    (-1)^{i+i'}
    (1 - \cos \mathbb{I})^{l + l' - \frac{i+i'}{2}}
    (1 + \cos \mathbb{I})^\frac{i+i'}{2}
    \sin \mathbb{I}
    \,
    \mathrm{d}\mathbb{I}
    \nonumber \\[0.5em]
    & =
    (-1)^{i+i'}
    \int_{0}^{1}
    (1 - \mathbb{x})^{l + l' - \frac{i+i'}{2}}
    (1 + \mathbb{x})^\frac{i+i'}{2}
    \,
    \mathrm{d}\mathbb{x}
    \nonumber \\[0.5em]
    &=
    \frac{(-1)^{i+i'}}{l+l'-\frac{i+i'}{2}+1}
    \,
    {_2F_1}\left(
    1, -\frac{i+i'}{2}; 2 + l + l' - \frac{i+i'}{2}; -1
    \right)
    \quad,
\end{proof}
which may again be computed recursively in a pre-processing step.

\subsection{Speeding up the computation}
\label{sec:inc-speedup}
The expressions in the previous section are a bit of a nightmare,
particularly because of the dimensionality of some of the linear
operators involved. The complexity of the expressions is due to
the fact that the second moment of the spherical harmonic vector
projected onto the sky (Equation~\ref{eq:Eyppkkp}) is time-dependent:
it changes as the star rotates. Computing the second moment of the
\emph{flux} requires computing the outer product of this tensor
with itself, leading to multi-indexed quantities like those in
Equation~(\ref{eq:nightmare}). In addition to being cumbersome to
evaluate, the full second moment matrix $\mathbf{E}_I$
(and hence the flux covariance matrix) is costly to compute.
It is helpful that Equation~(\ref{eq:QllpIiip}) does not depend
on any user inputs and thus may be pre-computed, but even still
we require evaluating the four nested sums
in Equation~(\ref{eq:Eyppkkp})
$\mathcal{O}(l_\mathrm{max}^3)$ times
\emph{for each entry} in the $(K \times K)$ matrix $\mathbf{E}_I$.

Fortunately, the inner two sums in Equation~(\ref{eq:Eyppkkp})
do not depend on user inputs, so those may be pre-computed,
and Equation~(\ref{eq:Eyppkkp}) may be cast as a straightforward
matrix dot product. In practice we also find it helpful to
take advantage of the phase independence (i.e., stationarity)
of the covariance of our GP, as we did in \S\ref{sec:inc-mom1}:
any two entries $(E_I)_{k,k'}$ and $(E_I)_{j,j'}$ are
the same if $t_{k} - t_{k'} = t_{j} - t_{j'}$. If the data
happen to be evenly sampled, such that the time difference between
adjacent cadences is constant, then we need only compute the
covariance at a total of $K$ points (as opposed to $K^2$), as the
covariance is a circulant matrix which is fully specified by a
single vector of length $K$.

In the more general case where the data are not evenly sampled,
we may still evaluate the covariance at a fixed number of points
$K' < K^2$ and approximate the full covariance matrix via
interpolation. As long as the data are roughly evenly sampled,
as is the case with \emph{Kepler} or \emph{TESS} light curves,
this approximation leads to negligible error when $K' \approx K$,
affording the same $\mathcal{O}(K)$ computational savings.
Note that even in the case where the flux is normalized
(see \S\ref{sec:gp-norm}), the non-stationary correction to the
covariance is applied \emph{after} the step where we marginalize
over the inclination, so this approach is still valid.

%
%
%
%
%
%

\clearpage
\pagebreak
\vspace*{\fill}
\section*{S. Supplementary figures}
Figures~\ref{fig:calibration_midlat}-\ref{fig:calibration_unnorm} below
are referenced in \S\ref{sec:inference-other} discussing additional
calibration runs for our GP.
\vspace*{\fill}
\clearpage
\pagebreak

\renewcommand{\thefigure}{S\arabic{figure}}

\begin{figure}[p!]
    \begin{centering}
        \includegraphics[width=\linewidth]{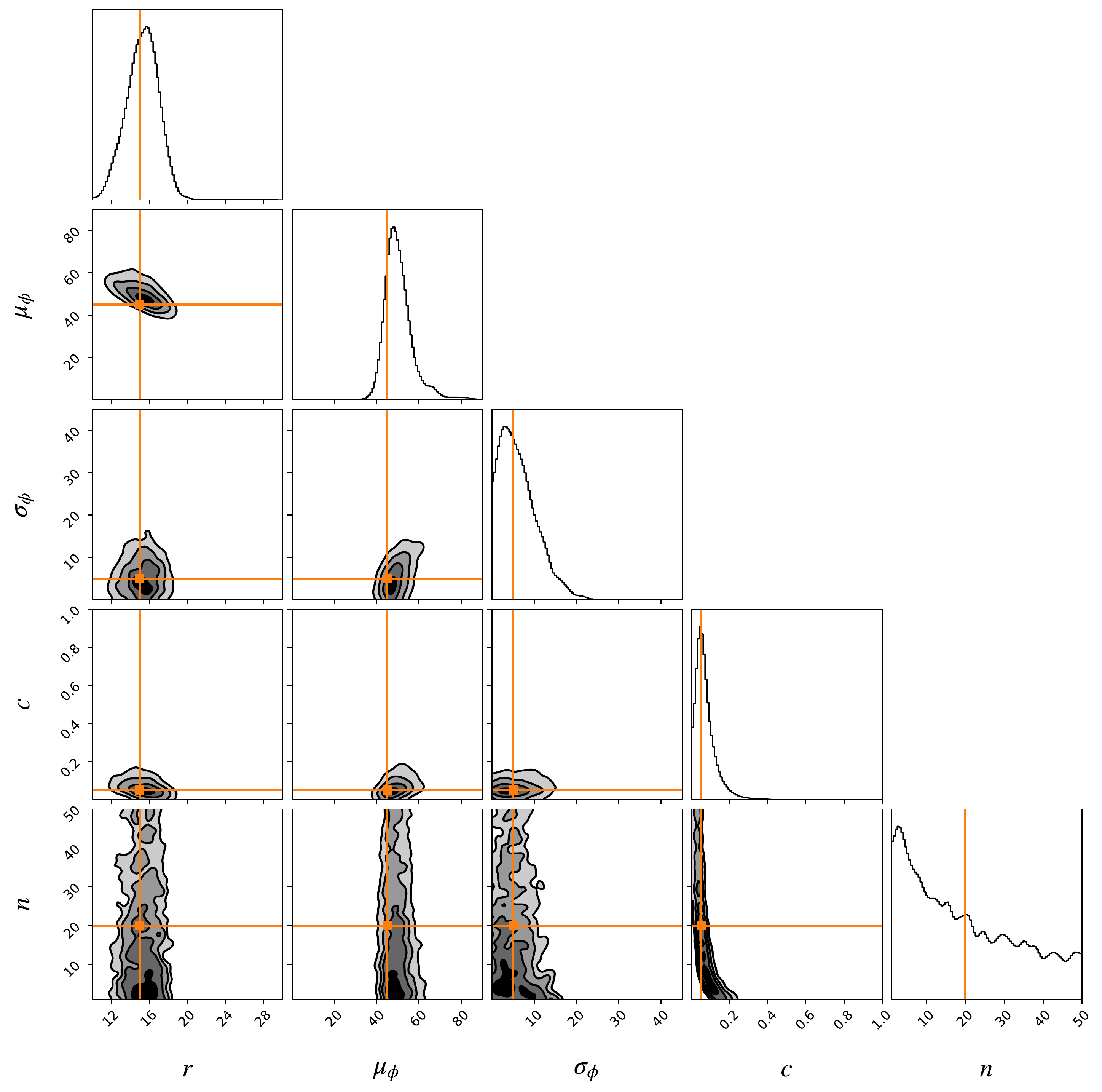}
        \\[1em]
        \includegraphics[width=\linewidth]{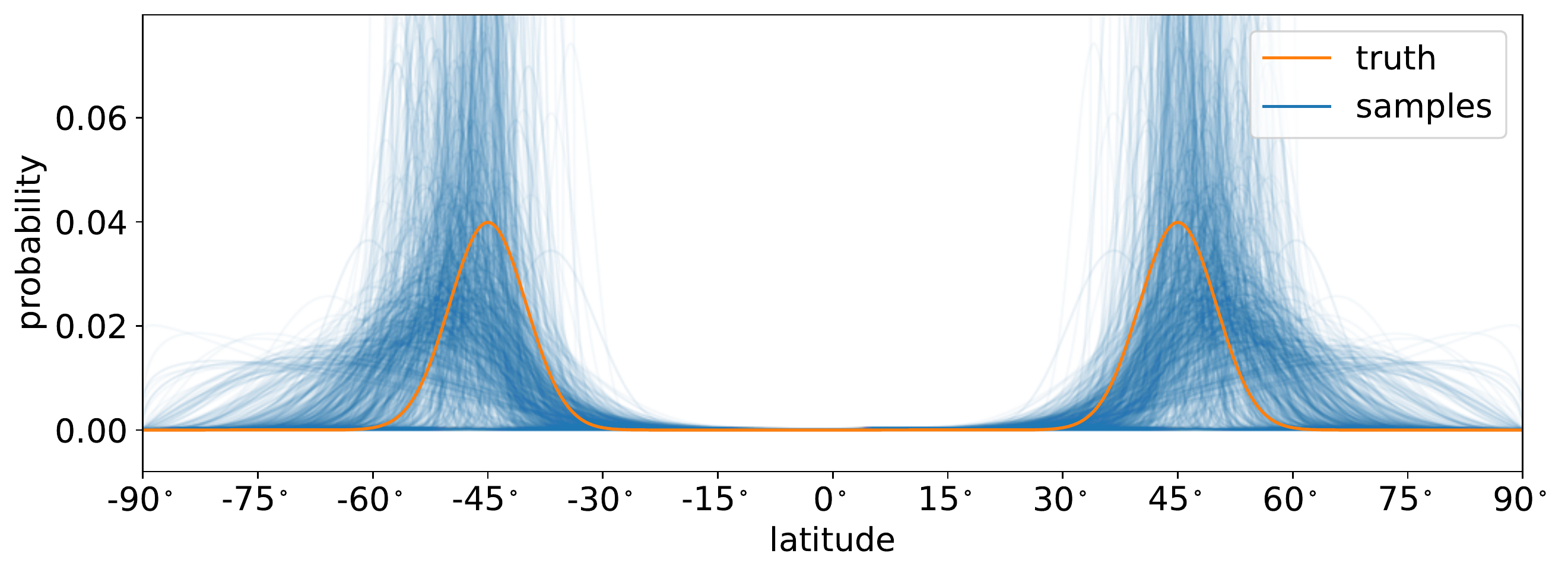}
        \oscaption{calibration_midlat}{%
            Same as Figures~\ref{fig:calibration_default_corner}
            and \ref{fig:calibration_default_latitude}, but for
            mid-latitude spots with $\mu_\phi = 45^\circ$.
            The radius and latitude parameters are again inferred correctly.
            \label{fig:calibration_midlat}
        }
    \end{centering}
\end{figure}

\begin{figure}[p!]
    \begin{centering}
        \includegraphics[width=\linewidth]{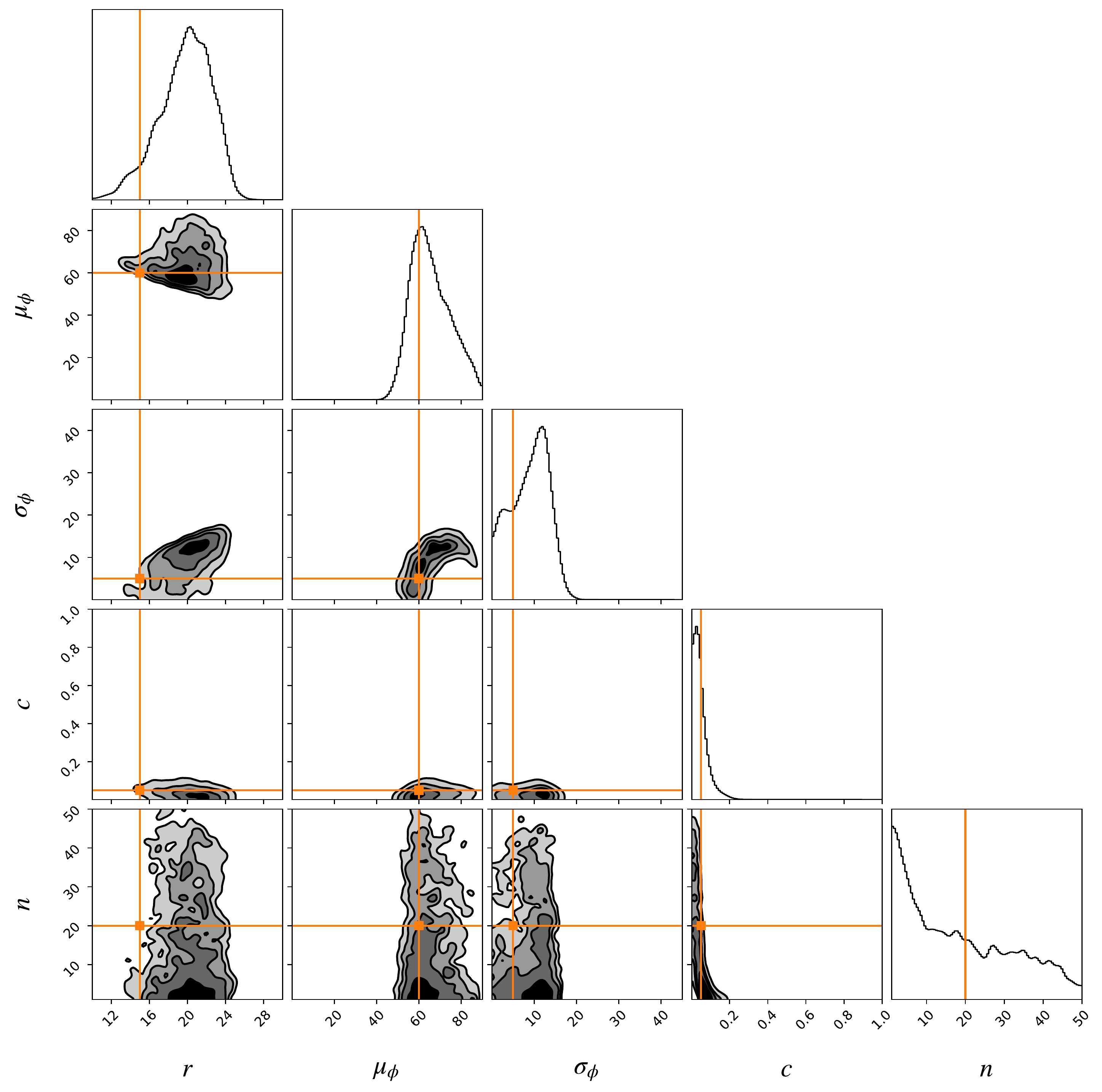}
        \\[1em]
        \includegraphics[width=\linewidth]{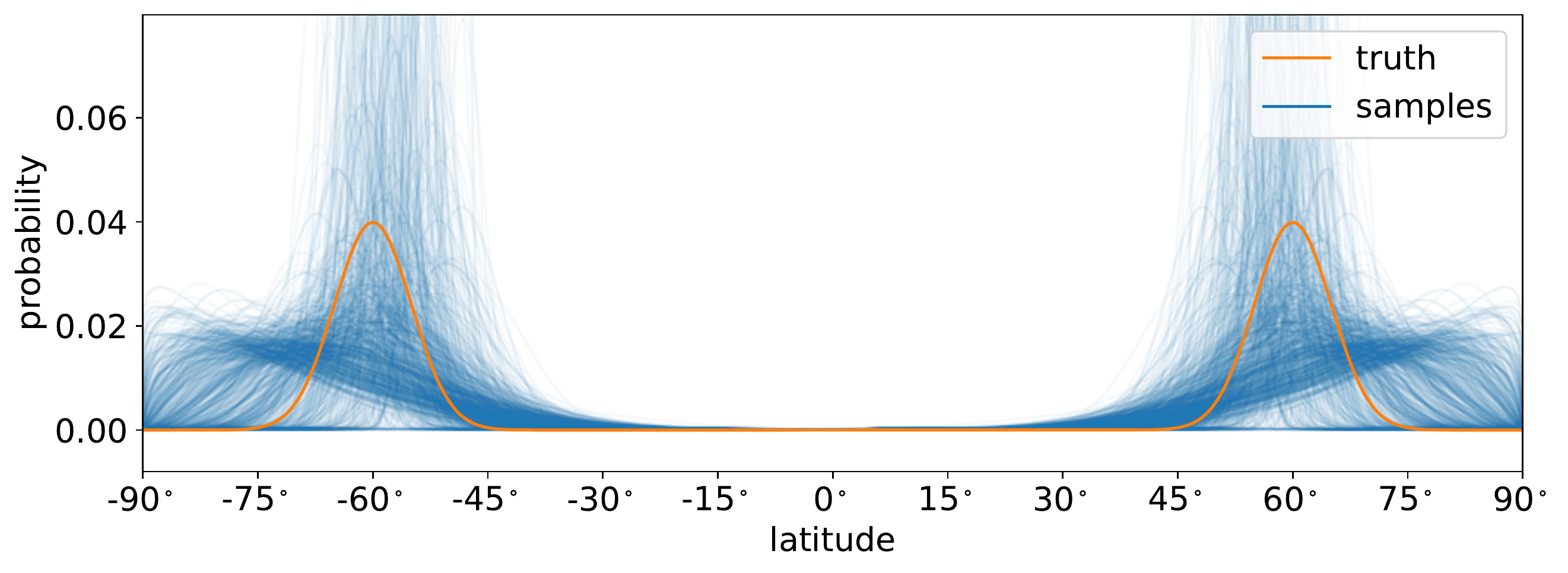}
        \oscaption{calibration_hilat}{%
            Same as Figure~\ref{fig:calibration_midlat}, but for
            high-latitude spots with $\mu_\phi = 60^\circ$.
            The radius and latitude parameters are again inferred correctly,
            although the model cannot rule out the presence of polar spots.
            \label{fig:calibration_hilat}
        }
    \end{centering}
\end{figure}

\begin{figure}[p!]
    \begin{centering}
        \includegraphics[width=\linewidth]{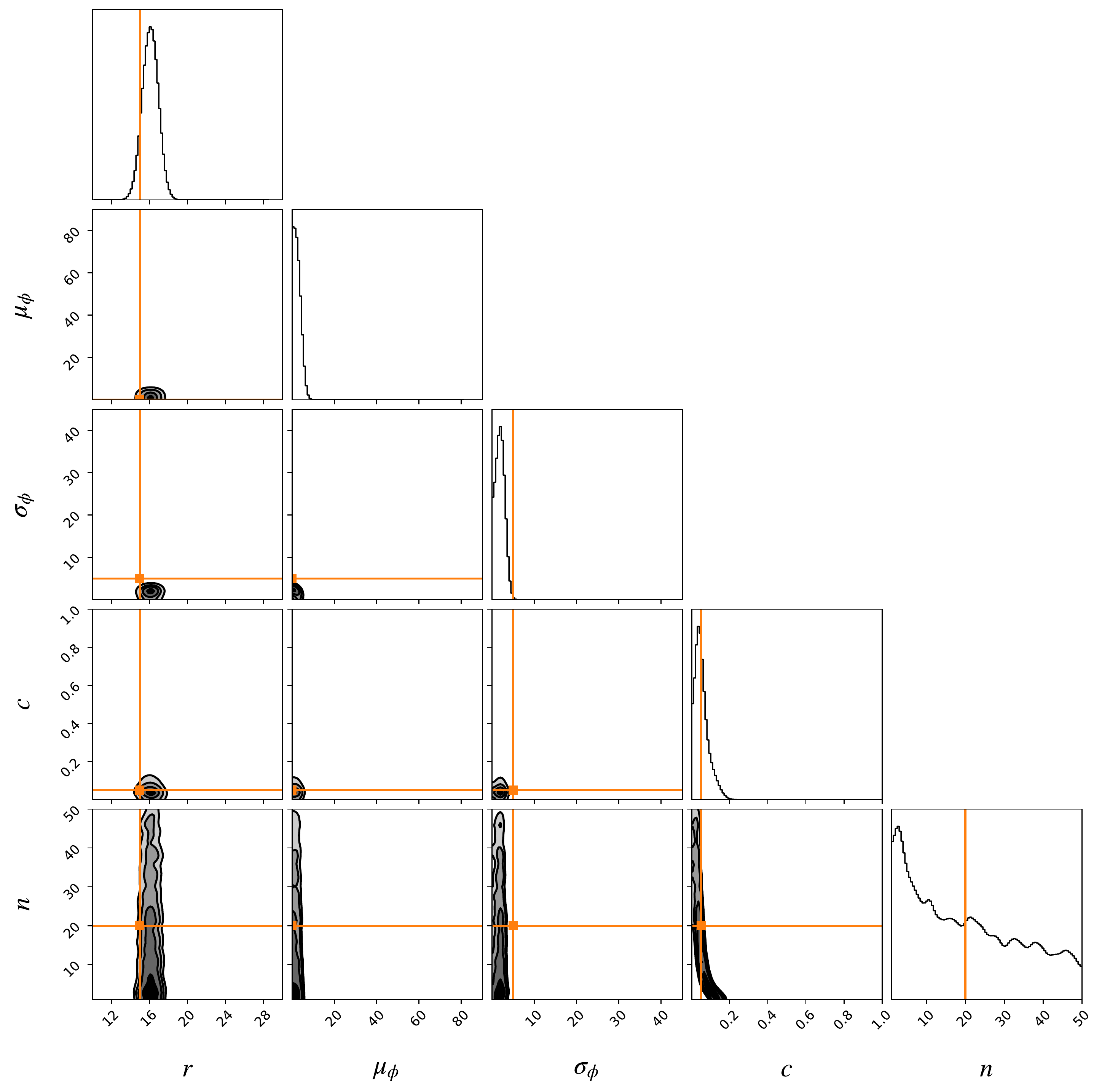}
        \\[1em]
        \includegraphics[width=\linewidth]{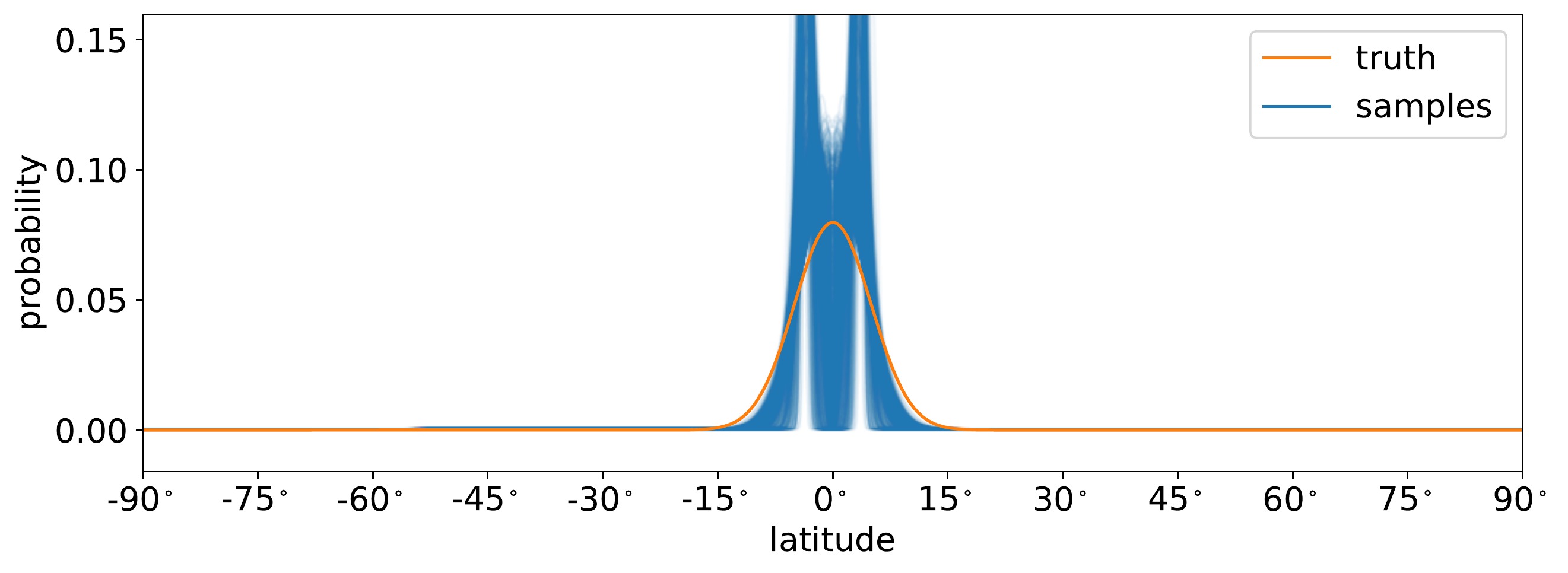}
        \oscaption{calibration_equatorial}{%
            Same as Figure~\ref{fig:calibration_midlat}, but for
            equatorial spots with $\mu_\phi = 0^\circ$. Even though the model
            favors a bimodal distribution at low latitudes, the
            posterior strongly supports the presence of equatorial spots.
            \label{fig:calibration_equatorial}
        }
    \end{centering}
\end{figure}

\begin{figure}[p!]
    \begin{centering}
        \includegraphics[width=\linewidth]{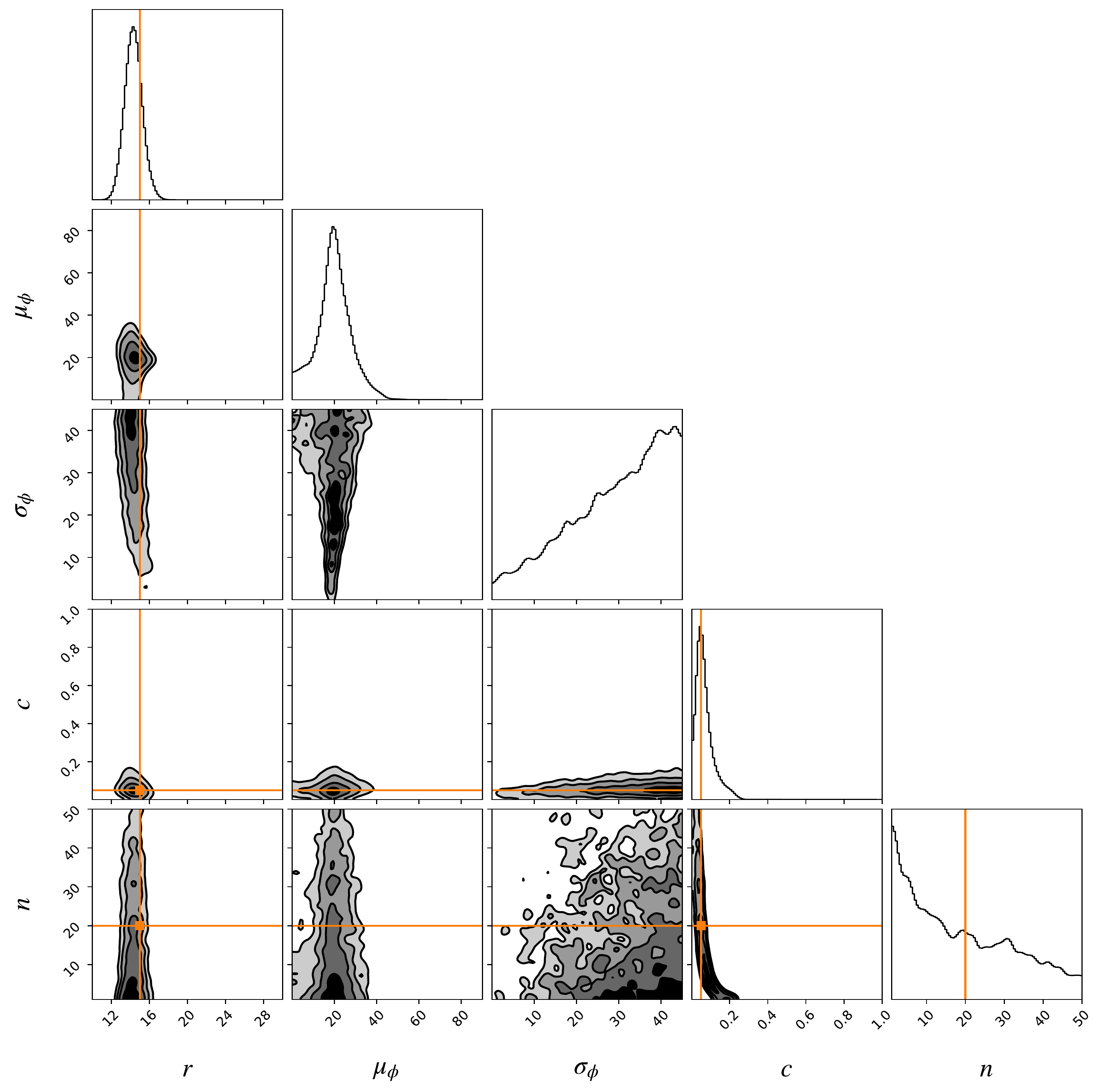}
        \\[1em]
        \includegraphics[width=\linewidth]{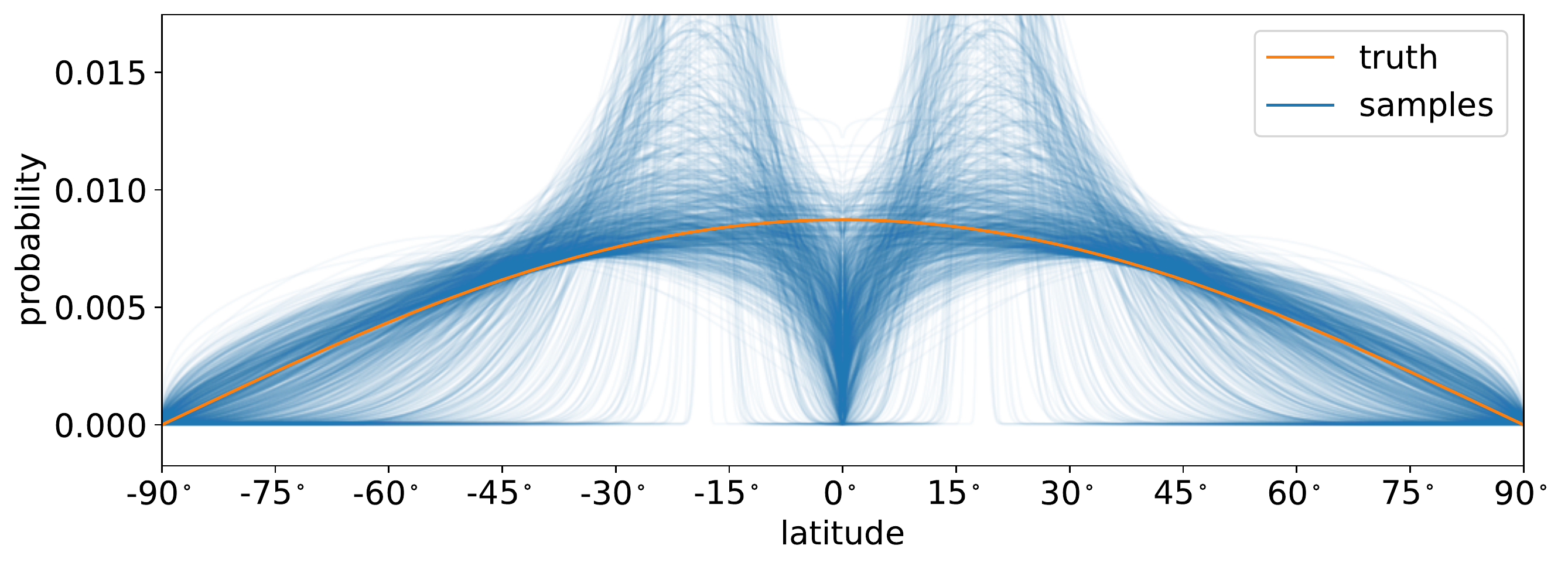}
        \oscaption{calibration_isotropic}{%
            Same as Figure~\ref{fig:calibration_midlat}, but for
            isotropically-distributed spots. The posterior accurately
            captures the cosine-like distribution of spot latitudes.
            \label{fig:calibration_isotropic}
        }
    \end{centering}
\end{figure}

\begin{figure}[p!]
    \begin{centering}
        \includegraphics[width=\linewidth]{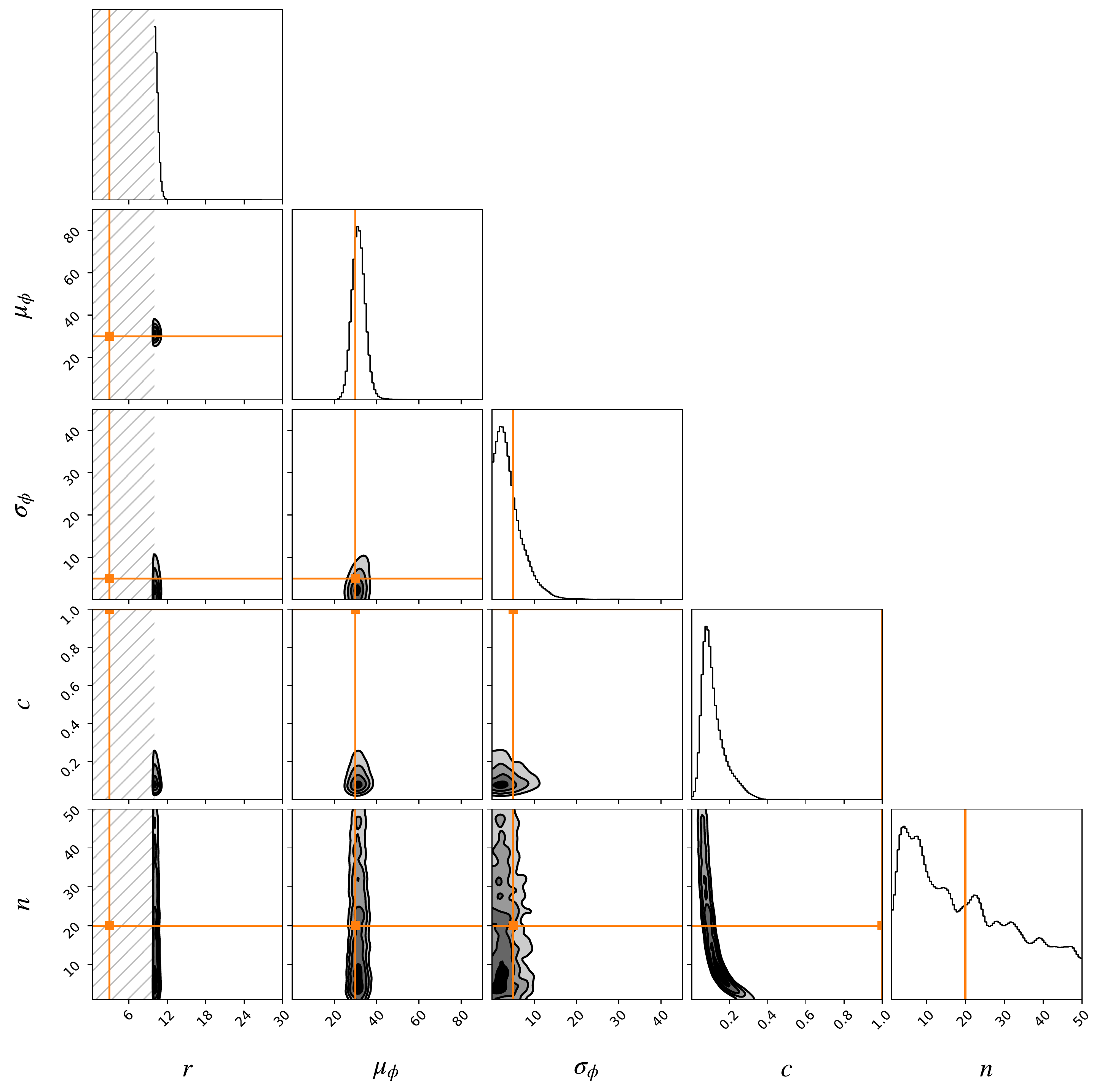}
        \\[1em]
        \includegraphics[width=\linewidth]{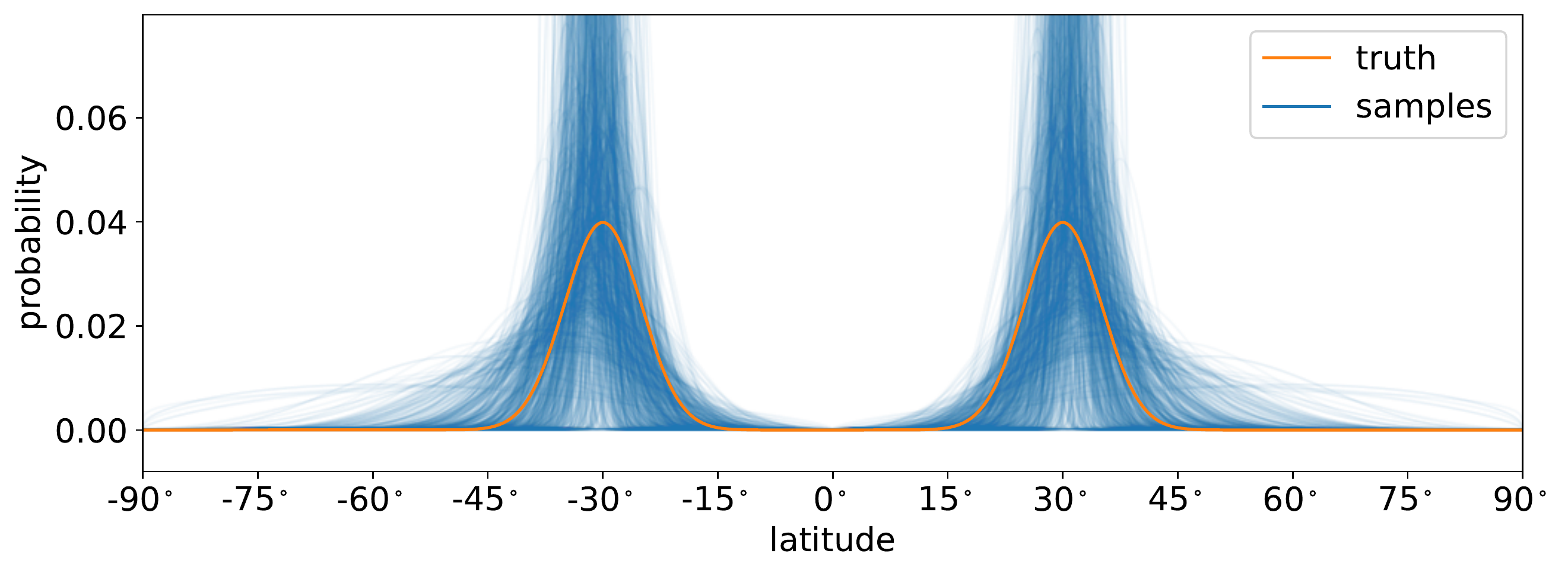}
        \oscaption{calibration_tinyspots}{%
            Same as Figures~\ref{fig:calibration_default_corner}
            and \ref{fig:calibration_default_latitude}, but for high contrast ($c = 1$)
            small ($r = 3^\circ$) spots (significantly lower than the
            effective resolution of the model).
            The hatched regions in the posterior
            plots for the radius ($r < 10^\circ$) are excluded by the prior, since
            the model cannot capture features that small.
            Despite this, the spot latitude distribution is still inferred correctly,
            although the spot contrast is off by more than $10\sigma$.
            \label{fig:calibration_tinyspots}
        }
    \end{centering}
\end{figure}

\begin{figure}[p!]
    \begin{centering}
        \includegraphics[width=\linewidth]{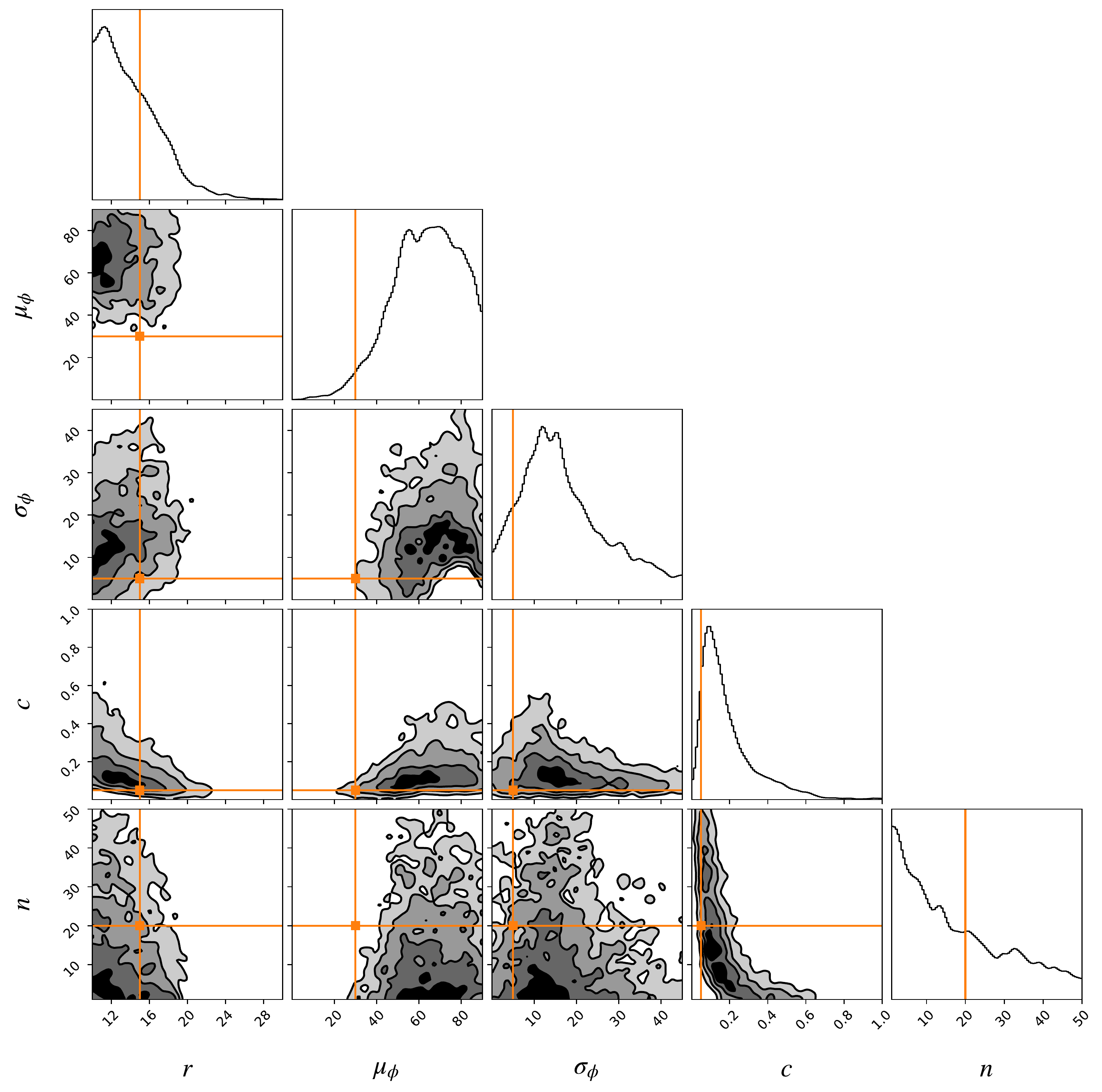}
        \\[1em]
        \includegraphics[width=\linewidth]{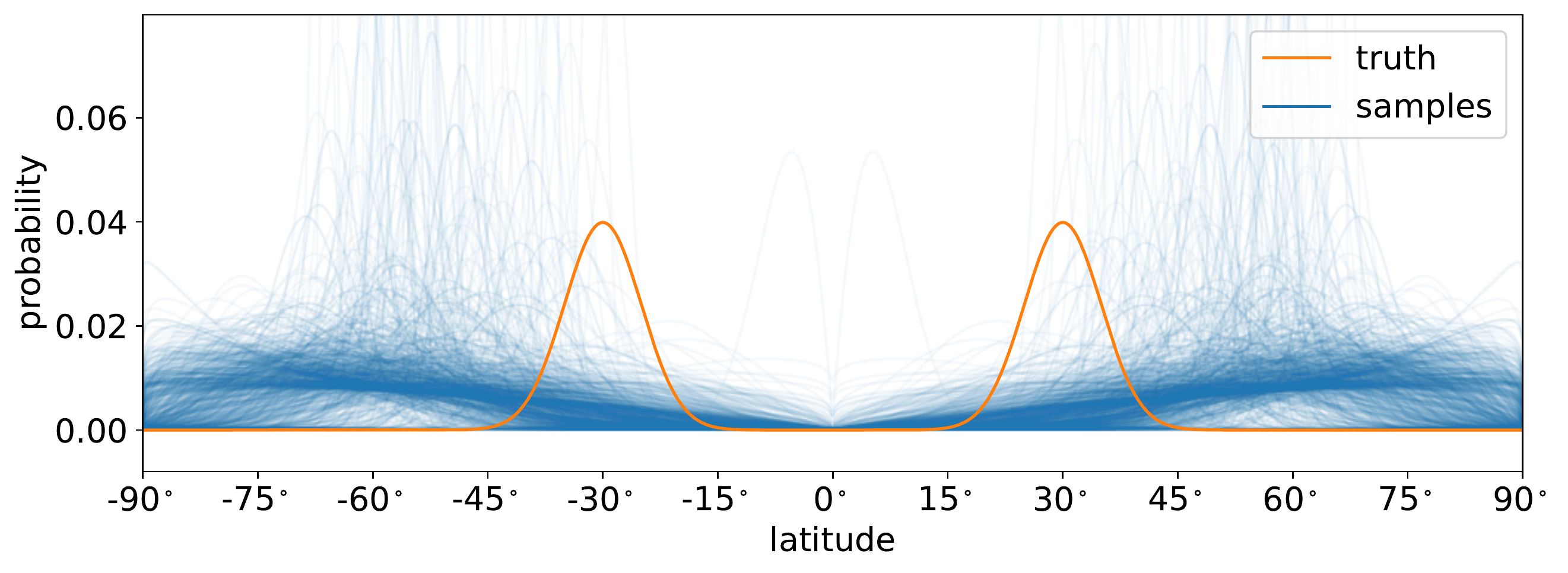}
        \oscaption{calibration_default_1}{%
            Same as Figures~\ref{fig:calibration_default_corner}
            and \ref{fig:calibration_default_latitude}, but for
            inference based on a single light curve ($M = 1$).
            The constraints on all of the parameters are dramatically weaker.
            \label{fig:calibration_default_1}
        }
    \end{centering}
\end{figure}

\begin{figure}[p!]
    \begin{centering}
        \includegraphics[width=\linewidth]{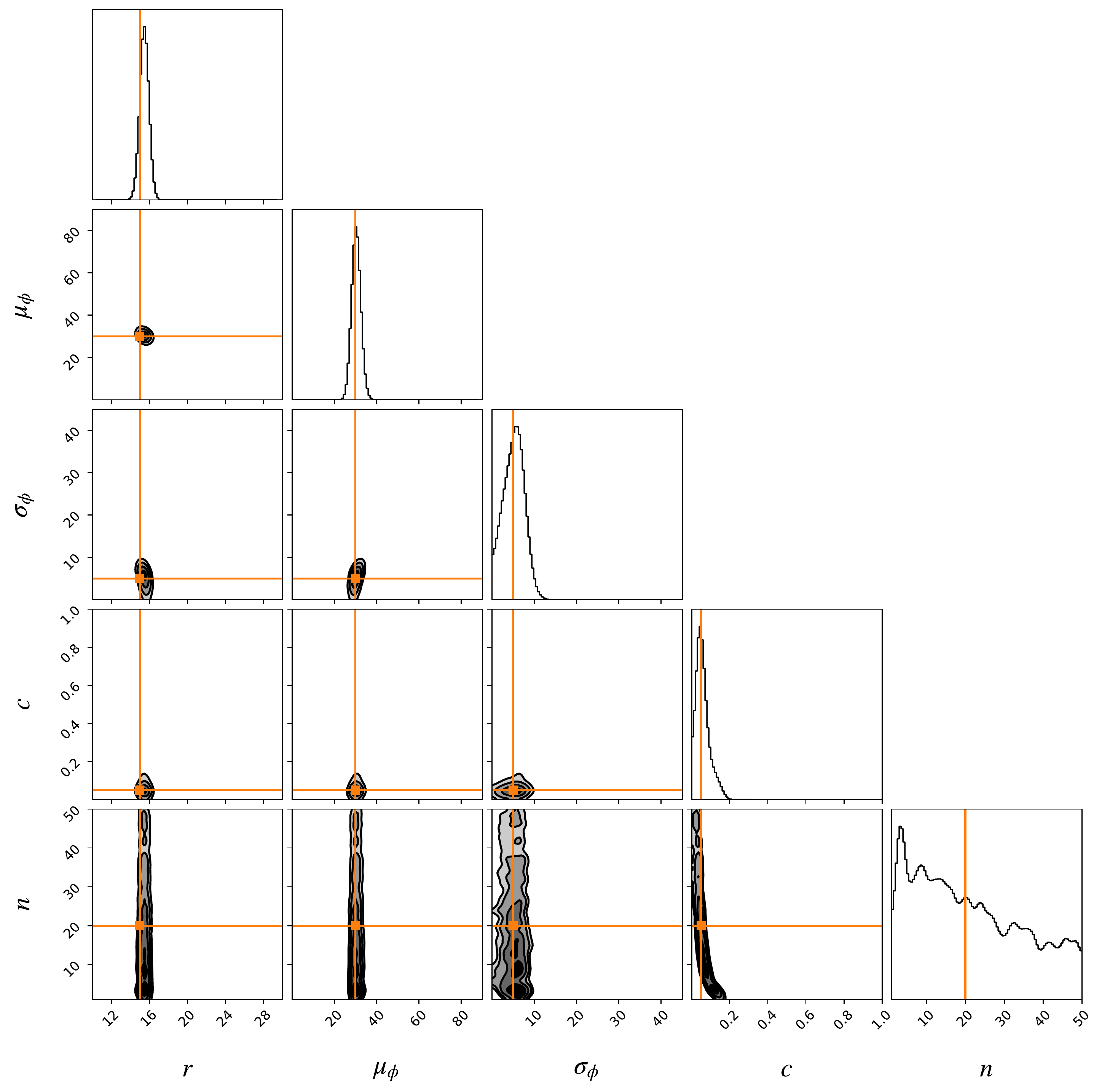}
        \\[1em]
        \includegraphics[width=\linewidth]{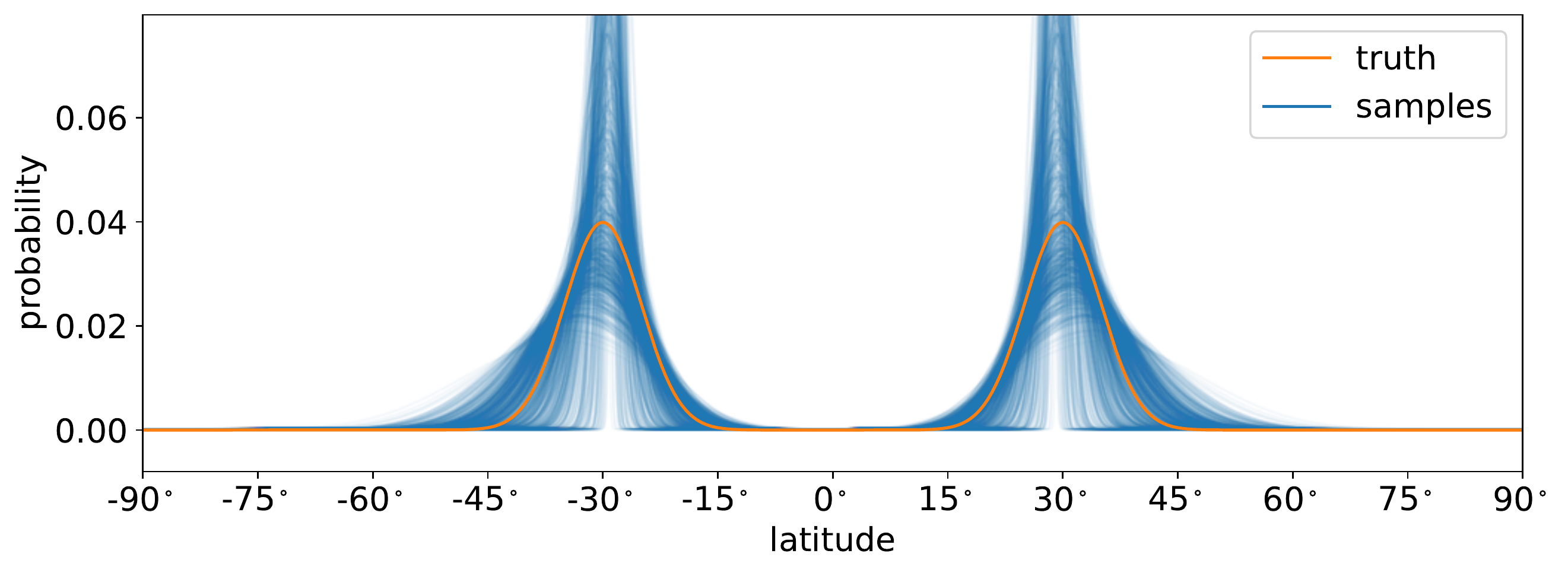}
        \oscaption{calibration_default_1000}{%
            Same as Figures~\ref{fig:calibration_default_corner}
            and \ref{fig:calibration_default_latitude}, but for
            inference based on one thousand light curves ($M = 1{,}000$).
            The constraints on the radius and the latitude parameters are
            dramatically tighter.
            \label{fig:calibration_default_1000}
        }
    \end{centering}
\end{figure}

\begin{figure}[p!]
    \begin{centering}
        \includegraphics[width=\linewidth]{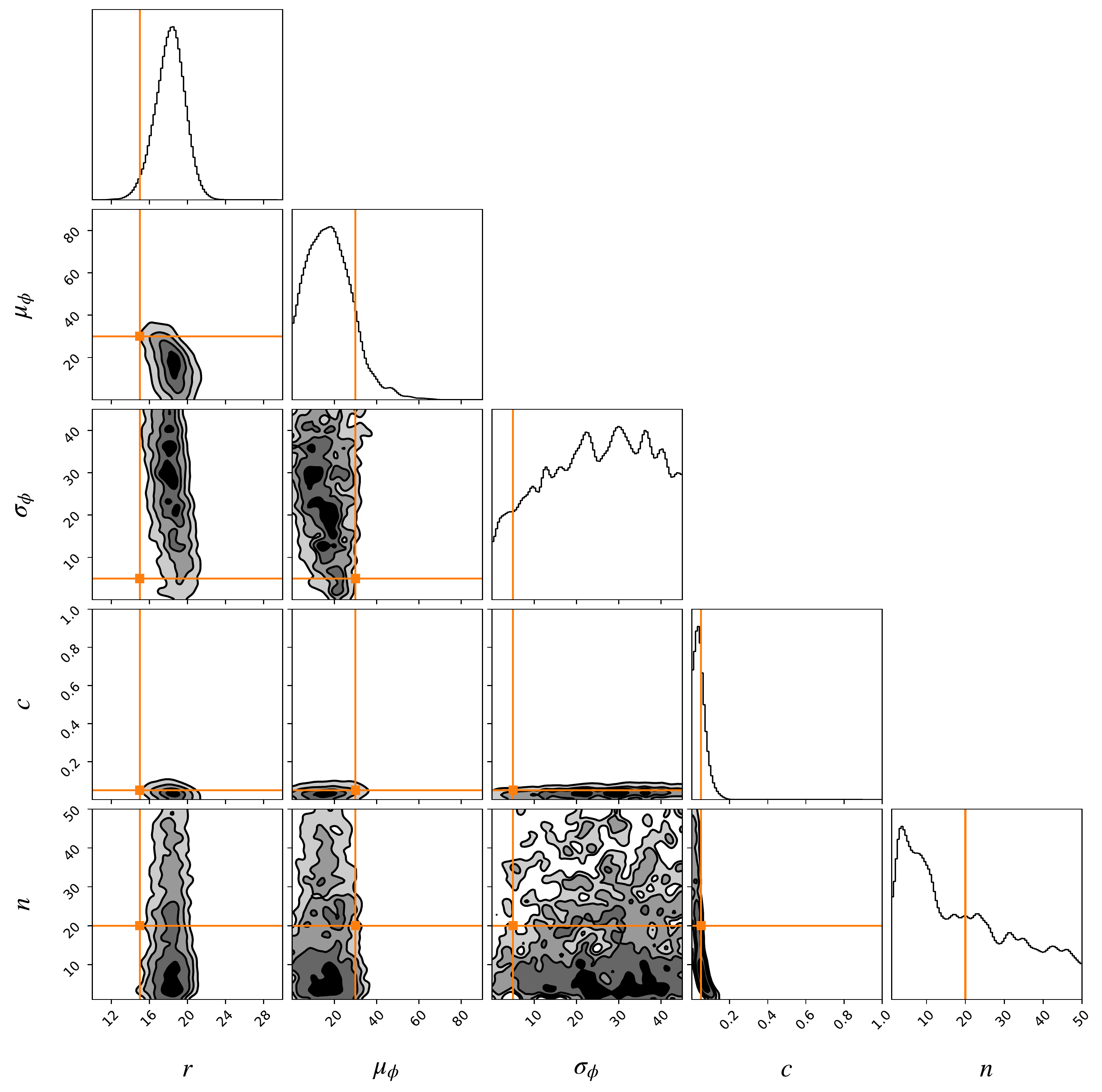}
        \\[1em]
        \includegraphics[width=\linewidth]{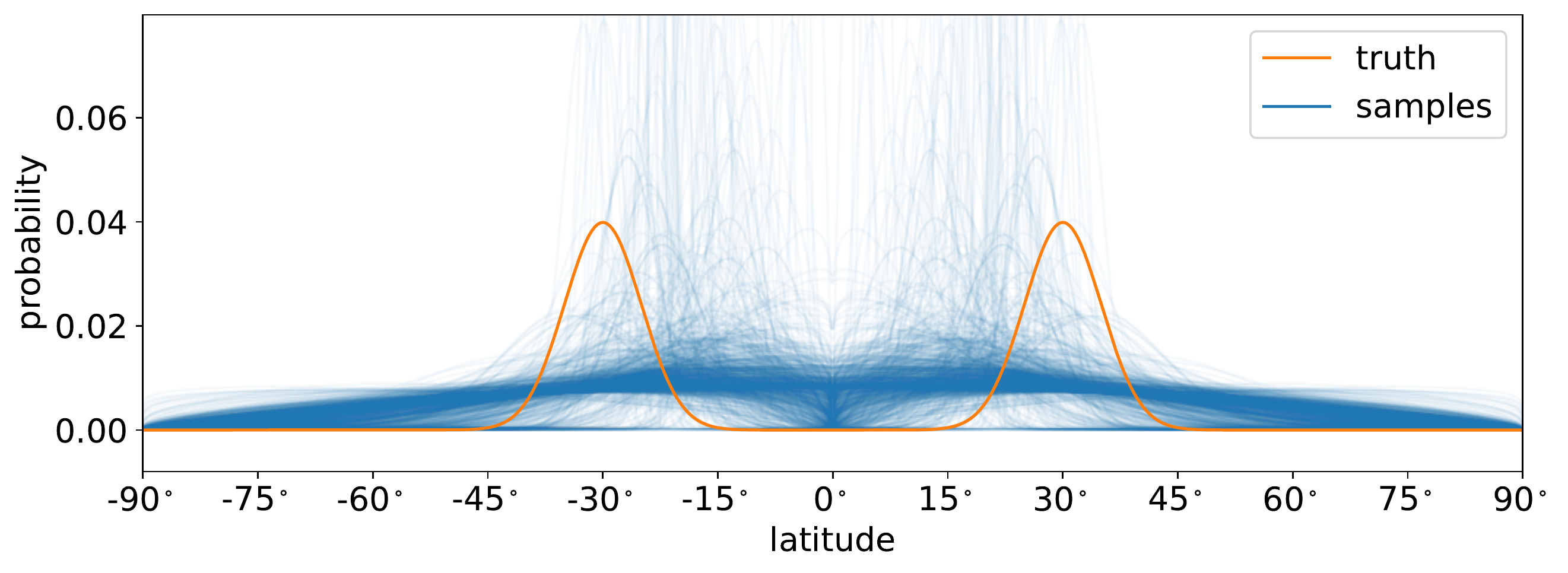}
        \oscaption{calibration_ld}{%
            Same as Figures~\ref{fig:calibration_default_corner}
            and \ref{fig:calibration_default_latitude}, but for
            stars with (assumed known) limb darkening coefficients
            $u_1 = 0.50$ and $u_2 = 0.25$. Limb darkening makes it
            much harder to infer the variance of the distribution of
            starspot latitudes.
            \label{fig:calibration_ld}
        }
    \end{centering}
\end{figure}

\begin{figure}[p!]
    \begin{centering}
        \includegraphics[width=\linewidth]{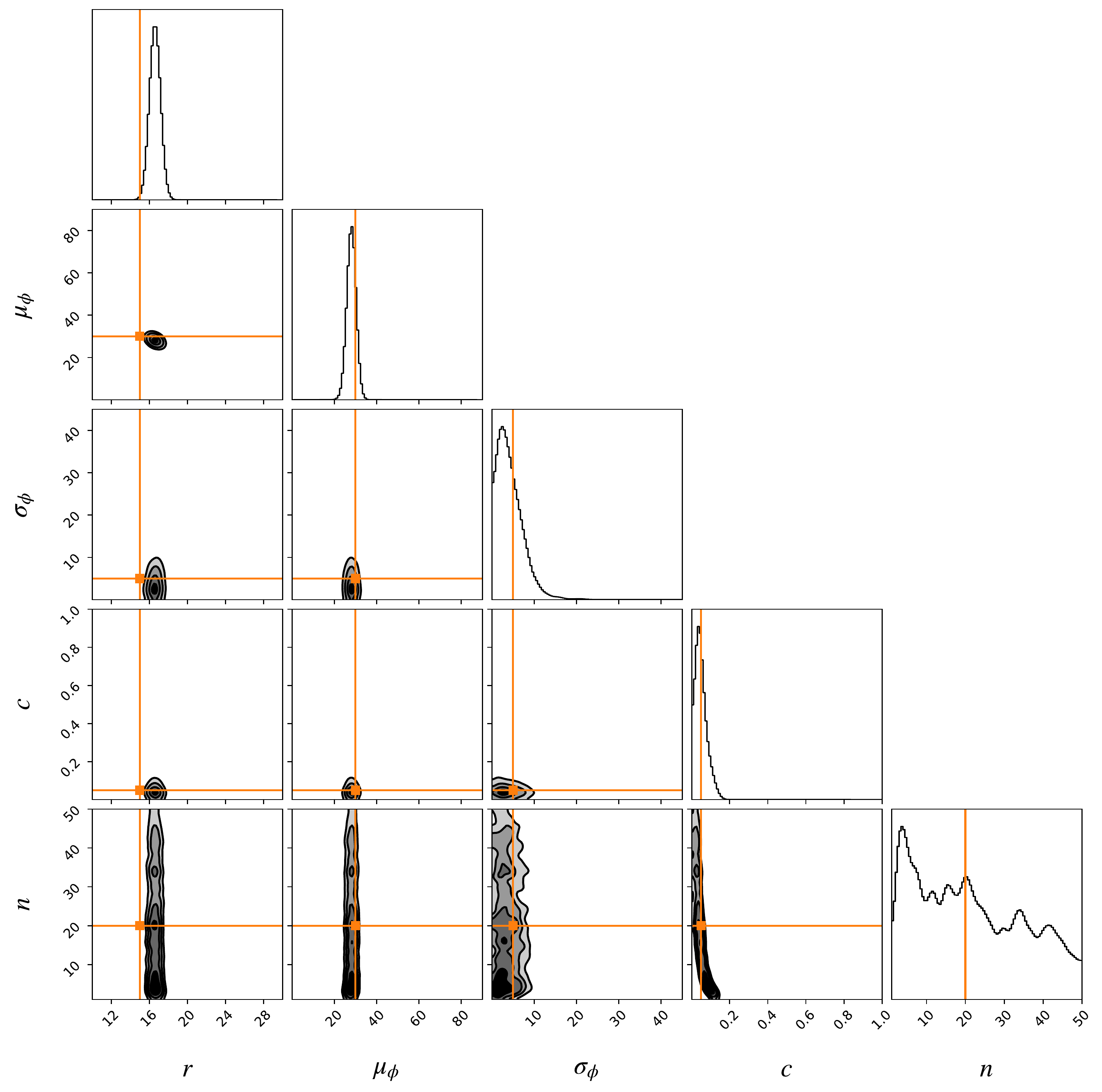}
        \\[1em]
        \includegraphics[width=\linewidth]{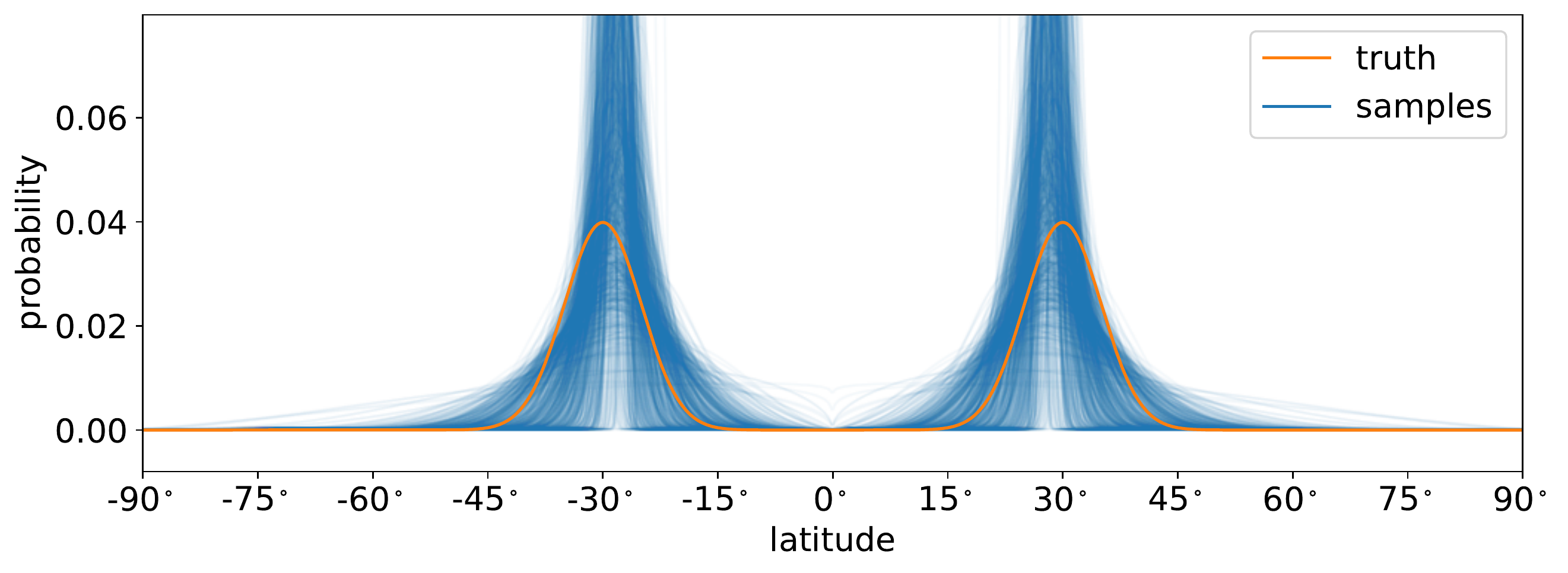}
        \oscaption{calibration_ld_1000}{%
            Same as Figure~\ref{fig:calibration_ld}, but for an ensemble
            consisting of $M = 1{,}000$ light curves. For a
            sufficiently large ensemble, it is possible to correctly infer the spot
            radii and latitudes in the presence of limb darkening.
            \label{fig:calibration_ld_1000}
        }
    \end{centering}
\end{figure}

\begin{figure}[p!]
    \begin{centering}
        \includegraphics[width=\linewidth]{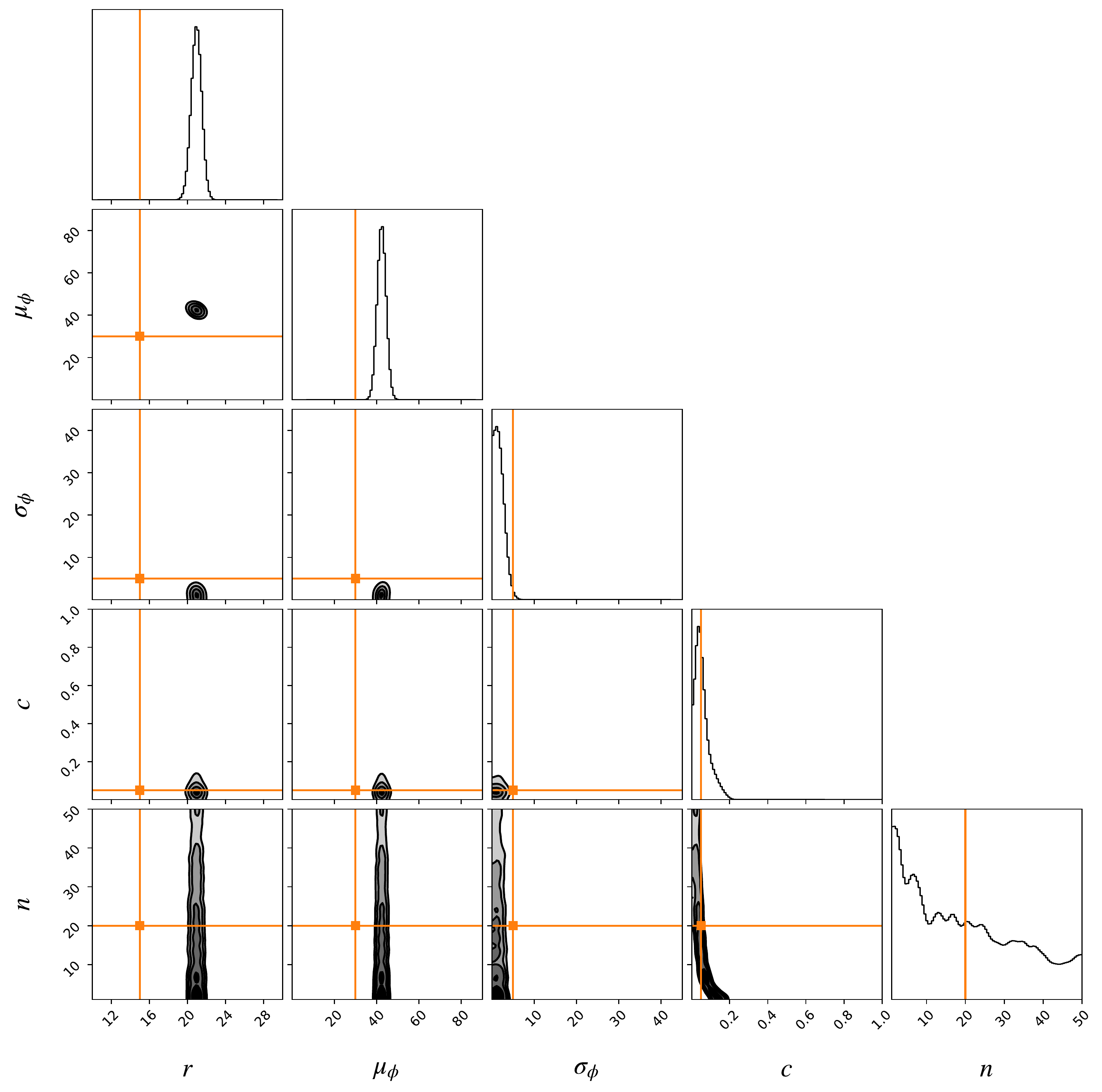}
        \\[1em]
        \includegraphics[width=\linewidth]{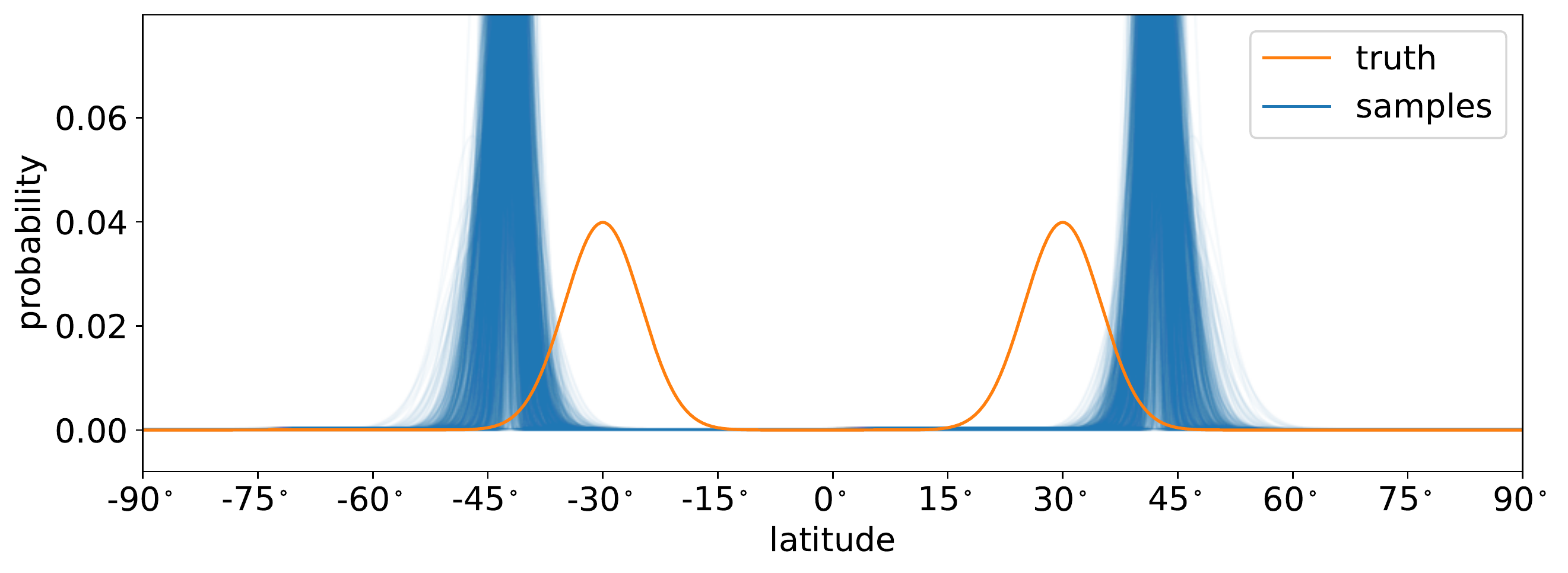}
        \oscaption{calibration_ld_500_no_ld}{%
            Same as Figure~\ref{fig:calibration_ld}, but for $M = 500$
            light curves and assuming no limb darkening when doing inference.
            Neglecting limb darkening leads to significant bias in the inferred
            spot radii and to a lesser extent in the mean spot latitude.
            \label{fig:calibration_ld_500_no_ld}
        }
    \end{centering}
\end{figure}

\begin{figure}[p!]
    \begin{centering}
        \includegraphics[width=\linewidth]{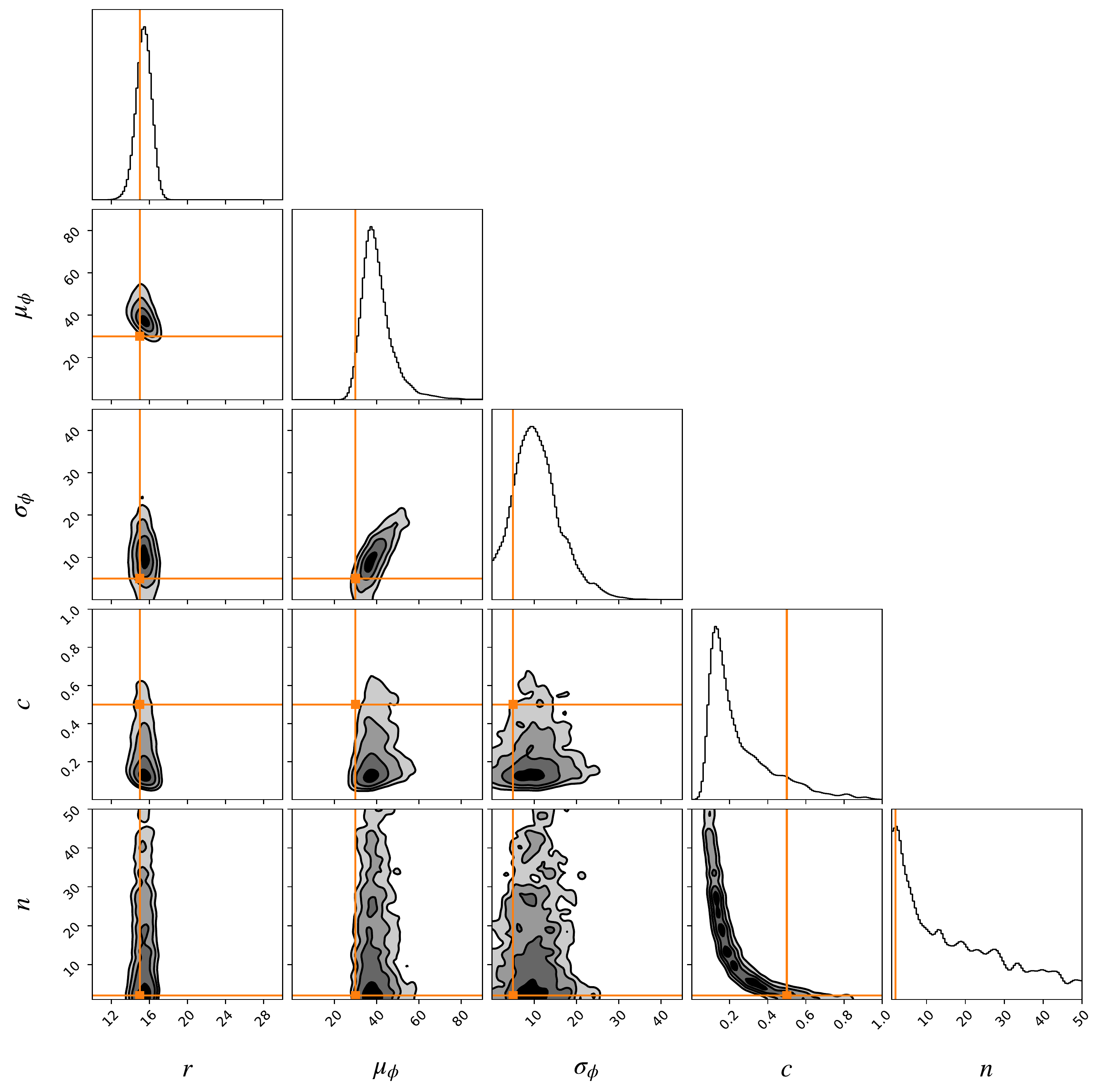}
        \\[1em]
        \includegraphics[width=\linewidth]{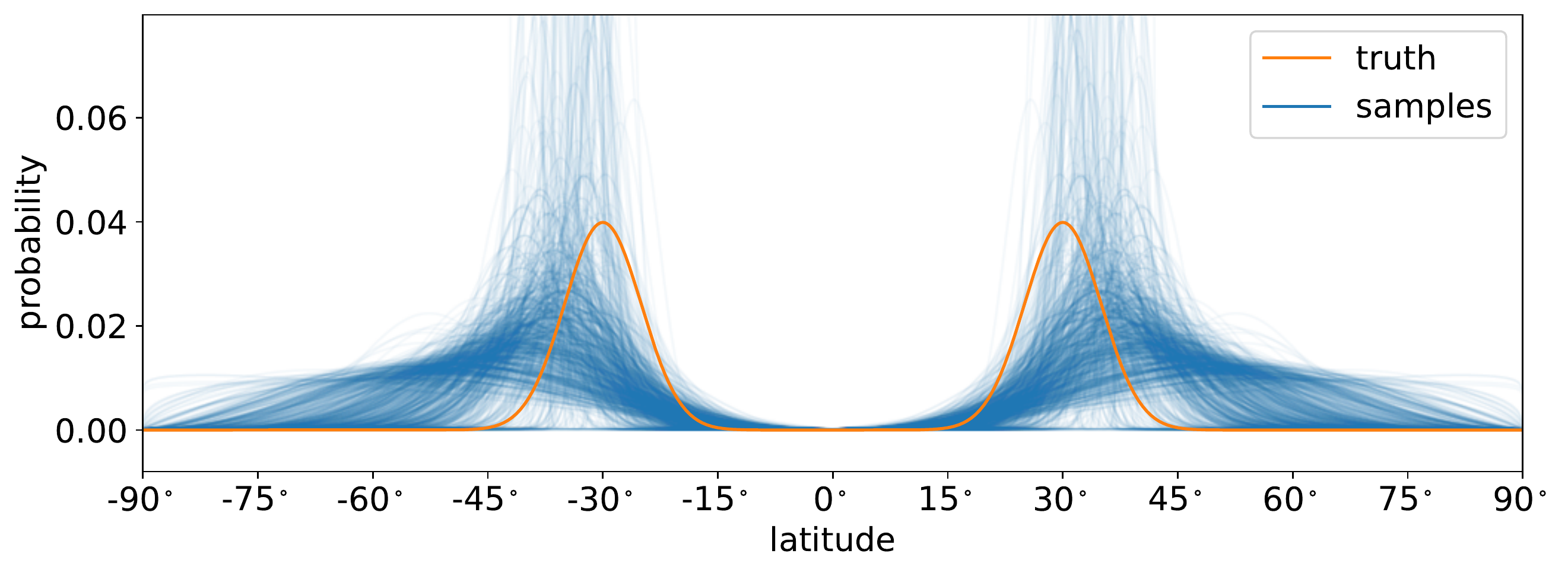}
        \oscaption{calibration_hicontrast}{%
            Same as Figures~\ref{fig:calibration_default_corner}
            and \ref{fig:calibration_default_latitude}, but for stars with
            $n = 2$ spots with high contrast $c = 0.5$. Our model correctly
            captures the increased contrast, but it is still strongly degenerate
            with the number of spots.
            \label{fig:calibration_hicontrast}
        }
    \end{centering}
\end{figure}

\begin{figure}[p!]
    \begin{centering}
        \includegraphics[width=\linewidth]{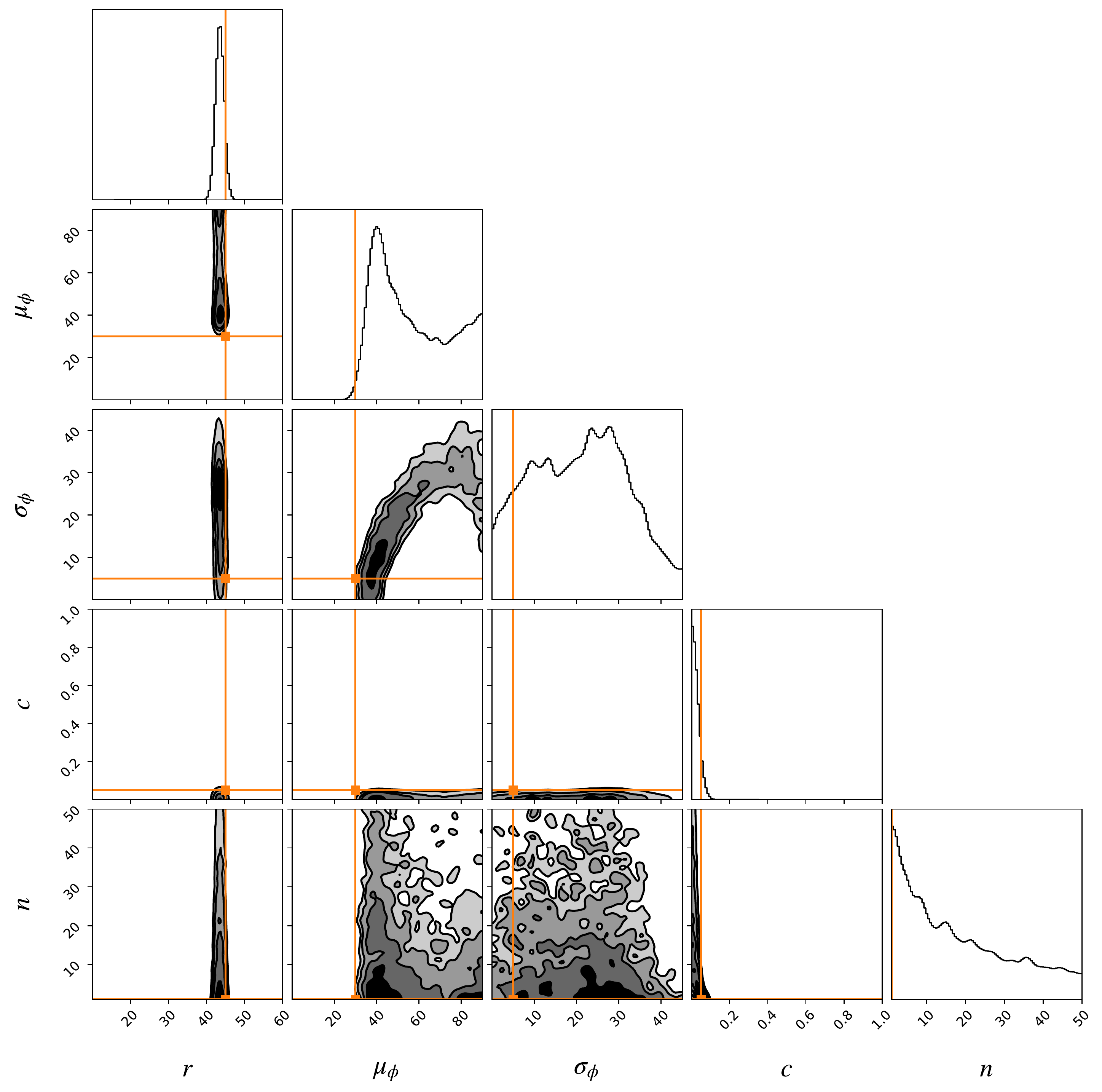}
        \\[1em]
        \includegraphics[width=\linewidth]{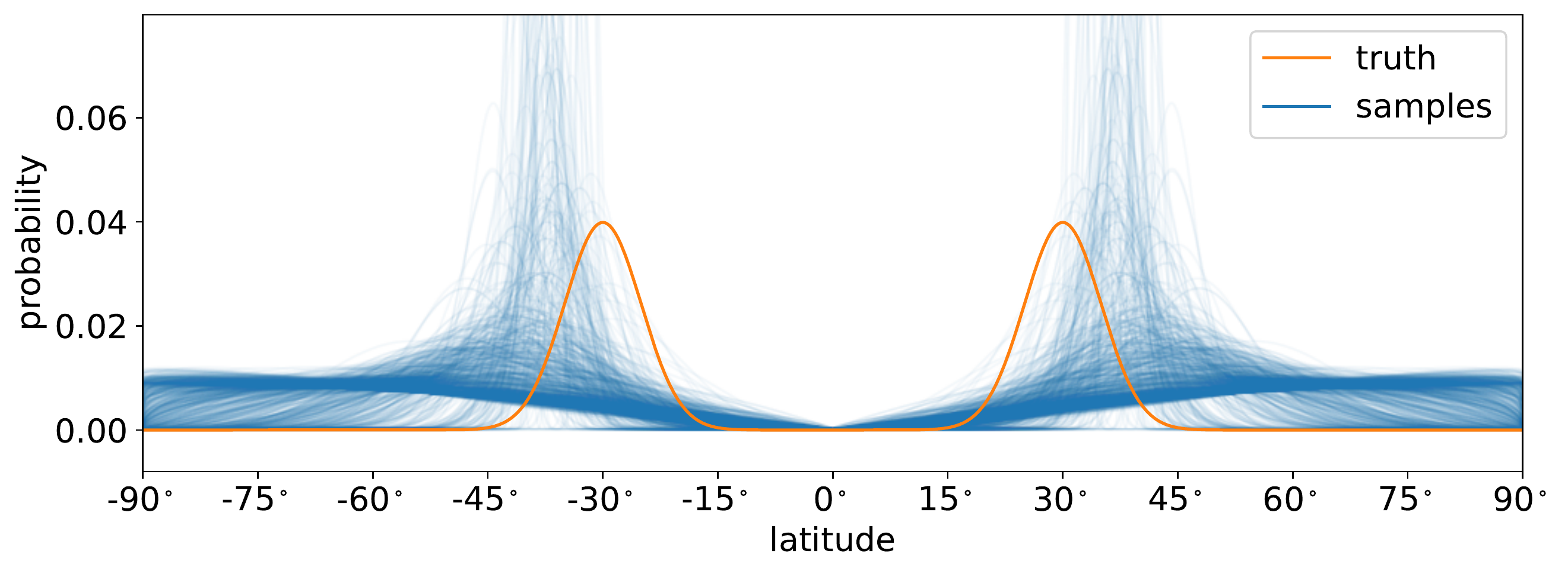}
        \oscaption{calibration_bigspots}{%
            Same as Figures~\ref{fig:calibration_default_corner}
            and \ref{fig:calibration_default_latitude}, but for a single ($n = 1$)
            large ($r \sim \mathcal{N}(45^\circ, {5^\circ}^2)$) spot on each star.
            Our model correctly infers the larger radius, and infers the
            latitude within $3\sigma$, albeit with large uncertainty.
            \label{fig:calibration_bigspots}
        }
    \end{centering}
\end{figure}

\begin{figure}[p!]
    \begin{centering}
        \includegraphics[width=\linewidth]{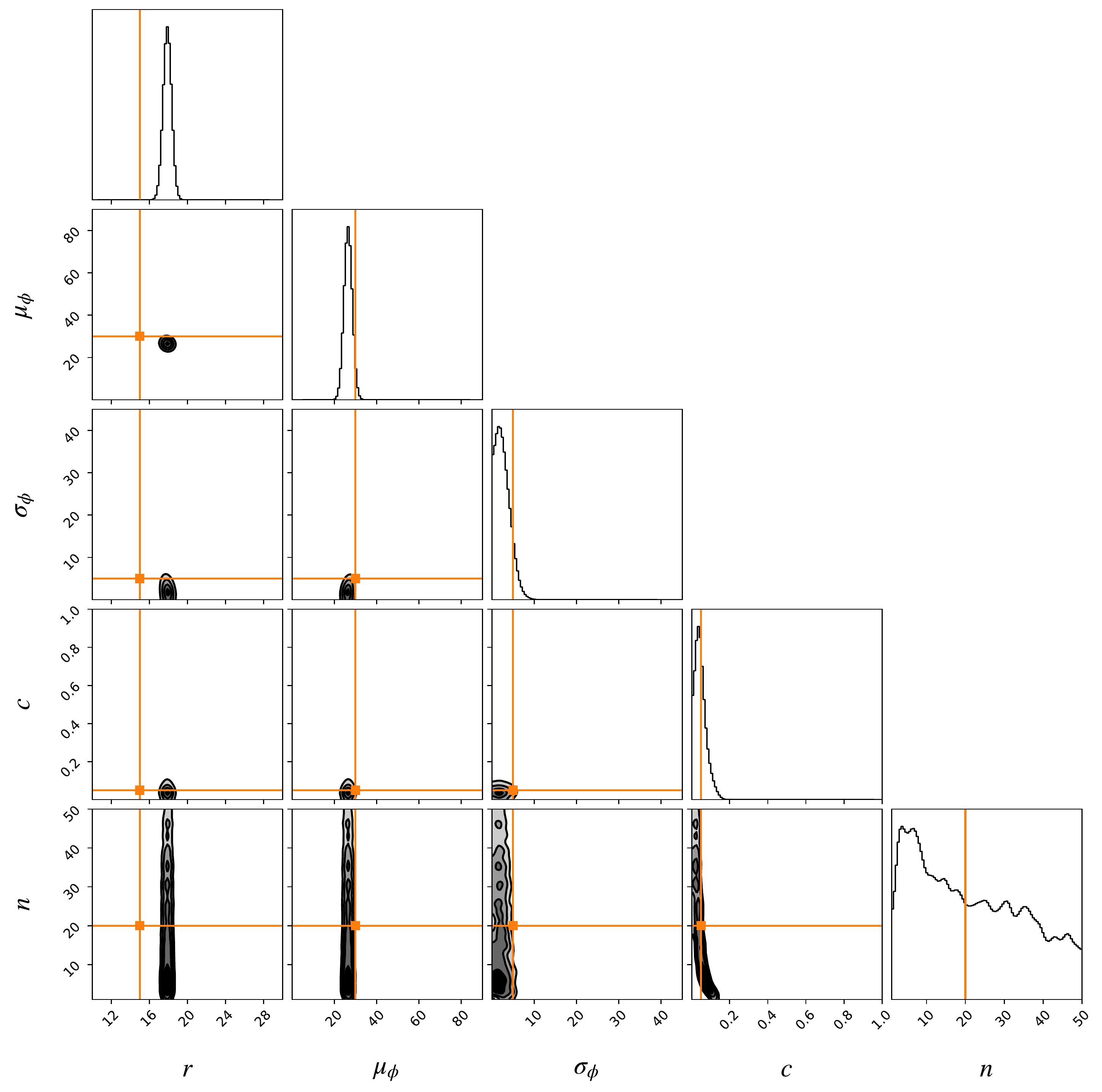}
        \\[1em]
        \includegraphics[width=\linewidth]{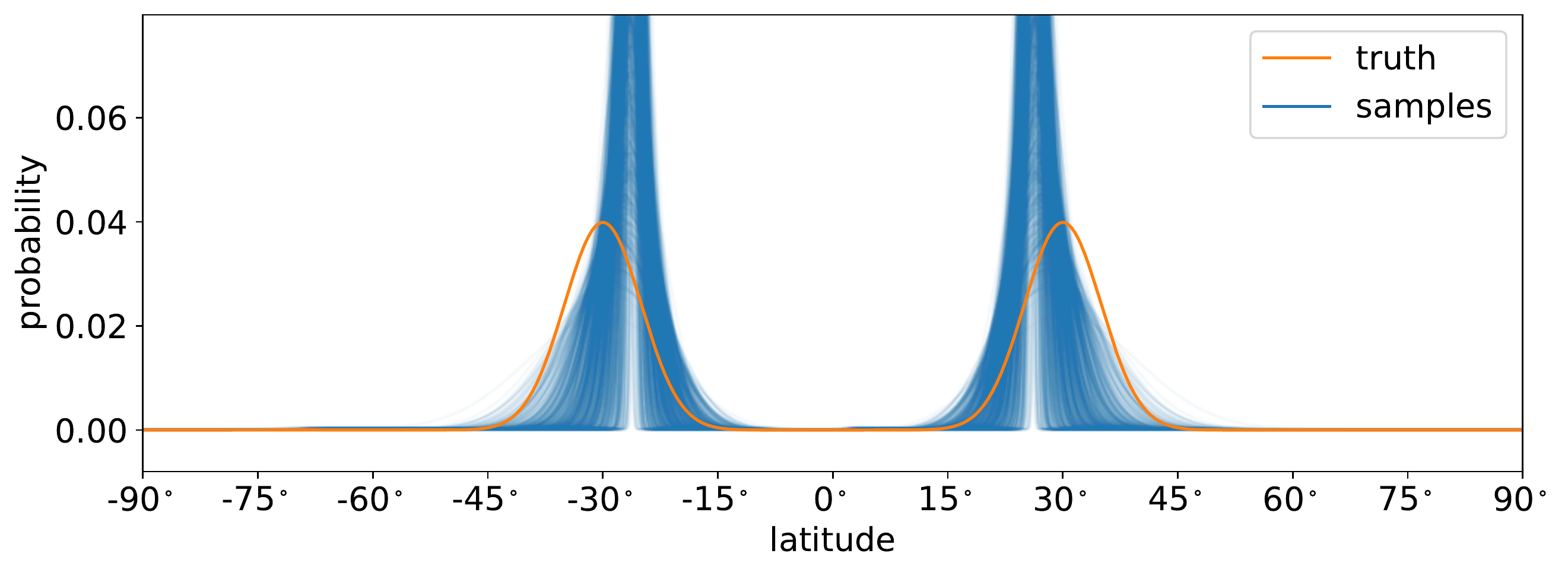}
        \oscaption{calibration_variance}{%
            Same as Figures~\ref{fig:calibration_default_corner}
            and \ref{fig:calibration_default_latitude}, but for stars
            with variance in their number of spots,
            $n \sim \mathcal{N}(20, 3^2)$,
            the spot radii, $r \sim \mathcal{N}(15^\circ, {3^\circ}^2)$,
            and the spot contrasts,
            $c \sim \mathcal{N}(0.05, 0.01^2)$.
            \label{fig:calibration_variance}
        }
    \end{centering}
\end{figure}

\begin{figure}[p!]
    \begin{centering}
        \includegraphics[width=\linewidth]{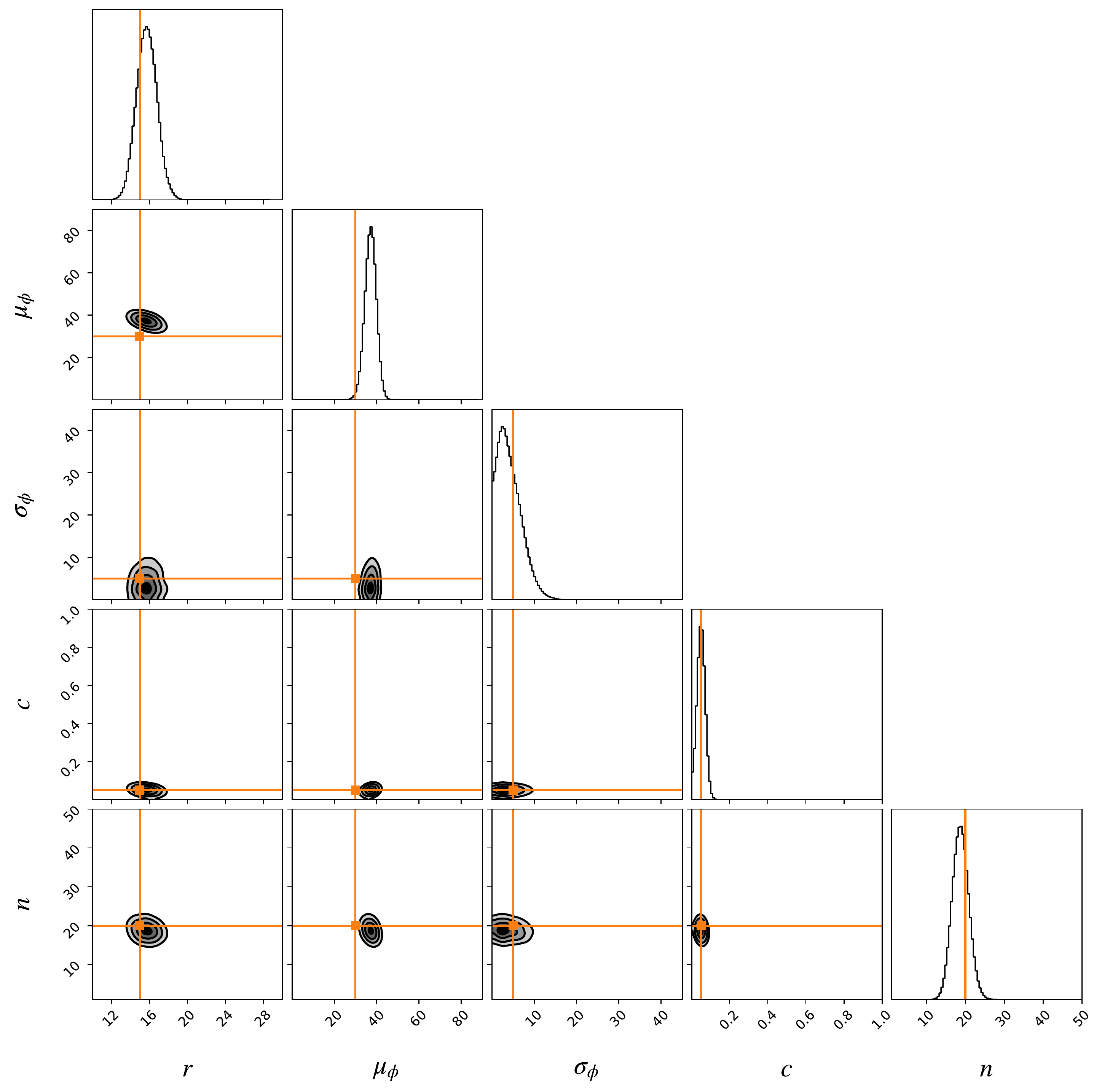}
        \\[1em]
        \includegraphics[width=\linewidth]{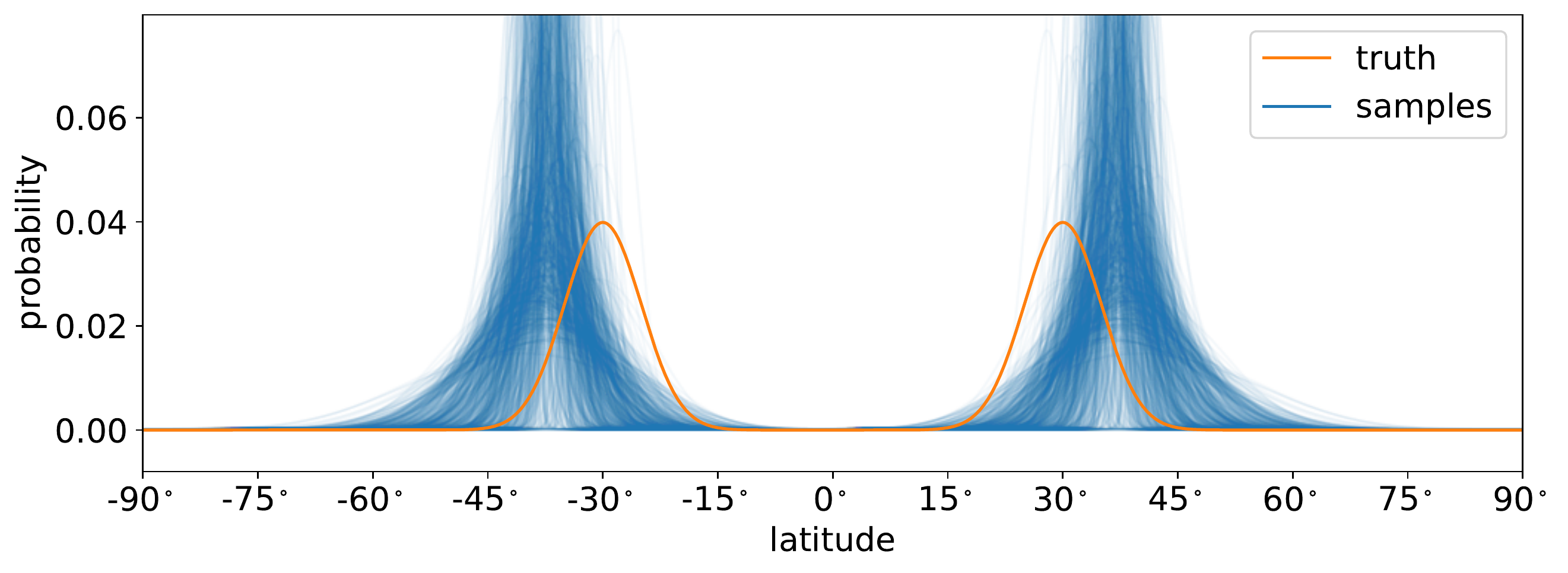}
        \oscaption{calibration_unnorm}{%
            Same as Figures~\ref{fig:calibration_default_corner}
            and \ref{fig:calibration_default_latitude}, but assuming
            the light curves are not normalized and the true amplitude is
            known. Knowledge of the normalization breaks the $c-n$ degeneracy
            and allows us to infer the total number of spots.
            \label{fig:calibration_unnorm}
        }
    \end{centering}
\end{figure}

\end{document}

%% file: cortex.tex
\usepackage{url}
\usepackage{amsmath}
\usepackage{mathtools}
\usepackage{amssymb}
\usepackage{natbib}
\usepackage{graphicx}
\usepackage{calc}
\usepackage{etoolbox}
\usepackage{xspace}
\usepackage[T1]{fontenc} 
\usepackage{textcomp}
\usepackage{ifxetex}
\ifxetex
  \usepackage{fontspec}
  \defaultfontfeatures{Extension = .otf}
\fi
\usepackage{fontawesome}
\usepackage{listings}
\usepackage{nicefrac}
\usepackage{booktabs}
\usepackage{longtable}

\renewcommand{\eqref}[1]{\ref{eq:#1}}

\definecolor{linkcolor}{rgb}{0.1216,0.4667,0.7059}
\definecolor{testpasscolor}{rgb}{0.13333333,0.5254902,0.22745098}
\definecolor{testmissingcolor}{rgb}{1.0,0.88,0.30}
\definecolor{testfailcolor}{rgb}{0.79607843,0.14117647,0.19215686}
\newcommand{\codeicon}{{\color{linkcolor}\faCloudDownload}}
\newcommand{\testmissingicon}{{\color{testmissingcolor}\faQuestion}}
\newcommand{\testpassicon}{{\color{testpasscolor}\faCheck}}
\newcommand{\testfailicon}{{\color{testfailcolor}\faTimes}}
\input{gitlinks}

\newtagform{eqtag}[]{(}{)}
\newcommand{\currentlabel}{None}
\newenvironment{proof}[1]{%
  \ifstrempty{#1}{%
    \renewtagform{eqtag}[]{\raisebox{-0.1em}{{\testmissingicon}}\,(}{)}%
  }{%
    \renewtagform{eqtag}[]{\href{https://github.com/rodluger/mapping_stellar_surfaces/blob/paper2-arxiv/paper2/tests/#1.py}{\raisebox{-0.1em}{\input{tests/#1.tex}}}\,(}{)}%
  }%
  \usetagform{eqtag}%
  \renewcommand{\currentlabel}{#1}
  \align%
}{%
  \endalign%
  \renewtagform{eqtag}[]{(}{)}%
  \usetagform{eqtag}%
  \message{<<<\currentlabel: \theequation>>>}%
}

\newcommand{\oscaption}[2]{\caption{#2 \codelink{#1}}}

\definecolor{codegreen}{rgb}{0,0.6,0}
\definecolor{codegray}{rgb}{0.5,0.5,0.5}
\definecolor{codepurple}{rgb}{0.58,0,0.82}
\definecolor{backcolour}{rgb}{0.95,0.95,0.95}
\lstdefinestyle{mystyle}{
  backgroundcolor=\color{backcolour},
  commentstyle=\color{codegreen},
  keywordstyle=\color{magenta},
  numberstyle=\tiny\color{codegray},
  stringstyle=\color{codepurple},
  basicstyle=\small\ttfamily,
  breakatwhitespace=false,
  breaklines=true,
  captionpos=b,
  keepspaces=true,
  numbers=left,
  numbersep=5pt,
  showspaces=false,
  showstringspaces=false,
  showtabs=false,
  tabsize=2,
  aboveskip=1em,
  belowskip=1em,
  keywords=[2]{map},
  keywordstyle=[2]{\color{black!80!black}},
  upquote=true
}
\renewcommand\quad{\hskip\fontdimen3\font}

\usepackage{stackengine,scalerel}
\def\lnlam{\ThisStyle{\ensurestackMath{\stackon[-2.4\LMpt]{%
        \SavedStyle\lambda}{\kern-.5pt\kern\LMpt\rule{1\LMex}{.25pt+.15\LMpt}}}}}

%% file: gitlinks.tex
\newcommand{\codelink}[1]{\href{https://github.com/rodluger/mapping_stellar_surfaces/blob/paper2-arxiv/paper2/figures/#1.py}{\codeicon}\,\,}
\newcommand{\animlink}[1]{\href{https://github.com/rodluger/mapping_stellar_surfaces/blob/paper2-arxiv/paper2/figures/#1.gif}{\animicon}\,\,}

%% file: style.tex
\usepackage{xifthen}
\usepackage{stackengine}
\usepackage{tabstackengine}
\usepackage{array}
\usepackage{upgreek}
\usepackage[bbgreekl]{mathbbol}
\usepackage{afterpage}
\usepackage[bb=boondox]{mathalpha}

\newcommand{\starry}{\textsf{starry}\xspace}

\newcommand{\cpp}{\textsf{C++}\xspace}
\newcommand{\theano}{\textsf{theano}\xspace}

\newcommand{\starryprocess}{\textsf{starry\_process}\xspace}
\newcommand{\Python}{\textsf{Python}\xspace}

\newcommand{\quadquad}{\quad\quad\quad\quad}

\newcommand{\LMAX}{15\xspace}

\DeclarePairedDelimiter\floor{\lfloor}{\rfloor}

\makeatletter
\def\Ddots{\mathinner{\mkern1mu\raise\p@
                \vbox{\kern7\p@\hbox{.}}\mkern2mu
                \raise4\p@\hbox{.}\mkern2mu\raise7\p@\hbox{.}\mkern1mu}}
\makeatother

\DeclareFontFamily{U}{mathc}{}
\DeclareFontShape{U}{mathc}{m}{it}{<->s*[1.03] mathc10}{}
\DeclareMathAlphabet{\mathscr}{U}{mathc}{m}{it}
\DeclareMathOperator{\imag}{\mathscr{i}}

%% file: tests/tally.tex
17 passed \testpassicon, 0 failed \testfailicon

%% file: table.tex
\begin{center}
    \begin{longtable}{W{c}{1cm} W{l}{9cm} W{l}{2cm}}
        \caption{%
            List of common variables and symbols used throughout this paper.
        }
        \label{tab:variables}
        \\
        \toprule
        \multicolumn{1}{c}{\textbf{Symbol}}
         &
        \multicolumn{1}{c}{\textbf{Description}}
         &
        \multicolumn{1}{c}{\textbf{Reference}}
        \\
        \midrule
        \endfirsthead
        \multicolumn{3}{c}%
        {{\bfseries \tablename\ \thetable{}}. (continued from previous page)}
        \\[0.5em]
        \toprule
        \multicolumn{1}{c}{\textbf{Symbol}}
         &
        \multicolumn{1}{c}{\textbf{Definition}}
         &
        \multicolumn{1}{c}{\textbf{Reference}}
        \\
        \midrule
        \endhead
        \bottomrule
        \endfoot
        \endlastfoot
        $\mathbf{1}$
         & Vector of ones
         & ---
        \\
        $\sim$
         & Denotes a normalized vector-valued random variable
         & \S\ref{sec:gp-norm}
        \\
        $\odot$
         & Elementwise product
         & \S\ref{sec:temporal}
        \\
        $\otimes$
         & Kronecker product
         & \S\ref{sec:temporal}
        \\
        $a$
         & GP hyperparameter: spot latitude shape parameter
         & Appendix~\ref{sec:lat}
        \\
        $\mathbf{a}^\top$
         & Row of the \starry design matrix
         & Equation~(\ref{eq:Arows})
        \\
        $\mathbf{A_1}$
         & \starry change of basis matrix
         & Appendix~\ref{sec:starry}
        \\
        $\pmb{\mathcal{A}}$
         & \starry design matrix
         & Equation~(\ref{eq:Arows})
        \\
        $\alpha$
         & GP hyperparameter: spot latitude shape parameter
         & Appendix~\ref{sec:lat}
        \\
        $b$
         & GP hyperparameter: spot latitude shape parameter
         & Appendix~\ref{sec:lat}
        \\
        $\beta$
         & GP hyperparameter: spot latitude shape parameter
         & Appendix~\ref{sec:lat}
        \\
        $c$
         & GP hyperparameter: spot contrast
         & Appendix~\ref{sec:contrast}
        \\
        $\mathbb{c}$
         & Spot contrast (random variable)
         & Appendix~\ref{sec:contrast}
        \\
        $\Gamma(\cdots)$
         & Gamma function
         & ---
        \\
        $\mathbf{D}_\mathbf{u}$
         & Complex Wigner rotation matrix about an axis $\mathbf{u}$
         & Appendix~\ref{sec:wigner}
        \\
        $\delta(\cdots)$
         & Delta function
         & ---
        \\
        $\delta_{ij}$
         & Kronecker delta
         & ---
        \\
        $\Delta r$
         & GP hyperparameter: spot radius spread
         & Appendix~\ref{sec:size}
        \\
        $\mathrm{E}\big[ \cdots \big]$
         & Expected value
         & Equation~(\ref{eq:mean})
        \\
        $\mathbf{e}_I$
         & First moment integral of the inclination
         & Equation~(\ref{eq:eI})
        \\
        $\mathbf{e}_r$
         & First moment integral of the radius
         & Equation~(\ref{eq:e1})
        \\
        $\mathbf{e}_\phi$
         & First moment integral of the latitude
         & Equation~(\ref{eq:e2})
        \\
        $\mathbf{e}_\lambda$
         & First moment integral of the longitude
         & Equation~(\ref{eq:e3})
        \\
        $\mathbf{e}_c$
         & First moment integral of the contrast
         & Equation~(\ref{eq:e4})
        \\
        $\mathbf{E}_I$
         & Second moment integral of the inclination
         & Equation~(\ref{eq:EI})
        \\
        $\mathbf{E}_r$
         & Second moment integral of the radius
         & Equation~(\ref{eq:E1})
        \\
        $\mathbf{E}_\phi$
         & Second moment integral of the latitude
         & Equation~(\ref{eq:E2})
        \\
        $\mathbf{E}_\lambda$
         & Second moment integral of the longitude
         & Equation~(\ref{eq:E3})
        \\
        $\mathbf{E}_c$
         & Second moment integral of the contrast
         & Equation~(\ref{eq:E4})
        \\
        $\mathbf{f}$
         & Flux vector
         & Equation~(\ref{eq:fAy})
        \\
        $\mathbbb{f}$
         & Flux vector (random variable)
         & Equation~(\ref{eq:fnormal})
        \\
        ${_2}F_1(\cdots)$
         & Gauss hypergeometric function
         & ---
        \\
        $\pmb{\theta}_\bullet$
         & Vector of GP hyperparameters
         & Equation~(\ref{eq:thetaspot})
        \\
        $I$
         & Stellar inclination
         & ---
        \\
        $\mathbb{I}$
         & Stellar inclination (random variable)
         & \S\ref{sec:gp-inc}
        \\
        $J$
         & Jacobian of the spot latitude transform
         & Equation~(\ref{eq:J})
        \\
        $k(\Delta t)$
         & GP kernel function
         & Equation~(\ref{eq:kernel})
        \\
        $K$
         & Number of points in light curve
         & ---
        \\
        $\mathbf{K}$
         & Temporal covariance matrix
         & \S\ref{sec:temporal}
        \\
        $l$
         & Spherical harmonic degree
         & Appendix~\ref{sec:notation}
        \\
        $\mathbf{L}$
         & Limb darkening operator
         & \S\ref{sec:ld}
        \\
        $\mathcal{L}$
         & Likelihood function
         & Equation~(\ref{eq:log-like})
        \\
        %
        %
        %
        %
        \pagebreak 
        $\bblambda$
         & Spot longitude (random variable)
         & Appendix~\ref{sec:lon}
        \\
        $m$
         & Spherical harmonic order
         & Appendix~\ref{sec:notation}
        \\
        $M$
         & Number of light curves in ensemble
         & ---
        \\
        $\mu$
         & Flux GP mean
         & Equation~(\ref{eq:scalar-mean})
        \\
        $\pmb{\mu}$
         & Flux GP mean vector
         & Equation~(\ref{eq:mean_f})
        \\
        $\pmb{\mu}_\mathbf{y}$
         & Spherical harmonic GP mean vector
         & Equation~(\ref{eq:mean_y})
        \\
        $n$
         & GP hyperparameter: number of spots
         & Appendix~\ref{sec:integrals}
        \\
        $\mathbb{n}$
         & Number of spots contrast (random variable)
         & Appendix~\ref{sec:integrals}
        \\
        $\mathcal{N}(\mu, \sigma^2)$
         & Normal distribution: mean $\mu$, variance $\sigma^2$
         & ---
        \\
        $p(\cdots)$
         & Probability, probability density
         & ---
        \\
        $P$
         & Stellar rotation period
         & ---
        \\
        $r$
         & GP hyperparameter: spot radius
         & Appendix~\ref{sec:size}
        \\
        $\mathbf{r}^\top$
         & Integral over unit disk of polynomial basis
         & Appendix~\ref{sec:starry}
        \\
        $\mathbb{r}$
         & Spot radius (random variable)
         & Appendix~\ref{sec:size}
        \\
        $\mathbf{R}_\mathbf{u}$
         & Real wigner rotation matrix about an axis $\mathbf{u}$
         & Appendix~\ref{sec:wigner}
        \\
        $\mathbf{s}$
         & Spherical harmonic expansion of spot
         & Equation~(\ref{eq:sofr})
        \\
        $\sigma_f$
         & Photometric uncertainty
         & \S\ref{sec:calibration}
        \\
        $\sigma_\phi$
         & GP hyperparameter: spot latitude standard deviation
         & Appendix~\ref{sec:lat}
        \\
        $\pmb{\Sigma}$
         & Flux GP covariance matrix
         & Equation~(\ref{eq:cov_f})
        \\
        $\pmb{\Sigma}^\mathbf{(t)}$
         & Flux GP covariance matrix w/ temporal evolution
         & Equation~(\ref{eq:cov_f})
        \\
        $\tilde{\pmb{\Sigma}}$
         & Flux GP covariance matrix (normalized processs)
         & Equation~(\ref{eq:SigmaTilde})
        \\
        $\pmb{\Sigma}_\mathbf{y}$
         & Spherical harmonic GP covariance matrix
         & Equation~(\ref{eq:cov_y})
        \\
        $\pmb{\Sigma}_\mathbf{y}^\mathbf{(t)}$
         & Spherical harmonic GP covariance w/ temporal evolution
         & Equation~(\ref{eq:cov_y})
        \\
        $t$
         & time
         & ---
        \\
        $\tau$
         & GP hyperparameter: timescale
         & \S\ref{sec:temporal}
        \\
        $\mathbf{u}$
         & Limb darkening coefficient vector
         & Appendix~\ref{sec:ld}
        \\
        $u_1, u_2$
         & Linear and quadratic limb darkening coefficients
         & Appendix~\ref{sec:ld}
        \\
        $\mathbf{U}$
         & Complex-to-real basis change operator (Wigner matrices)
         & Appendix~\ref{sec:wigner}
        \\
        $\mathcal{U}(a, b)$
         & Uniform distribution between $a$ and $b$
         & ---
        \\
        $\mathbbb{x}$
         & Random vector of spot properties
         & Equation~(\ref{eq:x})
        \\
        $\mathbf{y}$
         & Spherical harmonic coefficient vector
         & Equation~(\ref{eq:fAy})
        \\
        $\mathbbb{y}$
         & Spherical harmonic coefficient vector (random variable)
         & Equation~(\ref{eq:RRs})
        \\
        $\mu_\phi$
         & GP hyperparameter: spot latitude mode
         & Appendix~\ref{sec:lat}
        \\
        $\bbphi$
         & Spot latitude (random variable)
         & Appendix~\ref{sec:lat}
        \\
        $z$
         & GP normalization number
         & Equation~(\ref{eq:z})
        \\
    \end{longtable}
\end{center}

%% file: ms.bbl
\begin{thebibliography}{}
\expandafter\ifx\csname natexlab\endcsname\relax\def\natexlab#1{#1}\fi
\providecommand{\url}[1]{\href{#1}{#1}}

\bibitem[{{Aigrain} {et~al.}(2016){Aigrain}, {Parviainen}, \&
  {Pope}}]{Aigrain2016}
{Aigrain}, S., {et~al.} 2016, \mnras, 459, 2408

\bibitem[{{Ambikasaran} {et~al.}(2015){Ambikasaran}, {Foreman-Mackey},
  {Greengard}, {Hogg}, \& {O'Neil}}]{Ambikasaran2015}
{Ambikasaran}, S., {et~al.} 2015, IEEE Transactions on Pattern Analysis and
  Machine Intelligence, 38, 252

\bibitem[{{Angus} {et~al.}(2018){Angus}, {Morton}, {Aigrain}, {Foreman-Mackey},
  \& {Rajpaul}}]{Angus2018}
{Angus}, R., {et~al.} 2018, \mnras, 474, 2094

\bibitem[{{Angus} {et~al.}(2019){Angus}, {Morton}, {Foreman-Mackey}, {van
  Saders}, {Curtis}, {Kane}, {Bedell}, {Kiman}, {Hogg}, \&
  {Brewer}}]{Angus2019}
---. 2019, \aj, 158, 173

\bibitem[{{Barnes} {et~al.}(2001){Barnes}, {Sofia}, \&
  {Pinsonneault}}]{Barnes2001}
{Barnes}, S., {et~al.} 2001, \apj, 548, 1071

\bibitem[{{Basri} \& {Shah}(2020)}]{Basri2020}
{Basri}, G., \& {Shah}, R. 2020, \apj, 901, 14

\bibitem[{Beaumont(2019)}]{Beaumont2019}
Beaumont, M.~A. 2019, Annual Review of Statistics and Its Application, 6, 379

\bibitem[{{Blei} {et~al.}(2016){Blei}, {Kucukelbir}, \& {McAuliffe}}]{Blei2016}
{Blei}, D.~M., {et~al.} 2016, arXiv e-prints, arXiv:1601.00670

\bibitem[{{Brewer} \& {Stello}(2009)}]{BrewerStello2009}
{Brewer}, B.~J., \& {Stello}, D. 2009, \mnras, 395, 2226

\bibitem[{{Cantiello} \& {Braithwaite}(2019)}]{Cantiello2019}
{Cantiello}, M., \& {Braithwaite}, J. 2019, \apj, 883, 106

\bibitem[{Collado {et~al.}(1989)Collado, Rico, L\'opez, Paniagua, \&
  Ram\'irez}]{AlvarezCollado1989}
Collado, J. R.~A., {et~al.} 1989, Computer Physics Communications, 52, 323

\bibitem[{{Damasso} {et~al.}(2019){Damasso}, {Pinamonti}, {Scandariato}, \&
  {Sozzetti}}]{Damasso2019}
{Damasso}, M., {et~al.} 2019, \mnras, 489, 2555

\bibitem[{{Duane} {et~al.}(1987){Duane}, {Kennedy}, {Pendleton}, \&
  {Roweth}}]{Duane1987}
{Duane}, S., {et~al.} 1987, Physics Letters B, 195, 216

\bibitem[{{Feroz} {et~al.}(2009){Feroz}, {Hobson}, \& {Bridges}}]{Feroz2009}
{Feroz}, F., {et~al.} 2009, \mnras, 398, 1601

\bibitem[{{Foreman-Mackey} {et~al.}(2017){Foreman-Mackey}, {Agol},
  {Ambikasaran}, \& {Angus}}]{ForemanMackey2017}
{Foreman-Mackey}, D., {et~al.} 2017, \aj, 154, 220

\bibitem[{{Fuller} {et~al.}(2015){Fuller}, {Cantiello}, {Stello}, {Garcia}, \&
  {Bildsten}}]{Fuller2015}
{Fuller}, J., {et~al.} 2015, Science, 350, 423

\bibitem[{Gilbertson {et~al.}(2020)Gilbertson, Ford, Jones, \&
  Stenning}]{Gilbertson2020}
Gilbertson, C., {et~al.} 2020, The Astrophysical Journal, 905, 155.
\newblock \url{https://doi.org/10.3847/1538-4357/abc627}

\bibitem[{Gough \& Tayler(1966)}]{Gough1966}
Gough, D.~O., \& Tayler, R.~J. 1966, Monthly Notices of the Royal Astronomical
  Society, 133, 85.
\newblock \url{https://doi.org/10.1093/mnras/133.1.85}

\bibitem[{{Gully-Santiago} {et~al.}(2017){Gully-Santiago}, {Herczeg},
  {Czekala}, {Somers}, {Grankin}, {Covey}, {Donati}, {Alencar}, {Hussain},
  {Shappee}, {Mace}, {Lee}, {Holoien}, {Jose}, \& {Liu}}]{Gully2017}
{Gully-Santiago}, M.~A., {et~al.} 2017, \apj, 836, 200

\bibitem[{{Guo} {et~al.}(2018){Guo}, {Gully-Santiago}, \& {Herczeg}}]{Guo2018}
{Guo}, Z., {et~al.} 2018, \apj, 868, 143

\bibitem[{{Haywood} {et~al.}(2014){Haywood}, {Collier Cameron}, {Queloz},
  {Barros}, {Deleuil}, {Fares}, {Gillon}, {Lanza}, {Lovis}, {Moutou}, {Pepe},
  {Pollacco}, {Santerne}, {S{\'e}gransan}, \& {Unruh}}]{Haywood2014}
{Haywood}, R.~D., {et~al.} 2014, \mnras, 443, 2517

\bibitem[{{Hoffman} \& {Gelman}(2011)}]{Hoffman2011}
{Hoffman}, M.~D., \& {Gelman}, A. 2011, arXiv e-prints, arXiv:1111.4246

\bibitem[{{Ireland} \& {Browning}(2018)}]{Ireland2018}
{Ireland}, L.~G., \& {Browning}, M.~K. 2018, \apj, 856, 132

\bibitem[{{Jones} {et~al.}(2017){Jones}, {Stenning}, {Ford}, {Wolpert},
  {Loredo}, {Gilbertson}, \& {Dumusque}}]{Jones2017}
{Jones}, D.~E., {et~al.} 2017, arXiv e-prints, arXiv:1711.01318

\bibitem[{{Kervella} {et~al.}(2017){Kervella}, {Bigot}, {Gallenne}, \&
  {Th{\'e}venin}}]{Kervella2017}
{Kervella}, P., {et~al.} 2017, \aap, 597, A137

\bibitem[{{Kipping}(2013)}]{Kipping2013}
{Kipping}, D.~M. 2013, \mnras, 435, 2152

\bibitem[{{Kucukelbir} {et~al.}(2016){Kucukelbir}, {Tran}, {Ranganath},
  {Gelman}, \& {Blei}}]{Kucukelbir2016}
{Kucukelbir}, A., {et~al.} 2016, arXiv e-prints, arXiv:1603.00788

\bibitem[{{Luger}(2021)}]{Luger2021}
{Luger}, R. 2021, in preparation

\bibitem[{{Luger} {et~al.}(2019){Luger}, {Agol}, {Foreman-Mackey}, {Fleming},
  {Lustig-Yaeger}, \& {Deitrick}}]{Luger2019}
{Luger}, R., {et~al.} 2019, \aj, 157, 64

\bibitem[{{Luger} {et~al.}(2016){Luger}, {Agol}, {Kruse}, {Barnes}, {Becker},
  {Foreman-Mackey}, \& {Deming}}]{Luger2016}
---. 2016, \aj, 152, 100

\bibitem[{{Luger} {et~al.}(2021{\natexlab{a}}){Luger}, {Foreman-Mackey}, \&
  {Hedges}}]{JOSSPaper}
---. 2021{\natexlab{a}}, arXiv e-prints, arXiv:2102.01774

\bibitem[{{Luger} {et~al.}(2021{\natexlab{b}}){Luger}, {Foreman-Mackey},
  {Hedges}, \& {Hogg}}]{PaperI}
---. 2021{\natexlab{b}}, arXiv e-prints, arXiv:2102.00007

\bibitem[{{Luger} {et~al.}(2017{\natexlab{a}}){Luger}, {Lustig-Yaeger}, \&
  {Agol}}]{Luger2017}
---. 2017{\natexlab{a}}, \apj, 851, 94

\bibitem[{{Luger} {et~al.}(2017{\natexlab{b}}){Luger}, {Sestovic}, {Kruse},
  {Grimm}, {Demory}, {Agol}, {Bolmont}, {Fabrycky}, {Fernandes}, {Van Grootel},
  {Burgasser}, {Gillon}, {Ingalls}, {Jehin}, {Raymond}, {Selsis}, {Triaud},
  {Barclay}, {Barentsen}, {Howell}, {Delrez}, {de Wit}, {Foreman-Mackey},
  {Holdsworth}, {Leconte}, {Lederer}, {Turbet}, {Almleaky}, {Benkhaldoun},
  {Magain}, {Morris}, {Heng}, \& {Queloz}}]{Luger2017b}
---. 2017{\natexlab{b}}, Nature Astronomy, 1, 0129

\bibitem[{Miesch \& Toomre(2009)}]{Miesch2009}
Miesch, M.~S., \& Toomre, J. 2009, Annual Review of Fluid Mechanics, 41, 317.
\newblock \url{https://doi.org/10.1146/annurev.fluid.010908.165215}

\bibitem[{{Morris}(2020{\natexlab{a}})}]{Morris2020b}
{Morris}, B. 2020{\natexlab{a}}, The Journal of Open Source Software, 5, 2103

\bibitem[{{Morris}(2020{\natexlab{b}})}]{Morris2020}
{Morris}, B.~M. 2020{\natexlab{b}}, \apj, 893, 67

\bibitem[{{Morris} {et~al.}(2018){Morris}, {Agol}, {Davenport}, \&
  {Hawley}}]{Morris2018}
{Morris}, B.~M., {et~al.} 2018, \apj, 857, 39

\bibitem[{{Perger} {et~al.}(2020){Perger}, {Anglada-Escud{\'e}}, {Ribas},
  {Rosich}, {Herrero}, \& {Morales}}]{Perger2020}
{Perger}, M., {et~al.} 2020, arXiv e-prints, arXiv:2012.01862

\bibitem[{Price {et~al.}(2018)Price, Drovandi, Lee, \& Nott}]{Price2018}
Price, L.~F., {et~al.} 2018, Journal of Computational and Graphical Statistics,
  27, 1

\bibitem[{{Rackham} {et~al.}(2018){Rackham}, {Apai}, \&
  {Giampapa}}]{Rackham2018}
{Rackham}, B.~V., {et~al.} 2018, \apj, 853, 122

\bibitem[{{Rajpaul} {et~al.}(2015){Rajpaul}, {Aigrain}, {Osborne}, {Reece}, \&
  {Roberts}}]{Rajpaul2015}
{Rajpaul}, V., {et~al.} 2015, \mnras, 452, 2269

\bibitem[{Rasmussen \& Williams(2005)}]{RasmussenWilliams2005}
Rasmussen, C.~E., \& Williams, C. K.~I. 2005, Gaussian Processes for Machine
  Learning (Adaptive Computation and Machine Learning) (The MIT Press)

\bibitem[{{Robertson} {et~al.}(2020){Robertson}, {Stefansson}, {Mahadevan},
  {Endl}, {Cochran}, {Beard}, {Bender}, {Diddams}, {Duong}, {Ford}, {Fredrick},
  {Halverson}, {Hearty}, {Holcomb}, {Juan}, {Kanodia}, {Lubin}, {Metcalf},
  {Monson}, {Ninan}, {Palafoutas}, {Ramsey}, {Roy}, {Schwab}, {Terrien}, \&
  {Wright}}]{Robertson2020}
{Robertson}, P., {et~al.} 2020, \apj, 897, 125

\bibitem[{{Schuessler} {et~al.}(1996){Schuessler}, {Caligari}, {Ferriz-Mas},
  {Solanki}, \& {Stix}}]{Schuessler1996}
{Schuessler}, M., {et~al.} 1996, \aap, 314, 503

\bibitem[{Sikora {et~al.}(2018)Sikora, Wade, Power, \& Neiner}]{Sikora2018}
Sikora, J., {et~al.} 2018, Monthly Notices of the Royal Astronomical Society,
  483, 3127.
\newblock \url{https://doi.org/10.1093/mnras/sty2895}

\bibitem[{{Skilling}(2004)}]{Skilling2004}
{Skilling}, J. 2004, in American Institute of Physics Conference Series, Vol.
  735, Bayesian Inference and Maximum Entropy Methods in Science and
  Engineering: 24th International Workshop on Bayesian Inference and Maximum
  Entropy Methods in Science and Engineering, ed. R.~{Fischer}, R.~{Preuss}, \&
  U.~V. {Toussaint}, 395--405

\bibitem[{Skilling(2006)}]{Skilling2006}
Skilling, J. 2006, Bayesian Anal., 1, 833

\bibitem[{{Solanki} {et~al.}(2006){Solanki}, {Inhester}, \&
  {Sch{\"u}ssler}}]{Solanki2006}
{Solanki}, S.~K., {et~al.} 2006, Reports on Progress in Physics, 69, 563

\bibitem[{{Speagle}(2020)}]{Speagle2020}
{Speagle}, J.~S. 2020, \mnras, 493, 3132

\bibitem[{{Turcotte}(2003)}]{Turcotte2003}
{Turcotte}, S. 2003, in Astronomical Society of the Pacific Conference Series,
  Vol. 305, Magnetic Fields in O, B and A Stars: Origin and Connection to
  Pulsation, Rotation and Mass Loss, ed. L.~A. {Balona}, H.~F. {Henrichs}, \&
  R.~{Medupe}, 199

\bibitem[{{Vanderburg} {et~al.}(2016){Vanderburg}, {Plavchan}, {Johnson},
  {Ciardi}, {Swift}, \& {Kane}}]{Vanderburg2016}
{Vanderburg}, A., {et~al.} 2016, \mnras, 459, 3565

\bibitem[{Wandelt(2012)}]{Wandelt2012}
Wandelt, B. 2012, Gaussian Random Fields in Cosmostatistics

\bibitem[{{Weber} \& {Browning}(2016)}]{Weber2016}
{Weber}, M.~A., \& {Browning}, M.~K. 2016, \apj, 827, 95

\bibitem[{Wood(2010)}]{Wood2010}
Wood, S.~N. 2010, Nature, 466, 1102

\bibitem[{{Yadav} {et~al.}(2015){Yadav}, {Gastine}, {Christensen}, \&
  {Reiners}}]{Yadav2015}
{Yadav}, R.~K., {et~al.} 2015, \aap, 573, A68

\end{thebibliography}
